\documentclass[structabstract]{aa} 
\usepackage{natbib}
\usepackage{array}
\bibpunct{(}{)}{;}{a}{}{,} 
\usepackage{graphicx}
\usepackage{txfonts}
\usepackage{lscape}
\usepackage{caption}
\usepackage{longtable,amsmath} 
\usepackage{color}
\usepackage{amsmath}
\usepackage{amssymb}	
\usepackage{float}
\usepackage{float}
\usepackage{subfig}
\usepackage[titletoc]{appendix}
 \usepackage{booktabs,multirow,tabularx}
\usepackage{amsmath}
\usepackage{amsfonts}
\usepackage{amssymb}
\usepackage{titlesec}
\usepackage[dvipsnames]{xcolor}
\usepackage{fancyhdr}
\usepackage{amsmath,amssymb}
\usepackage{sfmath}
\usepackage[]{caption}
 \usepackage{epstopdf}
\usepackage{natbib,twoopt}
 \bibpunct{(}{)}{;}{a}{}{,} 
 \newcommandtwoopt{\citeads}[3][][]{\href{http://adsabs.harvard.edu/abs/#3}{\citealp[#1][#2]{#3}}}
 \newcommandtwoopt{\citepads}[3][][]{\href{http://adsabs.harvard.edu/abs/#3}{\citep[#1][#2]{#3}}}
 \newcommandtwoopt{\citetads}[3][][]{\href{http://adsabs.harvard.edu/abs/#3}{\citet[#1][#2]{#3}}}
 \newcommandtwoopt{\citeyearads}[3][][]{\href{http://adsabs.harvard.edu/abs/#3}{\citeyear[#1][#2]{#3}}}

\usepackage[ 
  breaklinks=true,
  colorlinks=true,         
   urlcolor=pdfurlcolor,    
   filecolor=pdffilecolor,  
   linkcolor=pdflinkcolor,  
   citecolor=pdfcitecolor,  %
   raiselinks=true,                     
   breaklinks,              
   backref=page,            
   pagebackref=false,       
   verbose,                                    
   hyperindex=true,         
   linktocpage=false,        
   hyperfootnotes=false,     
   bookmarks=true,          
   bookmarksopenlevel=1,    
   bookmarksopen=true,      
   bookmarksnumbered=true,  
   bookmarkstype=toc,       
   plainpages=false,        
   pageanchor=true,         
   pdfauthor={Sara Saeedi},      
   pdfdisplaydoctitle=true, 
   pdfstartview=FitV,       
   pdfpagemode=UseOutlines, 
   pdfpagelabels=true,           
   pdfpagelayout=TwoPageRight, 
]{hyperref} 
  
\definecolor{pdfurlcolor}{rgb}{0,0,0.6}
\definecolor{pdffilecolor}{rgb}{0.7,0,0}
\definecolor{pdflinkcolor}{rgb}{0,0,0.6}
\definecolor{pdfcitecolor}{rgb}{0,0,0.6}

\newcommand{\xmm}{{\it XMM-Newton}}
\newcommand{\erosita}{{\it eROSITA}}
\newcommand{\rosat}{{\it ROSAT}}
\newcommand{\einstein}{{\it EINSTEIN}}
\newcommand{\chandra}{{\it Chandra}}

\begin{document}

  \title{\erosita\, study of the 47~Tucanae globular cluster\thanks{Based on observations obtained with \erosita.}
}
   \subtitle{}
    \titlerunning{\erosita\, study of the 47~Tucanae globular cluster}
   \authorrunning{S.~Saeedi et al}
   \author{Sara Saeedi\inst{1}
          \and
          Teng Liu\inst{2}
          \and
           Jonathan Knies\inst{1}
          \and
           Manami Sasaki\inst{1}
           \and
           Werner Becker\inst{2}
          \and
           Esra Bulbul\inst{2}
           \and
          Konrad Dennerl\inst{2}
          \and
          Michael Freyberg\inst{2}
          \and
          Roman Laktionov\inst{1}
          \and
          Andrea Merloni\inst{2}
       }

\institute{\\\inst{1}  Dr. Karl Remeis-Sternwarte, Erlangen Centre for Astroparticle Physics, Friedrich-Alexander-Universit\"at Erlangen-N\"urnberg, Sternwartstrasse 7, 96049, Bamberg, Germany\\
    \email{sara.saeedi@fau.de}\\
    \inst{2} Max-Planck-Institut f\"ur extraterrestrische Physik, Giessenbachstraße 1, 85748 Garching bei M\"unchen, Germany}
 

  \date{Received DATE; accepted DATE}

\abstract{}{}{}{}{} 
 
  \abstract
   {}
   {We present the results of the analysis of five observations of the globular clutser 47~Tucanae\,(47~Tuc) with eROSITA (extended Roentgen Survey with an Imaging Telescope Array) on board Spektrum-Roentgen-Gamma (Spektr-RG, SRG). The aim of the work is the study of the X-ray population in the field of one of the most massive globular clusters in our Milky Way. We focused on the classification of point-like sources in the field of 47~Tuc. The unresolved dense core of 47~Tuc (1$\arcmin$.7 radius) and also the sources, which show extended emission are excluded in this study.}
   {We applied different methods of X-ray spectral and timing analysis together with multi wavelength studies for the classification of the X-rays sources in the field of 47~Tuc. }
   {We detected 888 point-like sources in the energy range of 0.2--5.0~keV. We identified  126 background AGNs and  25 foreground stars. One of the foreground stars is classified as a variable M~dwarf.  We also classified  14 X-ray sources as members of 47~Tuc, including   1 symbiotic stars,  2  quiescent low mass X-ray binaries, and   4 cataclysmic variable. There are also  5 X-ray sources, which can either be a cataclysmic variable or a contact binary, and also  1 X-ray sources which can be an active binary (Type RS CVn). We identified one X-ray binary, which belongs to the Small Magellanic Cloud. Moreover, we calculated the X-ray luminosity function of 47~Tuc.  No significant population that seems to belong to the globular cluster has been observed in the energy range of 0.5--2.0~keV using \erosita\, observations.}  
   {}


   \maketitle
%

\section{Introduction}
\label{intro}
Globular clusters\,(GCs) are known as spherical shaped, compact, old, and bright accumulations of stars, which are mainly observed in the halo, thick disk, and the bulge of the Galaxy, while they are not present in the thin disk \citep[e.g,][]{2019A&ARv..27....8G}. 
The dynamical structure of the GCs is ideal for the formation of a large number of binary systems, especially short-period close binaries.  Several studies \citep[e.g,][]{2003A&A...403L..11G,2003ApJ...591L.131P,2003ApJ...598..501H} have shown that there is a significant correlation between the number of low luminosity X-ray sources in GCs and the encounter frequency in GCs rather than with the mass of GCs. This means that X-ray sources with $L_{x}>10^{31}$ erg\,s$^{-1}$ in dense clusters are to a large extent dynamically formed. As the X-ray studies show the lower-density globular clusters are more dominated by BY Dra and RS CVn systems \citep[e.g,][]{2004ApJ...609..755B, 2008A&A...488..921B, 2018ApJ...869...52C, 2019MNRAS.483..315B, 2020MNRAS.492.5684H}. So far, various types of X-ray binary systems have been frequently observed in GCs. The observation of the bright persistent low max X-ray binaries\,(LMXBs; $L_{x}$>$10^{35}$\,erg\,s$^{-1}$) in GCs started since the earliest X-ray missions \citep[e.g, Uhuru,][]{1977ApJ...211L...9C}. Since then it was suggested that the mass-normalized formation rate of LMXBs in GCs is orders of magnitudes higher than in the Galactic disk due to the high stellar densities in the core of the GCs \citep{1975ApJ...199L.143C}. Later studies confirmed the presence of  eight persistently luminous LMXBs in the GCs  \citep[][]{2014ApJ...780..127B} and additional transient LMXBs have been detected in outbursts  \citep[e.g,][]{2008ApJ...674L..45A, 2010ApJ...714..894H, 2017A&A...598A..34S, 2018A&A...610L...2S, 2018ATel11598....1H}. The main population of X-ray sources in the GCs is that of the less luminous X-ray sources\,( $L_{x}\lessapprox10^{33}$\,erg\,s$^{-1}$), which are potentially a mixture of quiescent LMXBs, different types of accreting white dwarfs\,(AWDs), radio millisecond pulsars\,(MSP), and magnetically active binary systems. 

In this work, we study the population of X-ray sources in the field of the Galactic globular cluster 47~Tucanae (47 Tuc) observed with \erosita. 47~Tuc (also known as NGC\,104; RA: 00h24m05.36s, DEC: --72$^{\circ}$04$\arcmin$53.2$\arcsec$) with a half mass radius of 2.$\arcmin$76 and a mass of 7.10$\times10^{5}$\,M$_{\sun}$ is one the most massive GC in the Galaxy \citep{2010MNRAS.406.2000M}.  \citet{2013Natur.500...51H} measured an age of 9.7$\pm$0.4~Gyr and a metallicity of [Fe/H]=--0.75 for 47~Tuc. The most updated distance measurement using parallaxes from Gaia\,(2nd data release) yields a distance of 4.45$\pm$0.01$\pm$0.12 for 47~Tuc \citep{2018ApJ...867..132C}. The fist identification of the X-ray sources in the core of 47~Tuc has been performed using the data of \einstein\,\citep[e.g,][]{1983ApJ...267L..83H, 1992Natur.360...46P, 1989A&A...214..113A} and  \rosat, observatories \citep[][]{1994A&A...288..466H, 1998A&A...336..895V}. Later studies using the high resolution cameras of \chandra\,  identified more than one hundred X-ray sources in the core of 47 Tuc, including fifteen X-ray counterparts of the known radio millisecond pulsar at that time \citep[][]{2001Sci...292.2290G}. \citet{2003ApJ...596.1177E} performed the first deep optical/X-ray study using the data of \chandra\, and {\it Hubble Space Telescope}. In their study, optical counterparts of cataclysmic variables and active binaries have been found. Using deeper data of \chandra, \citet{2005ApJ...625..796H} published a catalogue of three hundred X-ray sources within the half-mass radius and  presented a classification of different types of X-ray sources. \citet{2017MNRAS.472.3706B} studied the sources of the core of 47~Tuc within a radius of 2$\arcmin$.7 using  \chandra\, observations and performed X-ray spectral analysis for known MSPs identified in a radio survey \citep{2016MNRAS.462.2918R}. They also reported the classification of five active binary systems in  47~Tuc. Moreover, recently, \citet{2019ApJ...876...59C} studied the distribution of both faint and bright X-ray sources within the radius of 7$\arcmin$.5. (see Fig.~\ref{rgb-image}). 
For the first time, \erosita\, has provided X-ray data of the field around 47~Tuc with in a large area of 40$\arcmin$ radius, which enables us to perform the analysis of the X-ray sources of 47~Tuc  in a noticeably larger region than in \chandra\, studies as  mentioned above. However, we had to ignore the central $1\arcmin.7$ circular region of 47~Tuc in the \erosita\, data since the emission is spatially unresolved. This paper reports the details of the X-ray analysis along with multi-wavelength studies (mainly in optical, infrared and near infrared), aiming at the classification of X-ray sources in the field of 47~Tuc. In Section \ref{data-ana}, we describe the data reduction, source detection, and the source catalogue preparation. In Sections \ref{multi-sec} and \ref{x-ana-sec}, we present the multi-wavelength studies and the X-ray analysis, respectively, which are used to classify the X-ray sources. In Section \ref{diss-sec}, we discuss the details of the classification of detected sources in the field of 47~Tuc.

\section{\erosita\, data analysis}
\label{data-ana}
\subsection{Data reduction and source detection}
We have analysed 5 observations of \erosita\, taken in the Calibration and Performance Verification \citep[CalPV,][]{2021A&A...647A...1P} phase in 2019. The details of the observations, which are sorted by  date are shown in Table~\ref{obs-data}. Data reduction and source detection were performed using the the \erosita\, Science Analysis Software System \texttt{eSASSusers$\_$201009} (Brunner et al.,2021, A$\&$A, submitted). 

\begin{figure*}[!ht]
\includegraphics[trim={1.cm 6.2cm 1.0cm 6.7cm},clip, width=1.0\textwidth]{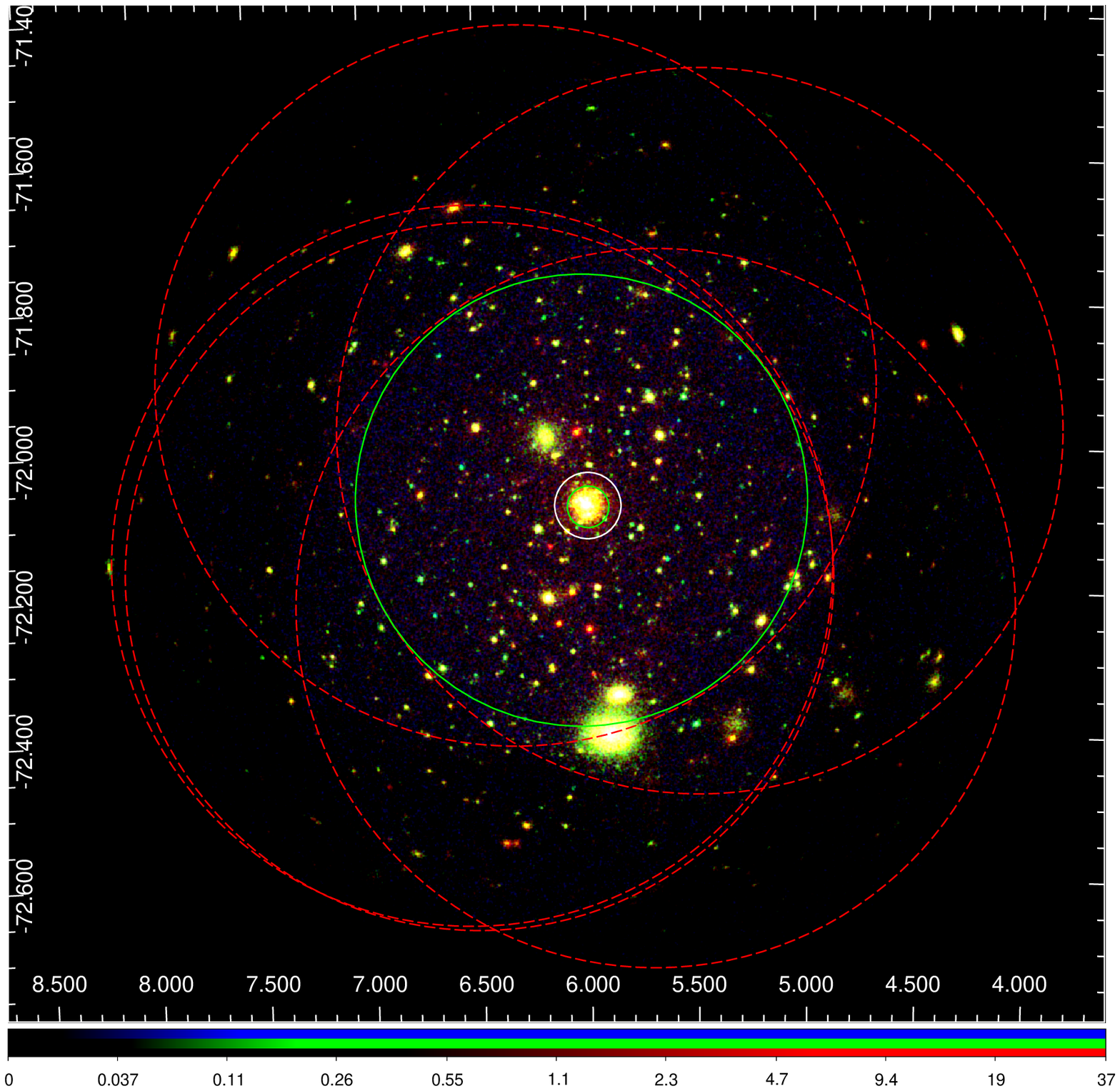}
\caption{Combined X-ray image of \erosita\, observations in the field of 47~Tuc with a total radius of 42$\arcmin$. In this work all the visible point-like sources (888 sources) are studied. The half-mass radius of 47~Tuc is shown within the hard white region. The dashed red circles show the regions observed by five \erosita\, observations. The larger and smaller hard green regions with radii of 18$\arcmin$.8 and 1$\arcmin$.7, show the area, which is covered by all observations and the extent of the unresolved emission from the center of 47~Tuc, respectively. The area between these two regions have been used to calculate the X-ray luminosity function (see Sect.~\ref{xlf-sec}).\label{rgb-image}}
\end{figure*}

 In this work, the source detection was run only on single observations and in each observation the data of all 7 telescope modules (TMs) of \erosita\, have been used in this work \citep{2021A&A...647A...1P}. The light curve of the event files have been used to filter the possible soft proton flares of the observations by a threshold of 30 cts\,s$^{-1}$\,deg$^{-2}$. Table~\ref{obs-data} shows the sum of good time intervals for each observation. The detection process has been run over the event files of the observations in four energy bands of 0.2--0.6~keV, 0.6--1.1~keV, 1.1--2.3~keV, and 2.3--5.0~keV, where the fourth band is considered as the hard band of \erosita\, \citep[e.g,][]{2021arXiv210614517B}.  Since the effective area of \erosita\, noticeably decreases above 2.3~keV  \citep{2012arXiv1209.3114M}, the majority of the sources can not be detected significantly above 2.3~keV.  We selected a minimum maximum likelihood\,($L$) of 10 for the source detection (\texttt{ermldet} task in {\tt eSASS}), which is equivalent to $>4\sigma$ significance according to the probability of Poisson random fluctuations of the counts\,($p$) detection minimum likelihood $L$=--ln$(p)$. 
\begin{table}
 \centering
\hspace{-0.5cm} \caption{\erosita\, observations of 47~Tuc \label{obs-data}}
     \begin{tabular}{cccc}
\hline\hline
OBS-N0 & OBS-ID & OBS-Date &  {EXP.T${^\ast}$ (ks)} \\
\hline
        1&  700012  &  2019-09-28  &    19.5 \\
       2&  700011  &  2019-11-01  &   25.8   \\
        3&  700163  &  2019-11-02  &  25.3    \\
       4&  700013  & 2019-11-02 &  25.2  \\
       5&  700014  &  2019-11-02  & 25.2     \\
      \hline
        \multicolumn{4}{l}{$\ast$:Net exposure time of observations.}\\
     \end{tabular}
\end{table}
 In this work, we study the point-like sources with an extent likelihood of 0 and exclude all the extended sources, which are probable candidates for, e.g., galaxy clusters, diffuse emission, bubble-like structures, etc. and will be studied in more details in future publications.
Figure.~\ref{rgb-image} shows the mosaic image of five observations of \erosita\, of the field of 47~Tuc.
With \erosita, one bright source is detected at the position of the centre of 47 Tuc, which was resolved into multiple sources with  
\chandra\, \citep{2001Sci...292.2290G}. This area was excluded in this study.

\subsection{Astrometric correction}
 The astrometric corrections of the 47~Tuc observations were calculated by Liu et al. (in prep.) in two steps.
First, sources were detected  in each observation and  the relative correction of the coordinates with respect to OBS3 was calculated for the other observations.
To calculate the corrections, only bright point sources with detection likelihood $>12$ within an off-axis angle of 25\arcmin\ were used, thus excluding  sources with poor positional uncertainties.
The uncertainties of the required corrections were calculated through bootstrapping. It was found that a coordinate shift ($\Delta$RA, $\Delta$DEC) is sufficient and a rotation correction is not needed.
The events of all the observations were corrected for the relative astrometry and  merged afterwards.
In the second step, the source catalogue from the merged data was matched to the CatWISE 2020 catalogue \citep{2021ApJS..253....8M}.
Based on the highly reliable positions of the CatWISE counterparts of the X-ray sources, a second-pass correction was calculated.
The total corrections applied to each of the 47~Tuc observations are listed in Table~\ref{offset}. We applied the same corrections to our  catalogue.

\subsection{Source Catalogue}
\label{source-cata-sec}
The final catalogue of point-like sources in the field of 47~Tuc is obtained by cross-checking all detected sources between the five observations. If sources, which have been detected in at least two observations, were closer to each other than the 3$\sigma$ positional errors, they are considered as the same source.  Sources, which were detected only in one of the observations near gaps or edges of the CCD chips or could be recognised as hot pixels, were removed from the source list. Table~\ref{catalogue-x-ray} presents the final list of 888 X-ray sources in the field of 47~Tuc. The catalogue lists source ID, RA, Dec, positional uncertainty, flux of the source in different observations, hardness ratio, variability factor, and the class if a source was classified. 
The ID of the sources in Table~\ref{catalogue-x-ray} is used to present the source in this work. 

\begin{table}
 \centering
  \hspace{-0.5cm} \caption{  Offsets of the \erosita\, observations \label{offset}}
     \begin{tabular}{ccc}
\hline\hline
OBS-N0 &   $\Delta$RA ($\arcsec$) & $\Delta$DEC ($\arcsec$) \\
\hline
\vspace{1mm}
OBS1        &  12.6$^{+1.27}_{-1.34}$  &  1.44$^{+0.49}_{-0.40}$  \\
\vspace{1mm}
OBS2        & 7.0$^{+1.44}_{-1.15}$  &   2.91$^{+0.47}_{-0.47}$  \\
\vspace{1mm}
OBS3        & 11.72$^{+0.30}_{-0.34}$  & 0.65$^{+0.15}_{-0.10}$  \\  
\vspace{1mm}
OBS4        &  11.47$^{+1.27}_{-1.29}$ &   4.02$^{+0.44}_{-0.30}$ \\
\vspace{1mm}
OBS5        &  7.21$^{+0.92}_{-1.06}$ &   0.84$^{+0.50}_{-0.50}$ \\  
      \hline
        \multicolumn{3}{l}{The astrometric correction from Liu et al.,\,(in prep.)}\\
   \end{tabular}
\end{table}

\section{X-ray analysis}
\label{x-ana-sec}
To extract  the light curves and the spectra of the sources we used \texttt{eSASS/srctool-V.1.61};  (Brunner et al.,2021, A$\&$A, submitted).
\subsection{X-ray timing analysis}
\label{X-time}
We carried out X-ray timing analysis for both short-term variability (periodicity and pulsation studies) and  long-term variability. For all unknown sources with  counts > 100, which have not been confirmed as foreground stars or AGNs in available catalogues, we searched for  pulsation signals using the pulsation  Z$^{2}_{n}$ test \citep{1983A&A...128..245B, 1988A&A...201..194B}. 
For unknown sources with counts\,>\,300 in each observation, we extracted the light curves of five observations and applied the Lomb-Scargle technique \citep{1982ApJ...263..835S} to find a signal of pulsation and or periodicity. We could not find any significant pulsation or periodicity in the X-ray data of bright sources,  which are candidates of neither foreground star nor background AGN.

\begin{figure}
  \centering
\includegraphics[clip, trim={1.6cm 0.cm 0.3cm 0.3cm},width=0.45\textwidth]{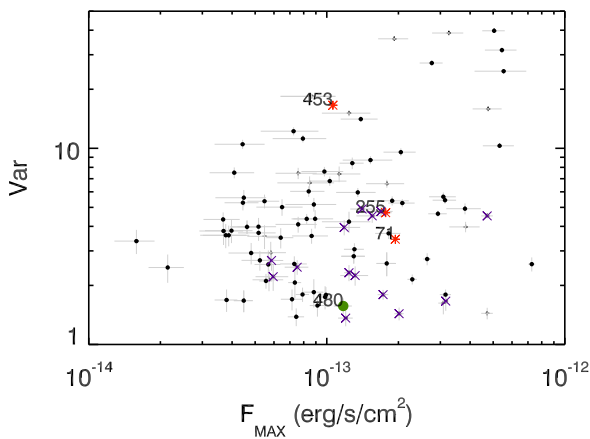}
\caption{Variability factor of sources with significant variability (S>3) in the energy band of 0.2–5.0 keV plotted versus the maximum flux. The symbols characterise foreground stars\,(\textcolor{red}{\large $\ast$}), background objects \,(\textcolor{violet}{\large $\times$}),  sources with counterpart classified as RGBs in 47~Tuc\,(\textcolor{blue}{\large $\bullet$}), main sequence member of 47~Tuc\,(\textcolor{Green}{\large $\bullet$}), members of SMC (\textcolor{violet}{\large $\bullet$}), and unclassified sources\,({\scriptsize $\bullet$}). \label{var-plot}}
\end{figure}

To study the long term variability, we checked the flux variation of sources over five observations. Flux variation and its significance were calculated using 
\begin{equation} 
Var=\frac{F_{\rm max}}{F_{\rm min}}~~ and ~~  S=\frac{F_{\rm max}-F_{\rm min}}{\sqrt{EF_{\rm max}^{2}+EF_{\rm min}^{2} }}, 
\end{equation}
respectively \citep{1993ApJ...410..615P}. Here,  $F_{\rm max}$ and $F_{\rm min}$ are the maximum and minimum X-ray flux, and $EF_{\rm min}$  and $EF_{\rm max}$ are their corresponding errors. For all source, which have been detected in both observations  the variability factor was calculated (see Table~\ref{catalogue-x-ray}). Sources with  $S$>3 are considered as sources with significant variability. Figure~\ref{var-plot} shows the significant variable sources versus the maximum flux of the source. As one cans see the nature of the most variable source in the  field of 47~Tuc remains unknown. The most variable known source in the field of 47~Tuc is classified as an M~dwarf foreground star (see Sect.~\ref{fg-sec}).

\subsection{Spectral analysis}
\label{spec-sec}
We performed an X-ray spectral analysis for the  bright sources in the field of 47~Tuc, which their  optical/infrared counterparts classified as a member of 47~Tuc (see Sect.~\ref{member-sec}). Also, the spectral analysis is performed for the most variable foreground star in the field of 47~Tuc (Src-No.\,453; see Sect.~\ref{fg-sec}). The spectra of the sources with a net source counts >500 in total have been extracted. We improved the statistics of the spectra by merging the spectra of all observations, in which the source was detected. Before merging the spectra of different observation, the variability of the source were checked to exclude the spectrum of the observation(s), in which the source shows a significantly different flux (see table~\ref{catalogue-x-ray}).  We were able to fit the spectrum of sources using models for power-law (po), black-body (bb),  collisionally-ionized thermal gas \citep[apec,][]{2001ApJ...556L..91S}, and X-ray emission from a hydrogen atmosphere of a neutron star \citep[nsa,][]{1996A&A...315..141Z} using Xspec\,(V.12.12.0). Figure~\ref{spec.fig} shows the spectrum of the X-ray sources and Table~\ref{spectral-Table} the details of the models fitted to the spectrum of sources. 
\subsection {Hardness ratio}
\label{hr-sec}
\begin{figure}[!htbp]
\includegraphics[clip, trim={0.0cm  0.0cm  0.0cm  0.cm},width=0.47\textwidth]{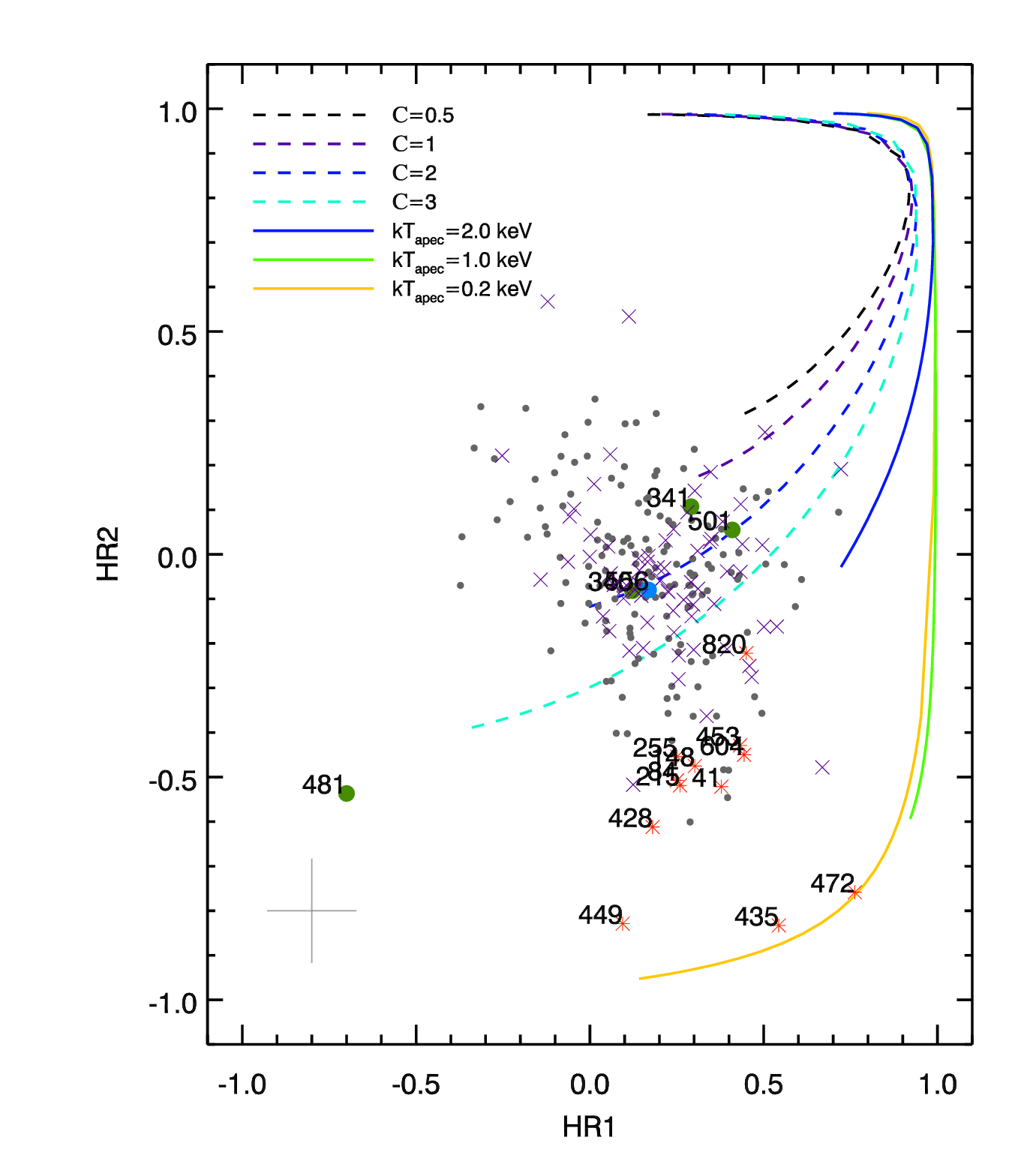}
\includegraphics[clip, trim={0.0cm  0.0cm  0.0cm  0.cm},width=0.47\textwidth]{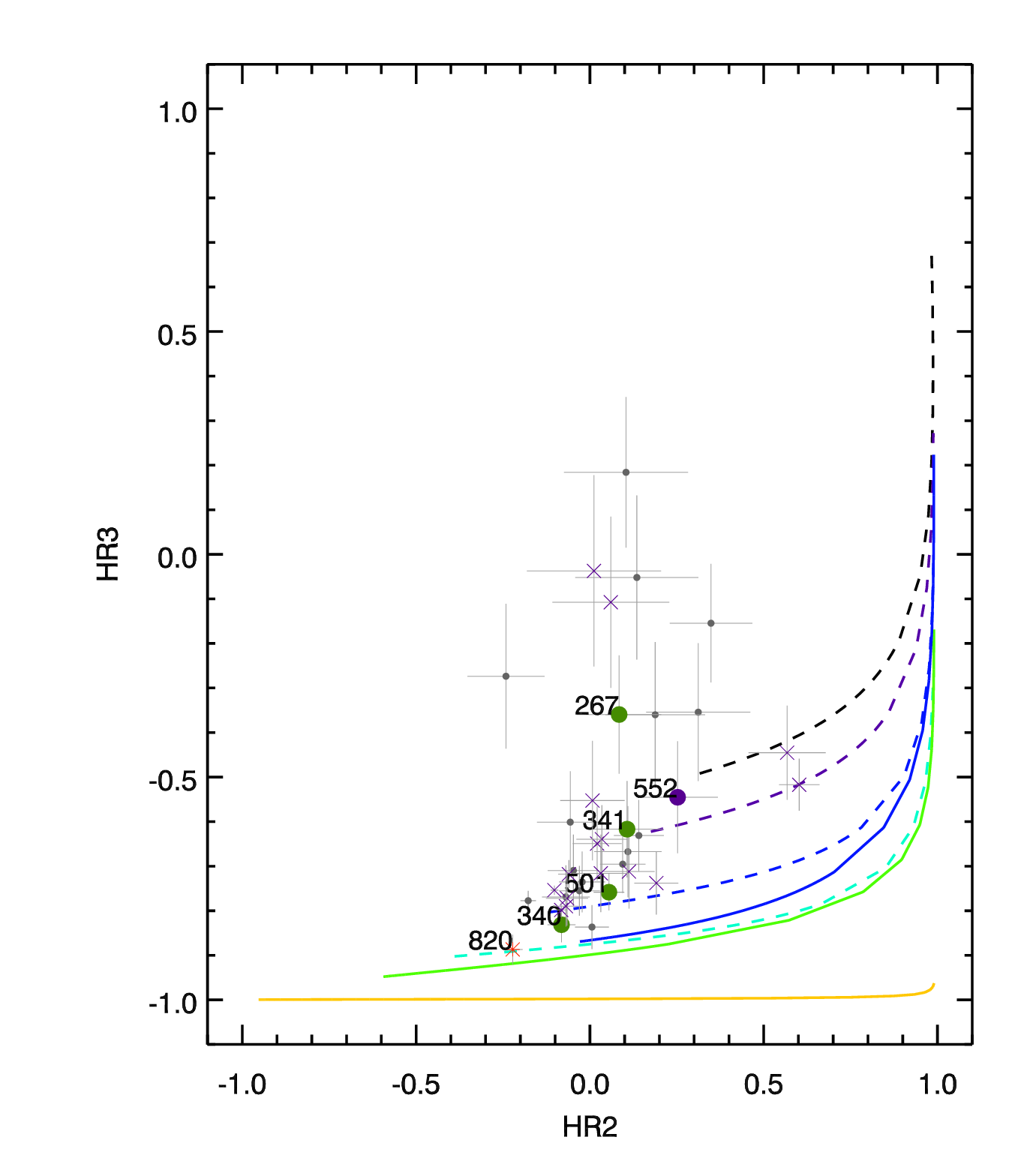}
\hspace{1cm}
\caption{Hardness ratio diagrams. The plotted hard lines are the hardness ratios calculated for different spectral models.  As one can see in the upper plot there are sources in the foreground and therefore with very low $N_{\rm H}$, they appear softer than the others. The symbols are the same as Fig.~\ref{var-plot}. \label{hrs-plot}}
\end{figure}

\begin{figure*}[!htbp]
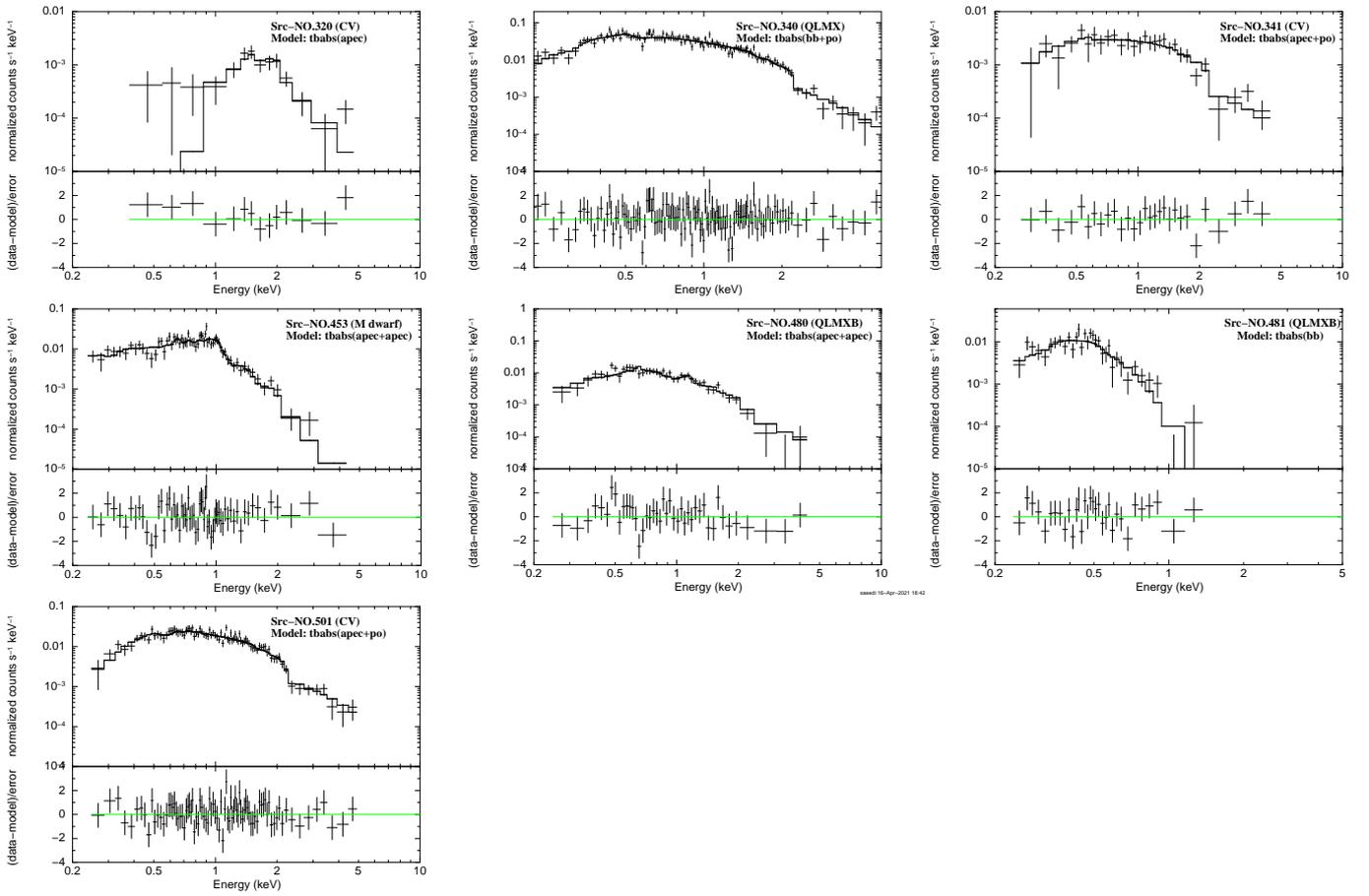


\includegraphics[angle=270, width=0.33\textwidth, trim=1.5cm 0.cm 0.cm 2.0cm]{spec-src-320.eps}
\includegraphics[angle=270, width=0.33\textwidth, trim=1.5cm 0.cm 0.cm 2.0cm]{spec-src-340.eps}
\includegraphics[angle=270, width=0.33\textwidth, trim=1.5cm 0.cm 0.cm 2.0cm]{spec-src-341.eps}\\
\includegraphics[angle=270, width=0.33\textwidth, trim=1.5cm 0.cm 0.cm 2.0cm]{spec-src-453.eps}
\includegraphics[angle=270, width=0.33\textwidth, trim=1.5cm 0.cm 0.cm 2.0cm]{spec-src-480.eps}
\includegraphics[angle=270, width=0.33\textwidth, trim=1.5cm 0.cm 0.cm 2.0cm]{spec-src-481.eps}\\
\includegraphics[angle=270, width=0.33\textwidth, trim=1.5cm 0.cm 0.cm 2.0cm]{spec-src-501.eps}
\caption{Combined spectrum of \erosita\, observations of the X-ray sources  \label{spec.fig}} 
\end{figure*}

Hardness ratios\,(HRs) are useful tool for the study of spectral properties of X-ray sources. The HR and its error are defined as:
\begin{equation} 
HR_\mathrm{i}=\frac{B_\mathrm{i+1}-B_\mathrm{i}}{B_\mathrm{i+1}+B_\mathrm{i}} ~~and~~ EHR_i=2\frac{\sqrt{(B_\mathrm{i+1}EB_\mathrm{i})^2+(B_\mathrm{i}EB_\mathrm{i+1})^2}} {(B_\mathrm{i+1}+B_\mathrm{i})^2}, 
\end{equation}
respectively, where $B_\mathrm{i}$ is the count rate and $EB_\mathrm{i}$ is the corresponding error in the energy band~$i$.  We calculated the hardness ratio from the observation, in which the source had the highest detection likelihood.  To increase the accuracy, we consider a HR measurement as significant only if the detection likelihood for the both corresponding energy bands was higher than 6\,(i.e,~>3$\sigma$). Table~\ref{catalogue-x-ray} lists the HRs for all sources. Figure~\ref{hrs-plot} shows the \erosita\ sources with significant HRs. To understand the spectrum of sources, we plotted the lines representing the hardness ratios of different spectral models with various column densities from $N_{\rm H}$=$10^{20}$\,cm$^{-2}$ to $N_{\rm H}$=$10^{23}$\,cm$^{-2}$. Four \texttt{power-law} models with photon-index $\Gamma$ of  0.5, 1, 2, 3 correspond to hardness ratio of the hard sources, e.g, X-ray binaries, AGNs, or galaxies. Three \texttt{apec} model with the temperature of $kT$ of  0.2, 1.0, and 2.0 keV represent the spectra of soft plasma emissions detected in  sources like supernova remnants\,(SNRs), foreground stars, and symbiotic stars.  As can be seen in the HR diagrams,  foreground  sources have a very soft spectrum. They have a very low $N_{\rm H}$, which also is the reason why they appear much softer than the others. The majority of the sources are located around \texttt{power-law} models with $\Gamma\sim2-3$ in the energy bands <2.3 keV. As expected from the sensitivity of the \erosita\, a few sources are significantly observed >2.3~keV. Considering these results, to calculate the X-ray flux of the sources, for which the spectrum have not been analysed in Sect.~\ref{spec-sec}, we assumed an absorbed \texttt{power-law} model with a $\Gamma$=3 and a Galactic absorption of 5.5$\times10^{20}$\,cm$^{-2}$ \citep[i.e, Galactic adsorption in the direction of 47~Tuc,][]{2016A&A...594A.116H}  (see Table~\ref{catalogue-x-ray}).
We consider a source as soft source if  it is detected significantly only in the first energy band. A  source is classified as a hard source is it has significant emission in  the highest energy band or in the last two higher energy bands. These sources can not be presented in the HR plots due to the lack of counts in bands necessary for the  HRs or due to very large EHR. In Table~\ref{catalogue-x-ray} these sources are classified as soft or hard X-ray sources. 

\begin{table*}
\centering
\caption{Best-fit parameters of the X-ray spectra. Errors are at the 90$\%$ confidence level. \label{spectral-Table}}
\small
\centering
\addtolength{\tabcolsep}{-0.1cm}   
\begin{tabular}{llcllcccc}
\hline\hline
Src-No & Model & $N_{\rm H}$&Photon index& $Temperature$ & Abundance&$\chi^2$ (d.o.f) &   Unabsorbed $F_{\rm X}$&Unabsorbed $L_{\rm X}^{\ast}$\\
&&$10^{22}$ cm$^{-2}$&&&&&$10^{-14}$erg\,s$^{-1}$\,cm$^{-2}$&erg\,s $^{-1}$\\
\hline
\vspace{2mm}
\vspace{2mm}
320&\texttt{tbabs$\times$(apec)}&$ 1.85_{-1.25}^{+0.59}$&--&>1.32 keV&--&$1.15\,(10)$&$5.20_{-0.87}^{+0.88}$&$1.23\times10^{32}$\\
\vspace{2mm}
& \texttt{tbabs$\times$(bb)}& 0.65$_{-0.48}^{+0.74}$&&   0.64$_{-0.23}^{+0.38}$ keV&& \ 0.950(11)&1.3$_{-0.66}^{+1.16}$&$2.90\times10^{31}$\\
\vspace{2mm}
& \texttt{tbabs$\times$(nsa)$^{\ast\ast}$}&0.78$_{-0.51}^{+0.81}$ &&   log(T):\,6.48~K&& 0.961(11)&  0.89$_{-0.33}^{+0.42}$& $2.11\times10^{31}$\\
\vspace{2mm}
\vspace{2mm}
340 &  \texttt{tbabs$\times$(bb+po)}&  $0.04_{-0.01}^{+0.015}$& 2.23$_{-0.30}^{+0.33}$& < 0.90 keV&& 1.05\,(116)& $12.38_{-0.98}^{+1.10}$& $2.80\times10^{32}$\\
\vspace{2mm}
& \texttt{tbabs$\times$(nsa+po)}&   <0.05&  2.06$_{-0.14}^{+0.08}$&  log(T): 5.8$_{-0.29}^{+0.13}$ K&&0.97\,(116)& $9.27_{-0.26}^{+0.27}$& $2.14\times10^{32}$\\
\vspace{2mm}
341&\texttt{tbabs$\times$(apec+po)}&$<0.19$&$1.22^{+1.01}_{-0.38}$ keV&$<0.2$&--&$ 0.80\,(24)$&$1.61_{-0.17}^{+0.14}$&$3.82\times10^{31}$\\
\vspace{2mm}
453&\texttt{tbabs$\times$(apec+apec)}&$0.020_{-0.03}^{+0.02}$&--&$T1: 0.28_{-0.05}^{+0.09}$ keV&$0.17^{+0.09}_{-0.06}$&$ 1.07\,(58)$&$2.34_{-0.11}^{+0.12}$& 1.12$\times10^{29\ast\ast\ast}$\\
\vspace{2mm}
&&&&$T2: 1.04_{-0.10}^{+0.09}$ keV&&&&\\
\vspace{2mm}
480&\texttt{tbabs$\times$(apec+apec)}&$0.03^{+0.02}_{-0.01}$&--& $T1: 0.25^{+0.05}_{-0.04}$ keV&--&$ 1.03\,(38)$&$  5.80_{-0.37}^{+0.36}$&$1.37\times10^{32}$\\
\vspace{2mm}
&&&&$T2: 4.27_{-1.03}^{+1.59}$ keV&&&&\\
\vspace{2mm}
&  \texttt{tbabs$\times$(bb+po)}&   $0.06_{-0.05}^{+0.06}$&  2.17$_{-0.96}^{+0.87}$& < 0.66 keV&& 0.68\,(38)& $3.92_{-0.31}^{+0.54}$& $9.01\times10^{31}$\\
\vspace{2mm}
481&\texttt{tbabs$\times$(bb)}&$0.05_{-0.02}^{+0.03}$&--&$0.075^{+0.009}_{-0.008}$ keV&--& $1.08\,(30)$& $1.65^{+0.18}_{-0.18}$&$3.90\times10^{31}$\\
\vspace{2mm}
&\texttt{tbabs$\times$(nsa)}& 0.07$_{-0.02}^{+0.03}$&--&log(T): 5.38$^{+0.12}_{-0.08}$ K&--& $1.09\,(30)$&3.63$^{+0.10}_{-0.11}$& 8.81$\times10^{31}$\\
\vspace{2mm}
501& \texttt{tbabs$\times$(apec+apec)}&$0.05^{+0.03}_{-0.02}$&--&$  T1: 0.30_{+0.60}^{-0.08}$ keV&--&$1.27\,(80)$&$6.14_{-0.25}^{+0.26}$&$1.49\times10^{32}$\\
\vspace{2mm}
&&&&$T2: 5.41_{-2.08}^{+13.2}$ keV&&&&\\
&  \texttt{tbabs$\times$(nsa+po)}&   0.04$^{+0.04}_{-0.03}$&  1.45$_{-0.51}^{+0.41}$&  log(T): 6.24$_{-0.28}^{+0.42}$ K&& 1.43\,(81)&  $6.27_{-0.30}^{+0.34}$&  $1.45\times10^{32}$\\
\vspace{1mm}\\
\hline
\multicolumn{9}{l}{$\ast$: We assumed a distance of 4.45~kpc to estimate the X-ray luminosity of sources in 47~Tuc (see Sect.~\ref{intro}).}\\
\multicolumn{9}{l}{{$\ast\ast$: Nonmagnetic neutron star atmosphere\,(\texttt{nsa}) model \citep{1996A&A...315..141Z}  has been applied for the systems, which have been candidate to have}}\\
\multicolumn{9}{l}{{~~~~~~ neutron star as their compact object. We assumed a mass and radius of $M_{NS}$=1.4M$_{\sun}$ and R$_{NS}$=10.~km for the neutron star as it is has been}}\\ 
 \multicolumn{9}{l}{{~~~~~~ used in \texttt{nsa} model for the QLMXBs and MSPs spectra \citep[e.g,][]{2006ApJ...646.1104B,2005ApJ...618..883W}.}} \\
\multicolumn{9}{l}{$\ast\ast\ast$: For Src-No.\,453, the distance of the counterpart, which is a foreground star located at $\sim$200\,pc is considered.}\\
\vspace{2mm}
\end{tabular}
\end{table*}
\section{Multi-wavelength studies of counterparts}
\label{multi-sec}
We have searched for counterparts of the X-ray sources ( within their 3$\sigma$ positional error) in optical and infrared using the NWAY code, which is a Bayesian algorithm for cross-matching multiple catalogues provided by \citet{2018MNRAS.473.4937S}. In order to yield high accuracy, we only considered an infrared/optical counterpart for the X-ray source if the distance posterior probability, which is the probability computed using the Bayesian approach considering asymmetric parameters (e.g, positional uncertainties, distance of a counterpart from the X-ray source, and number densities) as it is explained in 
\citet{2018MNRAS.473.4937S} in their Appendix B5, was higher than 50$\%$. If both infrared and optical counterparts exist for a source, we checked if the position of optical and infrared counterparts are the same and it is from one source. Otherwise only the counterpart, which had a higher match distance probability and was closer to the position of the X-ray source is reported.  To estimate the possibility of spurious matches we calculated the chance coincidence probability for the counterparts of X-ray sources in the field of 47~Tuc. We assumed a shift of 10.$\arcsec$ in a random direction for the position of each source. The shifts are repeated four times. The probability of finding a new counterpart is considered as chance coincidence probability, which was (8.11$\pm$1.15)$\%$ for the whole observational area. We  also calculated the chance coincidence probability for those sources, which are located inside the 18.8$\arcmin$ circle area (where we have the highest exposure time of all observations) and for the sources outside this circle. It was (5.73$\pm$1.55)$\%$ for the inner region and (8.47$\pm$1.89)$\%$ for the outer region. This result shows that by using the positional error of the merged observations (as explained in Sect.~\ref{source-cata-sec}) we have lowered the chance coincidence probability in the crowded region of 47~Tuc. In order to decrease the consequences of spurious matches we applied additional criteria: we only accepted a counterpart of background galaxies/AGN or foreground star if the distance posterior probability of match was >70$\%$. For sources, which are classified as members of 47~Tuc the counterpart is accepted if it was located witin 2$\sigma$ X-ray positional error. Moreover, the \chandra\, position has been used to search for infrared/optical counterpart (in the case that the source has a \chandra\, counterpart). These details are discussed in Sect.~\ref{diss-sec}. In the next following sections, we discuss the multi-wavelength photometry used to uncover the stellar nature of our sources. 
\subsection{Infrared counterparts of the sources}
\label{infra-sec}
\begin{figure}[!htb]
\includegraphics[trim={1.2cm 0.0cm 0.0cm 0.cm},width=0.45\textwidth]{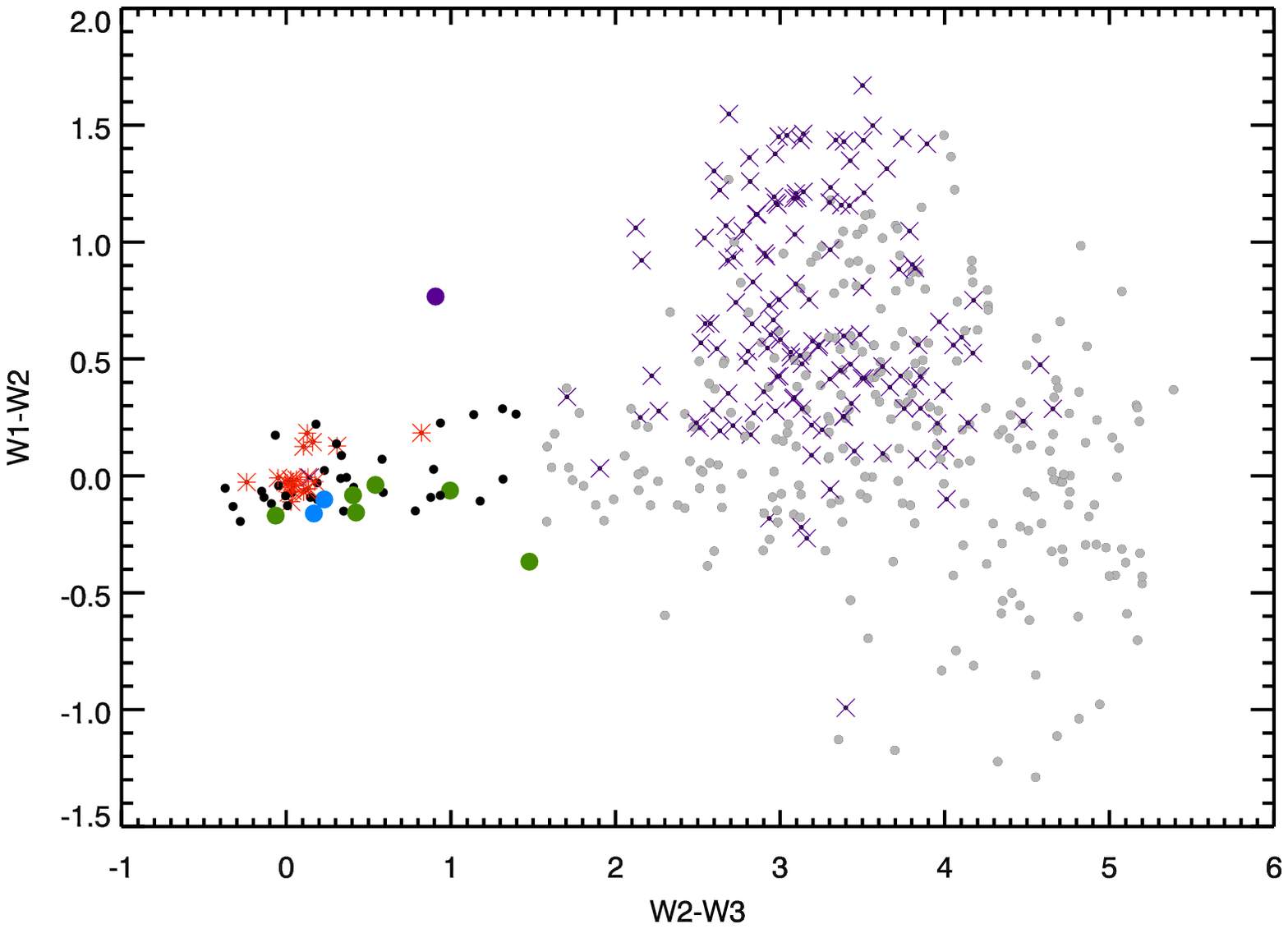}
\caption{Colour-colour diagram of mid-infrared WISE\,($W1$-$W2$ versus $W2$-$W3$). The symbols are the same as Fig.~\ref{var-plot}. The sources, which had a WISE counterpart with an upper limit in $W2$, or $W3$, or both are shown with gray circles.}\label{wise-plot}
\end{figure}

\begin{figure}[!htb]
\centering
\includegraphics[trim={1.2cm 0.0cm 1.0cm 0.cm},width=0.45\textwidth]{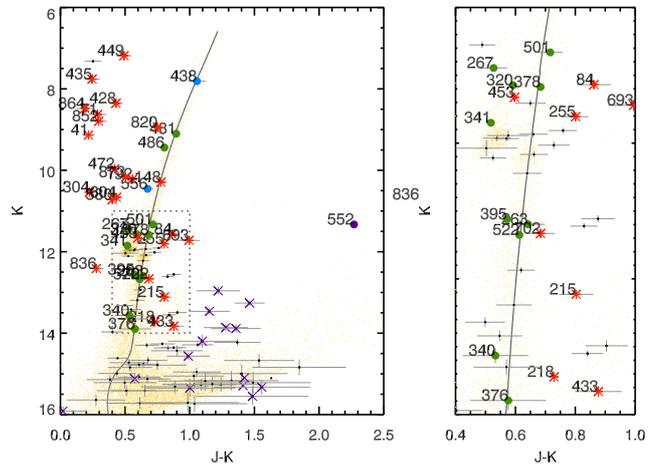}
\caption{\raggedright  The colour-magnitude diagram of 2MASS counterparts of the X-ray sources in the field of 47~Tuc. To have a better look into the crowded region the left plot shows a zoom to the dashed square in the right plot. The yellow dots are all 2MASS sources detected in the field of 47 Tuc.
The symbols are the same as Fig.~\ref{var-plot}. \label{2mass-plot}}
\end{figure}

We searched for mid-infrared counterparts in the WISE All-Sky Survey in four energy bands \citep[3.4, 4.6, 12, and 22 $\mathrm \mu$m, named $W1$, $W2$, $W3$, and $W4$, respectively;][]{2014yCat.2328....0C}. The extinction for the infrared WISE bands in the direction of 47~Tuc was negligible \citep{2011ApJ...737..103S}.  Table~\ref{inf-count} lists the magnitudes of WISE counterparts of the X-ray sources. Figure~\ref{wise-plot} shows the colour-colour diagram of the WISE counterparts of the X-ray sources in the field of 47~Tuc. The infrared colours shown in this plot can give us information about the nature of the counterpart, i.e., whether the counterpart is a stellar object, an AGN, or a galaxy. The study of \citet{2010AJ....140.1868W} shows that  background objects are usually expected to be red (i.e, $W2-W3$>1.5) in WISE colour, while stellar objects show a different colour (i.e, $W2-W3$<1.5). One can see that the counterparts of known background objects are separated from the X-ray sources with known stellar counterparts (see Sect.~\ref{diss-sec} and Fig.~\ref{wise-plot}). We also checked if the X-ray sources have near-infrared counterparts in the 2MASS All-Sky Survey Catalogue in the three standard bands of $J$, $H$, $K_{s}$ standard bands \citep{2003yCat.2246....0C}. In the direction of 47~Tuc, we applied the extinction of  0.03, 0.02, 0.01 for the $J$, $H$, $K_{s}$ bands, respectively \citep{2011ApJ...737..103S}. In the colour magnitude diagram of the 2MASS counterparts (Fig.~\ref{2mass-plot}) we also show the position of the main isochrone of 47~Tuc, which was obtained using the theoretical models of the Dartmouth stellar evolution database \citep{2008ApJS..178...89D} for the age, metallicity, and distance of 47~Tuc as discussed in Sect.~\ref{intro}.  As another key for the classification  of symbiotic stars we considered the results of the machine learning method of \citet{2019MNRAS.483.5077A}. They show that in the population of known symbiotic stars, the majority of systems appears to have  $J-H>0.78$ and only a small fraction of S-type symbiotics behaves differently. The second criterion is  $K-W3<1.18$, which separates the symbiotic stars significantly form the other types of sources. The second criterion might not apply for  dusty symbiotic stars. In this case, there are two other criteria on the colours of $H-W2>3.80$ and $W1-W4<4.72$. We applied these criteria for the classification of symbiotic stars as we discuss in Sect~\ref{diss-sec}.  


\subsection{Optical counterparts of the sources}
\label{optical-sec}
\begin{figure*}[!htb]
\centering
\includegraphics[trim={1.2cm 0.0cm 1.0cm 0.cm},width=0.43\textwidth]{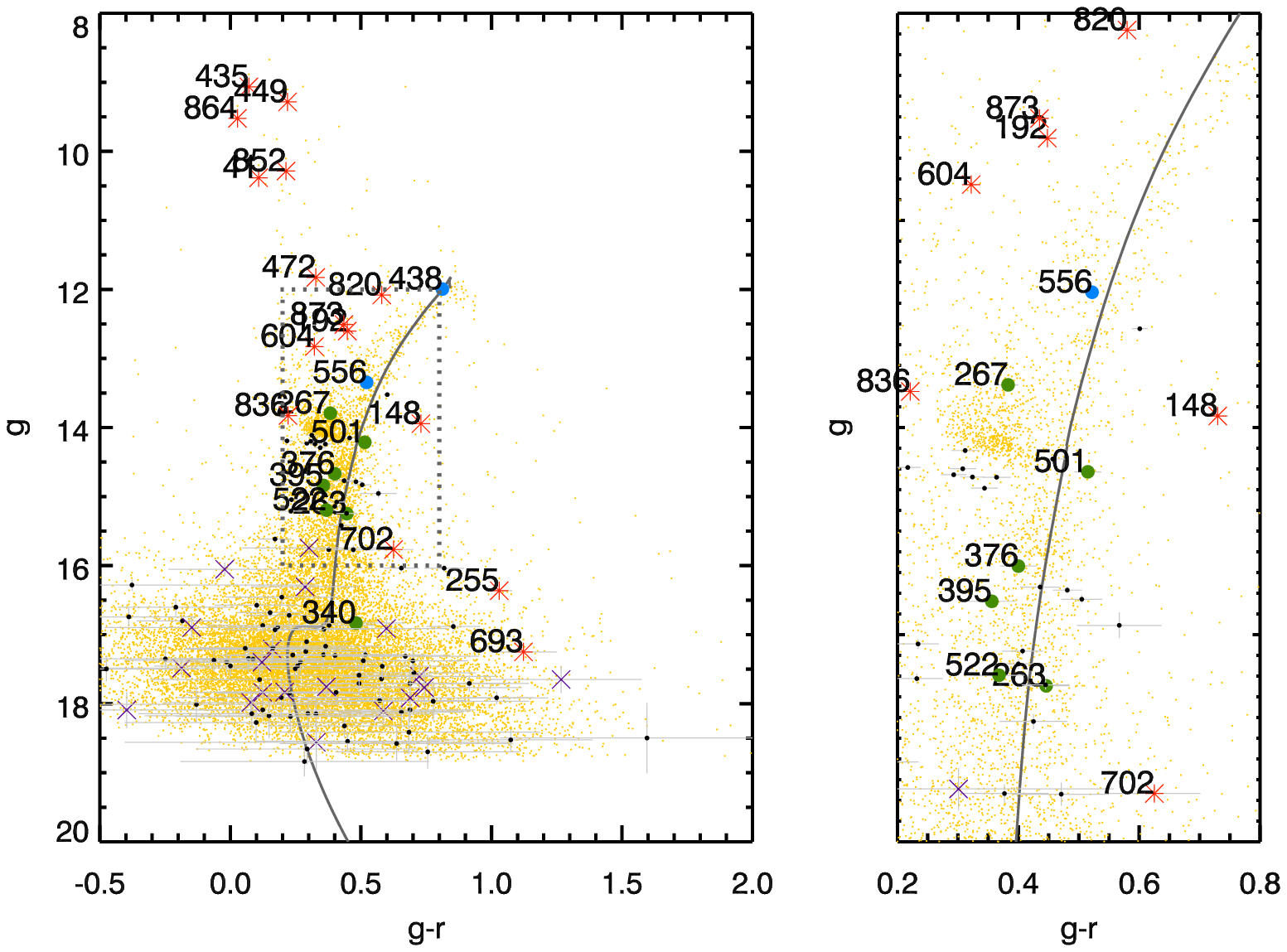}
\hspace{1cm}
\includegraphics[trim={1.2cm 0.1cm 1.0cm 0.cm},width=0.43\textwidth]{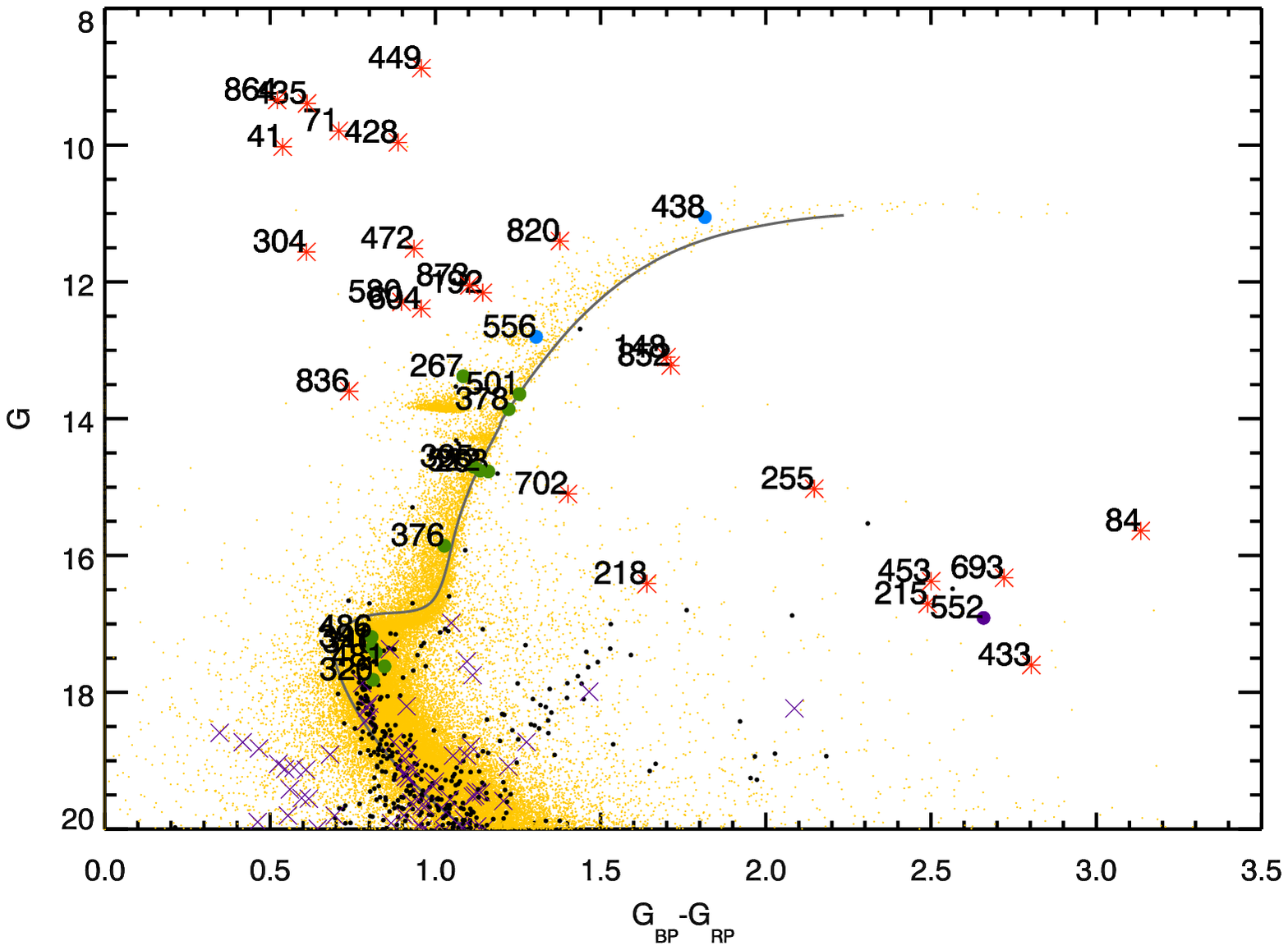}
\caption{The optical counterpart of the X-ray sources in the field of 47~Tuc observed by SkyMapper\citep{2018PASA...35...10W} (\textbf{left panel}) and Gaia third data released\citep{2020yCat.1350....0G}\,(\textbf{Right panel}). The yellow dots are all optical sources detected in the field of 47~Tuc dSph and the gray hard lines are the theoretical isochrone of The Dartmouth stellar evolution database \citep{2008ApJS..178...89D} for the age, metallicity, and distance of 47~Tuc\,(see Sect.~\ref{intro}). The rest of the symbols are the same as Fig.~\ref{var-plot}. To have a better look into the crowded region of Skymapper colour-magnitude diagram a zoom to the dashed square is shown.} \label{opt-counterpart}
\end{figure*}

The most recent all-sky optical surveys, third Gaia Data Release \citep{2020yCat.1350....0G} and the first data release of the SkyMapper southern survey \citep{2018PASA...35...10W}, have been used to search for the optical counterparts of the eROSITA sources in the field of 47~Tuc. Table~\ref{opt-count-table} presents the Gaia and SkyMapper magnitudes of the optical counterparts of the X-ray sources. The SkyMapper catalogue includes photometric data in the energy bands from the optical to the near infrared. In SkyMapper survey we have mainly used the two known optical Petrostian magnitude bands \citep{2018PASA...35...10W} of $g$\,($\lambda_{eff}$=467~nm) and $r$\,($\lambda_{eff}$=616~nm) to plot the colour magnitude diagram of the optical counterparts (see left diagram of  Fig.~\ref{opt-counterpart}). The Gaia surveys also report the magnitudes of the sources in three filter of $G$ mag (roughly $\lambda$=300 nm), $G_{BP}$\,($\lambda$=400-500~nm) mag, and  $G_{RP}$\,($\lambda$=600-750~nm) \citep{2018A&A...616A...1G}, which have been considered in our study for the comparison with the SkyMapper magnitudes and for the Gaia colour magnitude diagram (see right diagram of Fig.~\ref{opt-counterpart}). We also considered the Gaia-parallax measurement to identify foreground stars 
as presented in the work of \citet{2021yCat.1352....0B}. 

The extinction of 0.12 and 0.09 has been applied for the $g$ and $r$ bands of SkyMapper. Also 0.18, 0.13, and 0.07 for $G$, $G_{BP}$, and  $G_{RP}$ for the Gaia bands, respectively \citep{2011ApJ...737..103S}. The theoretical isochrone line of 47~Tuc is also plotted for colour magnitude diagrams of SkyMapper and Gaia counterparts \,(Fig.~\ref{opt-counterpart}) as it is explained in Sect.~\ref{infra-sec}.

The logarithmic X-ray to optical flux ratio log$(\frac{F_\text{X}}{F_\text{opt}})$, versus the X-ray flux and also HR2

\begin{figure*}
\includegraphics[clip, trim={1.6cm 0.5cm 0.5cm 0.3cm},width=0.48\textwidth]{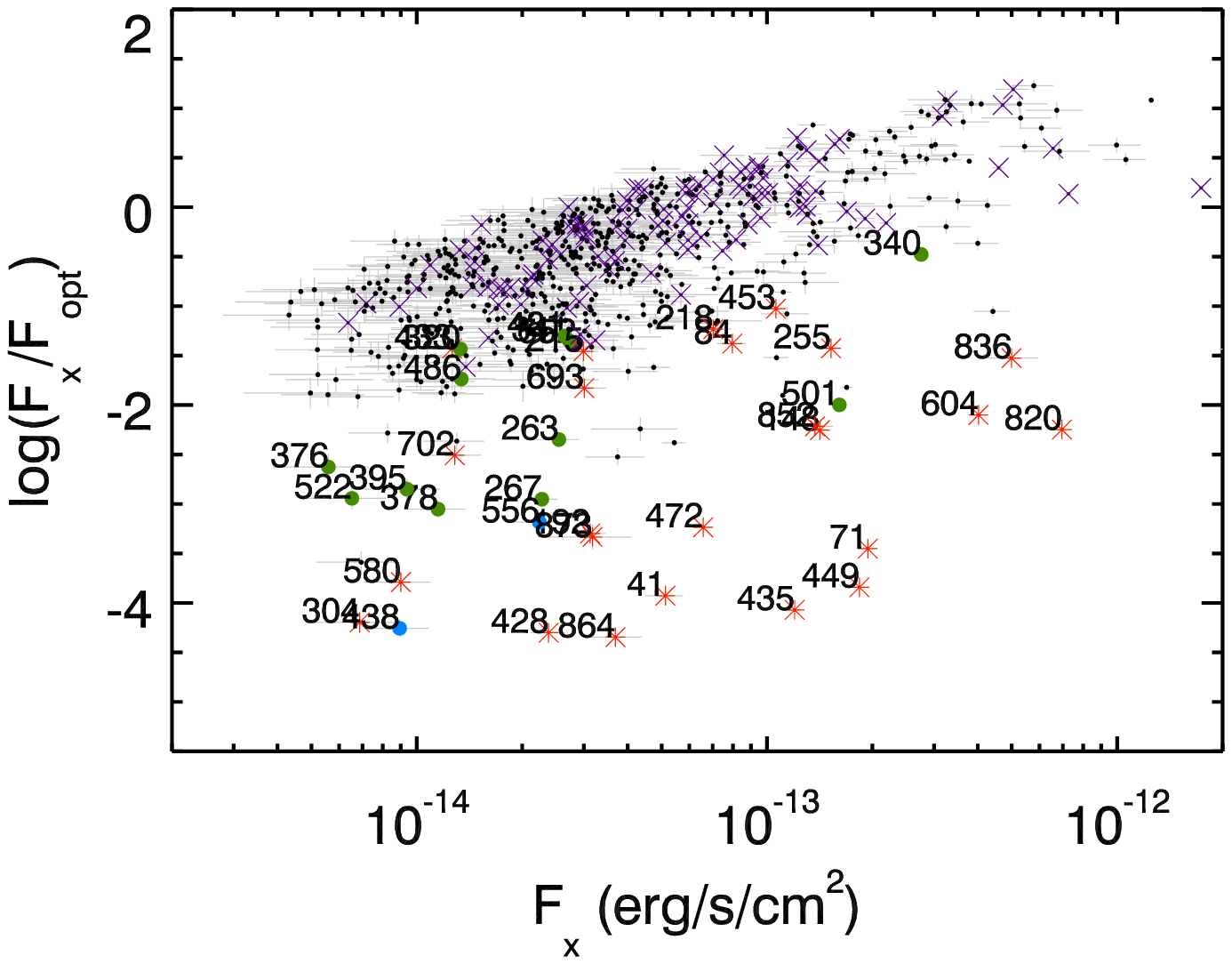}
\includegraphics[clip, trim={1.6cm 0.5cm 0.5cm 0.3cm},width=0.48\textwidth]{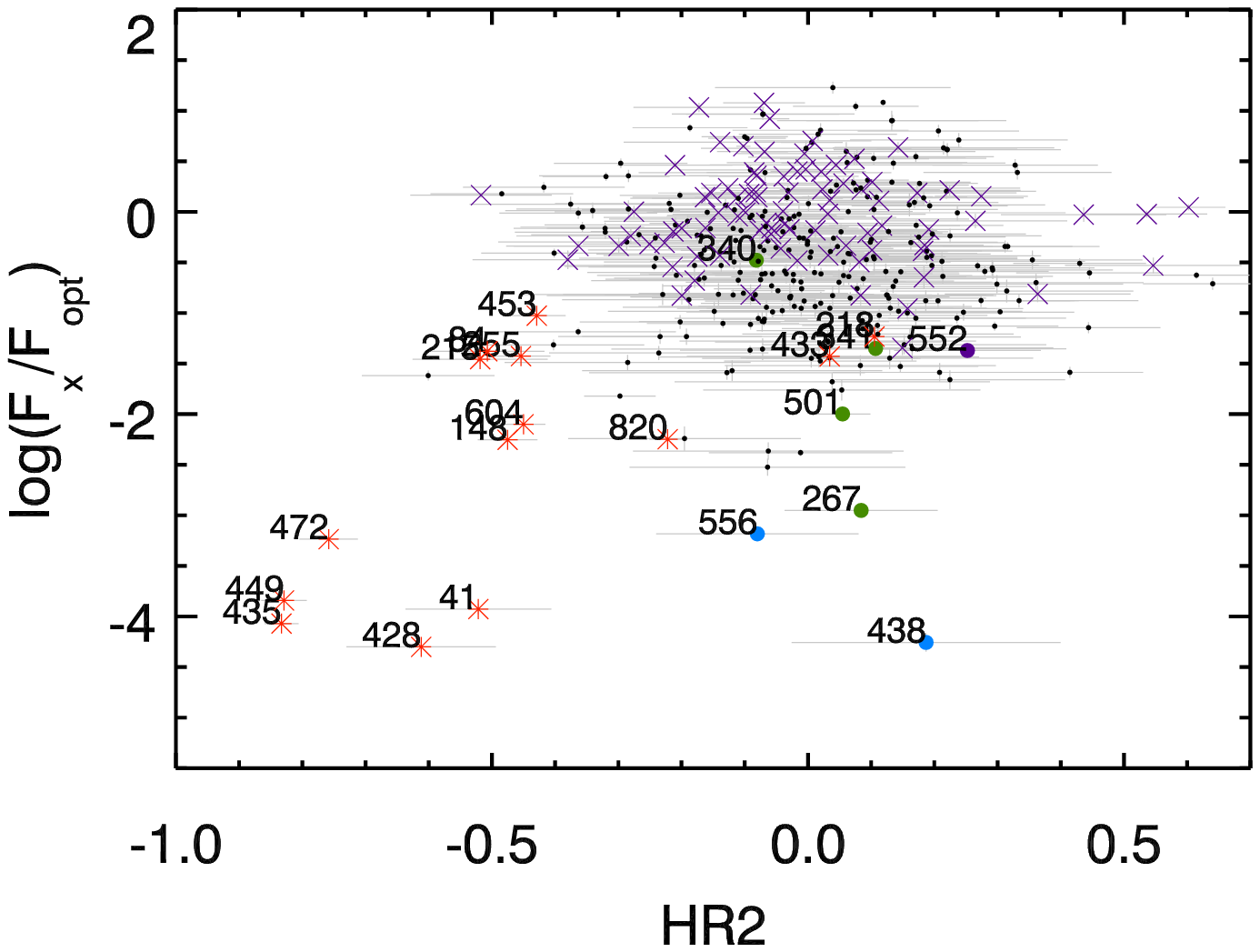}
\caption{Logarithmic X-ray to optical flux ratio  log$(\frac{F_\text{X}}{F_\text{opt}})$ versus the maximum X-ray flux\,(left) and HR2\,(right) for the sources in the field of 47 Tuc. The symbols are the same as Fig.~\ref{var-plot}. \label{log-x-opt} }
\end{figure*}

(see Sect.~\ref{hr-sec}) are shown in Figure~\ref{log-x-opt}. The modified version of the flux ratio equation log$(\frac{F_\text{X}}{F_\text{opt}})$  \citep{1988ApJ...326..680M} with an average of $G_{BP}$ and $G_{RP}$ Gaia magnitudes is applied: 
\begin{equation}
{\rm log}\bigg(\frac{F_ \text{X}}{F_\text{opt}}\bigg)={{\rm log}_{10}(F_\text{X})}+\frac{G_{BP}+G_{RP}}{2\times2.5}+5.37,
\end{equation}
where $F_\text{X}$ is the X-ray flux and $g$ and $r$ are the SkyMapper magnitudes of the optical counterpart associated with the X-ray source. As Figure~\ref{log-x-opt} shows, the main part of the classified sources of 47~Tuc are more dominant in optical radiations. 

\subsection{Catalogues of AGNs and galaxies}
\label{AGN-cata}
The following catalogues were cross-checked with all X-ray sources to find the classified background objects in available catalogues: \begin{itemize}
    \item The Million Quasars (Milliquas) Catalogue \citep{2019arXiv191205614F}
    \item Quasar and galaxy classification in 2nd Gaia Data Release \citep{2019MNRAS.490.5615B}
    \item The UV-bright Quasar Survey \citep{2016AJ....152...25M}
    \item The SWIFT AGN and Cluster Survey \citep{2015ApJS..218....8D}
  \item Identification of 1.4 Million AGNs in the mid-Infrared using WISE Data \citep{2015ApJS..221...12S}
  \item Identifications of AGNs from the WISE, 2MASS, and \rosat\, All-Sky Surveys \citep{2012ApJ...751...52E}
\end{itemize}

\subsection{Catalogues of members of the 47~Tuc}
\label{47-Tuc-member}
Sources, which are located on the  principal sequence of optical and infrared colour magnitude diagrams (see Fig.~\ref{2mass-plot} and Fig.\ref{opt-counterpart}) and are listed in the following catalogues, were confirmed as members of 47~Tuc: 
\begin{itemize}
    \item Catalogue of \citet{2014ApJ...780...94C} present the detailed of the  abundances of 164 RGB Stars in the 47 Tuc. 
    \item  \citet{2013A&A...550A..34C} provides the  aluminium abundances for a sample of about 100 RGBs in 47 Tuc and M4 GCs.
    \item The work of \citet{2013A&A...549A..41G} presents analysis of the composition  110 red horizontal branch stars in 47~Tuc.
    \item Based on  various parameters e.g, the metallicity and  radial velocity, the membership of more than  43000 sources in the field of GCs including 47 Tuc  are investigated. \citet{2011A&A...530A..31L}. 
\end{itemize}

\section{Discussion}
\label{diss-sec}
The procedure for the classification of the X-ray sources using the above results is explained the the following:
\subsection{Classification of background sources in the field of 47~Tuc}
\label{bkg-sec}
\begin{figure}
\includegraphics[clip, trim={0.0cm 1.6cm 1.0cm 1.cm},width=0.40\textwidth]{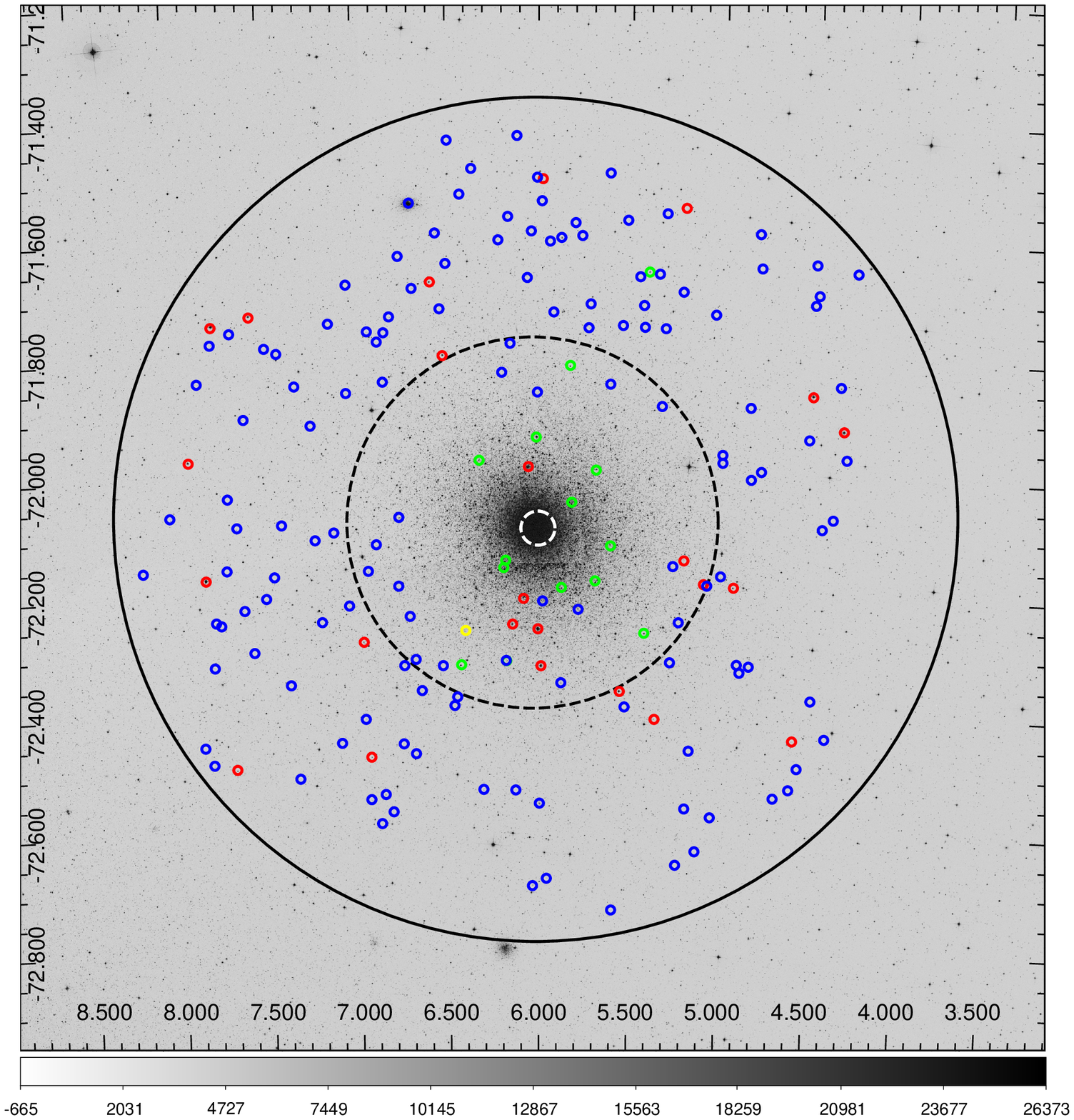}\\
\includegraphics[clip, trim={1.0cm 0.0cm 0.50cm 0.cm},width=0.40\textwidth]{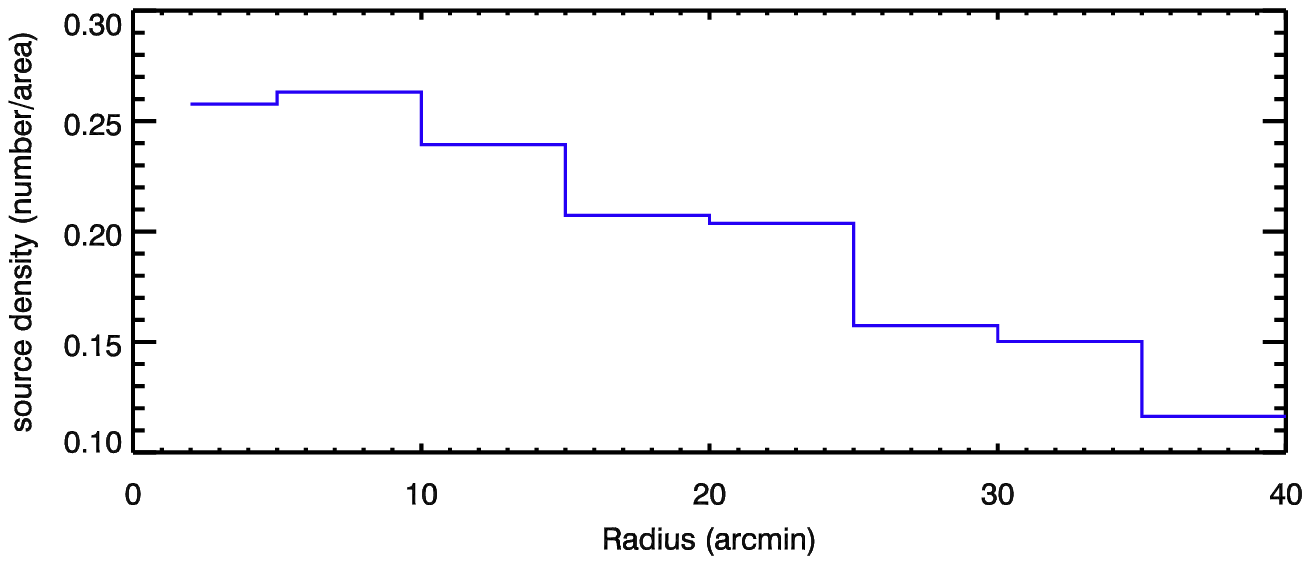}
\caption{{\bf Upper image:}The optical image of the field of 47~Tuc observed by DSS survey\,(red filter) \citep{2005MNRAS.362..542B}. Hard black region shows the total area which have been observed by \erosita\,(radius of 42$\arcmin$.). Dashed black region is the area which was if the field of view of all observations and have been used for the XLF calculation (radius of 18$\arcmin$.8). The dashed white region is the area, where the observation of \erosita\, was unresolved and have been excluded in this study (radius of 1$\arcmin$.7). The position of classified AGNs, foreground stars, and accreting binaries are shown by blue, red, and green circles, respectively.  The single source classified as a XRB in SMC is in yellow. {\bf Lower plot:} Normalized source density versus the radius of the field of view of \erosita\, observation.  No sources are detected for radius  <2.$\arcmin$ due to the large unresolved source in the center. \label{optical-image}}
\end{figure}

For the classification of AGNs in the field of 47~Tuc, we mainly used the criteria, which have been defined in the study of \citet{2010AJ....140.1868W}\,(see Sect.~\ref{infra-sec}). The X-ray sources with a WISE counterpart (with >70$\%$ probability to be a match to the X-ray source), which have significant magnitudes in $W1$, $W2$, $W3$ bands and fulfil the condition of  $W2-W3>1.5$ are considered as background sources. In this classification, we excluded the sources, which have only an upper limit magnitude in the bands $W2$ and/or $W3$ of their WISE counterpart. These sources remain unclassified since WISE counterpart in the colour of   $W2-W3$ could not be significantly considered as a background object. Moreover, we cross-correlated all available AGN/quasar/galaxy catalogues\,(see Sect.~\ref{AGN-cata}) with our catalogue to classify the other known X-ray background sources.  Almost all the classified background objects in other available catalogues had a WISE counterpart satisfying the condition of $W2-W3>1.5$ as well. We ended up with the classification of ninety-two AGNs/galaxies in the field of 47~Tuc. Fig.~\ref{optical-image} shows the distribution of classified background sources. They are mainly located outside the region, where most of the X-ray members of 47~Tuc are detected (i.e, $\gtrapprox$12$\arcmin$.0). Fig.~\ref{log-x-opt} also shows that the classified AGNS have a higher relative X-ray flux than the X-ray sources in 47~Tuc and the foreground stars. 
\subsection{Classification of foreground stars/systems in the field of 47~Tuc}
\label{fg-sec}
\begin{figure}
\centering
\includegraphics[clip, trim={0.cm 2.0cm 0.cm 0.cm},width=0.30\textwidth]{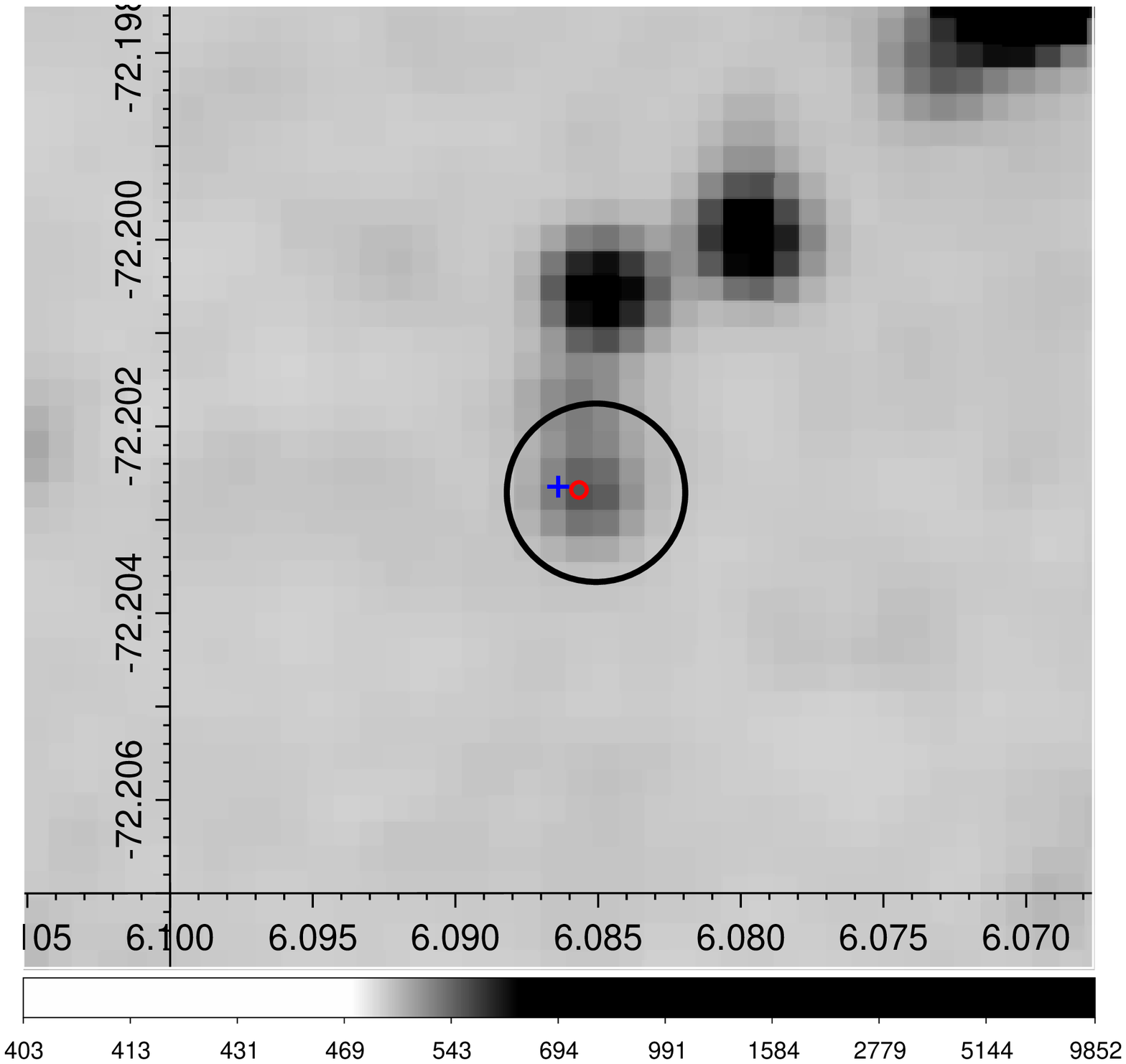}
\caption{ The Infrared 2MASS\,($k_{s}$ band) image of the position of Src-No.\,453, which is classified as a foreground M~dwarf. regions show 3$\sigma$ positional error of X-ray source\,(black) and infrared 2MASS counterpart\,(red). The blue cross shows the position of optical Gaia counterpart. \label{mdwarf-image}}
\end{figure}

For the classification of the foreground stars/systems we consider three main criteria for the infrared/optical counterpart of the X-ray source: (having >70$\%$ distance match probability to the X-ray source), being a stellar object according  to the WISE colours\,(see Sect.~\ref{infra-sec}), and/or the distance of optical counterpart according to the Gaia parallax measurement shows that the source is a foreground object in the field of 47~Tuc  \citep[][]{2018AJ....156...58B}.
The colour magnitude diagrams of the counterparts of 2MASS (Fig.~\ref{2mass-plot}), SkyMapper, and Gaia\,(Fig.~\ref{opt-counterpart}) show that the position of the classified foreground stars  are located outside the main sequence of the 47~Tuc. As Fig.~\ref{var-plot} shows, \textbf{Src-No.\,453} is the most variable foreground stars. The source seemed to be in the flaring state in four first observations, as its count rate drops down in OBS\,5 (see Table~\ref{catalogue-x-ray}). We combined all the data of four observations and fit the spectrum with two absorbed  plasma models (see Table~\ref{spectral-Table} and Fig.~\ref{spec.fig}). The X-ray spectrum  is very similar to that of stellar object, also the measured column density is lower than the Galactic absorption in the direction of 47~Tuc (i.e, 5.5$\times10^{20}$\,cm$^{-2}$), which is expected from a foreground star. Considering the infrared and optical colours of the source counterpart ($J-H$= 0.44$\pm$0.12, $H-K_{s}$=0.25$\pm$0.13, $i-z$=0.49$\pm$0.05, and  $z-J$=1.52$\pm$0.07), it can be classified as an early type M\,dwarf \citep[$M0$--$M3$,][]{2011AJ....141...97W}. Figure~\ref{mdwarf-image} shows the infrared 2MASS\,($k_{s}$ band) image of the counterpart of Src-No.\,453.
\begin{table}
 \centering
\hspace{-0.5cm} \caption{Characteristic of low luminosity X-ray sources\,(AWDs, XRBs, and contact binaries) \label{properties}}
     \begin{tabular}{lll}
\hline\hline
Source&  Spectral emission  & Luminosity\\
class&keV& erg\,s$^{-1}$\\
\hline
    Symbiotic:\,$\alpha$-type& <0.5&$L_{bol}>10^{36}$\\
    Symbiotic:\,$\beta$-type& <2.4&$L_{x}\sim10^{30-32}$\\
    Symbiotic:\,$\delta$-type&>2.4& $L_{x}\sim10^{31-34}$\\
    Symbiotic:\,$\gamma$-type&>2.4&$L_{x}>10^{34}$\\
    CV:\,Non-magnetic& 2.--5.&$L_{x}\sim10^{29-32}$\\
    CV:\,Polars&<5.0&$L_{bol}\sim10^{30-31}$\\
    CV:\,Intermediate polars&5.--50.&  $L_{x}<10^{34}$\\
    Quiescent LMXBs&<5.&$L_{x}\sim10^{31-33}$\\
    Millisecond pulsars&0.2--2.5&$L_{x}\sim10^{30-31}$\\
    Active binaries& <2.5&$L_{x}\sim10^{29-32}$\\
      \hline
     \end{tabular}
\end{table}
\subsection{X-ray sources in 47~Tuc}
\label{member-sec}
The majority of X-rays sources of GCs are expected to be low luminosity (<$10^{33}$\,erg\,s$^{-1}$) X-ray sources, which can be quiescent LMXBs, different types of AWDs, MSPs, and magnetically active binary systems. In the following, we briefly review the X-ray spectra and luminosity of these sources. Also, Table~\ref{properties} summarises the properties of X-ray spectrum and luminosity of all types of the sources. 

One of the major class of X-ray sources in GCs is AWDs. They can be symbiotic stars, which have a red giant branch (RGB) star as the companion of the white dwarf, or cataclysmic variables\,(CVs), which are systems made of a white dwarf with a main sequence companion \citep{2017PASP..129f2001M}.

{\bf Symbiotic stars:} These systems are categorised into the four types of $\alpha$, $\beta$, $\gamma$, and $\delta$ \citep{2013A&A...559A...6L}. In 
the $\alpha$-type symbiotics the  X-ray emission  originates from the quasi-steady  burning of the material transferred from the red giant via Roche lobe overflow  onto the surface of the WDs. The emission is detected <0.5~keV, which is the reason to call them  supersoft sources. Supersoft sources are  bright  mainly in the UVs  and very soft X-rays with a bolometric luminosity >$10^{36}$\,erg\,s$^{-1}$. $\beta$-type symbiotics have the main X-ray emission <2.4~keV ($L_{x}$$\sim$$10^{30-31}$\,erg\,s$^{-1}$), which most likely arises from the collision of the wind of the WD with the wind of the red giant. Observations show that the X-ray luminosity can be around two orders of magnitudes higher, when the system is in the outburst \citep{2013A&A...559A...6L}.  $\delta$-type symbiotics are highly absorbed hard X-ray sources\,(>2.4~keV, $L_{x}$$\sim$$10^{31-34}$\,erg\,s$^{-1}$). Theoretical models suggest that the X-ray emission originates from the boundary layer between the accretion disk and the WD \citep[e.g,][]{2013A&A...559A...6L}. There is also a class of $\beta/\delta$-type symbiotics, which have both soft and hard components. For these types of symbiotic stars usually one or two ionised plasma model(s) have been used to fit to the spectra \citep[e.g,][]{2013A&A...559A...6L}. The $\gamma$-type symbiotic stars (symbiotic X-ray binaries) are actually a subclass of LMXBs, in which the binary system consists of a red giant and a neutron star (as the mass accretor). The main part of the emission of these sources is hard (>2.4~keV) and can have a high X-ray luminosity ($L_{x}$>$10^{34}$\,erg\,s$^{-1}$).

{\bf CVs:} There are two main classes of the CVs, magnetic and non-magnetic CVs. In the non-magnetic CVs, a disk forms around the WD, however  the disk is not hot enough to produce X-ray emission, therefore, the disk, by itself, is not visible in X-rays. On the other hand, the boundary layer between the disk and the surface of the WD produces X-ray emission with temperatures of few keV and X-ray luminosities between  $L_{x}$$\sim$$10^{29-32}$\,erg\,s$^{-1}$ \citep[e.g,][]{2017PASP..129f2001M,2006csxs.book..421K,  1996A&A...315..467V}. Magnetic CVs can be divided into two sub-classes of polars and intermediate polars. Polars are systems without an accretion disk, where the materials reach to the surface of WD following the magnetic field lines \citep[][]{2017PASP..129f2001M}. \xmm\, observations show that polars are usually soft (<5.0~keV) and very faint X-ray sources with a bolometric luminosity of $\sim10^{30}$\,erg\,s$^{-1}$ \citep[e.g,][]{2004MNRAS.350.1373R}.  Intermediate polars have the dominant emission in hard X-rays \citep[5-50~keV, $L_{x}$<$10^{34}$\,erg\,s$^{-1}$; e.g,][]{2019MNRAS.482.3622S}, which is produced by a strong shock above the poles of WDs, where a noticeable amount of materials from the inner part of the truncated accretion disk follow the magnetic field lines. They also  can show soft X-ray emission, which are orders of magnitude fainter  than that of hard X-rays \citep[][]{2017PASP..129f2001M, 2012MmSAI..83..585B}. \citet{2017PASP..129f2001M} shows that in general CVs have log$(\frac{F_\text{X}}{F_\text{opt}})$<1.0 and magnetic CVs usually have larger log$(\frac{F_\text{X}}{F_\text{opt}})$ than non-magnetic CVs. The temperature of the soft X-ray emission of CVs can be estimated using the plasma atmosphere models \citep[e.g,][]{2016A&A...592A.114W, 2002A&A...387..201H}.

{\bf Quiescent LMXBs:} A black hole or a neutron star forms a binary systems with a late type low mass star\,(<1.0M$_{\sun}$) \citep{2014MNRAS.439.2771B}. Observational  studies  show  that quiescent LMXBs  are X-ray sources with soft X-ray mission (<5.0 keV) and X-ray luminosities  $\sim10^{30-33}$\,erg\,s$^{-1}$ \citep[e.g,][]{1998A&ARv...8..279C, 2000ApJ...539..191Y, 2007ApJ...665L.147J}. The X-ray spectrum of QLMXBs  can be fitted with a black-body or a neutron star atmosphere model or has a hard power-law component \citep[e.g, ][]{2005ApJ...618..883W}. A radius of 1--2~km \citep[fitting with a black-body model,][]{1998ApJ...504L..95B} or 10--12~km \citep[fitting with neutron star atmosphere models, e.g,][]{2002ApJ...578..405R, 2003ApJ...598..501H}  are expected to be inferred from the models.

{\bf  MSPs:} They are known to be fast spinning neutron stars in a binary system with a low mass companion (<1. M$_{\sun}$) mainly detected in GCs. According to the recycling scenario, an old neutron star in an LMXBs spins up by accreting matter from the companion. The  spun-up neutron star is still visible in X-rays when the accretion phase ends due to the detachment of the companion and/or when the companion has lost its atmosphere \citep[e.g,][]{2020arXiv201009005D}.  Detached MSPs  in general have soft X-ray emission of 0.5--2.5~keV and $L_{x}\sim10^{30-31}$\,erg\,s$^{-1}$ \citep[e.g,][]{2017MNRAS.472.3706B, 1999A&A...341..803B}. There are rare cases of isolated pulsars, which show a slightly harder spectrum \citep[e.g,][]{2016ApJ...831...21M}. Therefore, in general, they can be distinguished from the quiescent LMXBs which have $L_{x}\sim10^{31-33}$\,erg\,s$^{-1}$. The emission of the MSPs comes from a smaller region than that of QLMXBs:  the radius inferred from a black-body model, or neutron star atmosphere model, fit to the X-ray spectrum of MSPs is $\sim$0.1--0.3~km, or $\sim$0.5--0.2~km, respectively.  \citep[e.g,][]{2006ApJ...646.1104B}.

{\bf Active binary systems:} Magnetically active binary systems\,(e.g, RS Canum Venaticorum, or BY Draconis, etc)  consist of  at least one (sub)-giant (RS CVn systems), or two main sequence stars (BY Dra systems) \citep[e.g][]{1993ApJS...86..599D}. The X-ray emission in these systems is caused by the strong magnetic  activity, induced by rapid rotation in close binary system \citep[e.g,][]{1997ApJ...478..358D} and tend to have soft spectrum often with KT<2.5 keV.  BY Draconis are in general observed with a low X-ray luminosity of $10^{29-31}$\,erg\,s$^{-1}$ \citep[e.g,][]{1997ApJ...478..358D, 2005ApJ...625..796H}. RS CVn systems have several cases with luminosity $<10^{31}$\,erg\,s$^{-1}$, however, they can be observed at higher luminosities as well \citep[e.g,][]{1983MNRAS.205..447R, 1993ApJ...413..333D}. It also has to be considered that in these \erosita\, observations, we only detected sources with X-ray luminosities $L_{x}\gtrsim10^{31}$\,erg\,s$^{-1}$. Therefore, only luminous and high temperature active binaries are observed, which can show higher temperatures of $\sim$3.5 keV  \citep[e.g,][]{1993ApJ...413..333D}. 

The details of the spectrum and luminosity of X-ray sources together with the multi-wavelength information of their counterpart help to classify low luminosity X-ray sources. 
As an example, \citet{2003ApJ...591L.131P} have suggested a way to distinguish the quiescent LMXBs from the other types of low luminosity X-ray sources in GCs: only AWDs and quiescent LMXBs have $L_{x}$>$10^{32}$\,erg\,s$^{-1}$, while the quiescent LMXBs show much softer X-ray spectrum than that of AWDs. Therefore, quiescent LMXBs can be distinguished from the rest of the sources.  Moreover, the spectral model is useful for the classification of MSPs and QLMXBs. If the spectrum of a source is fitted with either a black-body and/or a neutron star atmosphere model component in such way that a reasonable radius can be inferred from the model(s), the source is rather a QLMXB or a MSPs candidate than a CV, for which we expect its soft spectrum to be fitted better with a plasma atmosphere model. The radii of black-body model and/or neutron star atmosphere model would help to characterise the neutron star. 

The faintest object detected by \erosita\, in the field of 47~Tuc is Src-No.\,330 with a flux of $\sim$4.6$\times10^{-15}$\,erg\,s$^{-1}$\,cm$^{-2}$\,(assuming located at the distance of 47~Tuc: $L_{x}$=9.9$\times10^{30}$\,erg\,s$^{-1}$). Therefore, in principle  we are able  to observe different types of low luminosity X-ray sources of 47~Tuc with $L_{x}$>$10^{31}$\,erg\,s$^{-1}$. 
In order to be more precise in the classification of members of 47~Tuc, we only confirm an optical/infrared source as the counterpart for the X-ray source if it is within the 2$\sigma$ X-ray positional error circle and if there are no multiple candidates as counterparts in the error circle.
Fig.~\ref{opt-counterpart} and Fig.~\ref{2mass-plot} show diagrams of the properties of the near-infrared and optical counterparts of the X-ray. The sources, which have a counterpart as a star on the main sequence or on the RGB in  47~Tuc are marked in blue and green circles, respectively. All these counterparts are confirmed as members of 47~Tuc \citep{2011A&A...530A..31L} and the blue sources are also classified as red giants \citep[e.g,][]{2014ApJ...780...94C} (see Sect.~\ref{47-Tuc-member}). We double checked the position of the X-ray source with the \chandra\, position in case the sources was detected with \chandra\, as well to improve the positional accuracy for the counterpart selection. Figure~\ref{infrared-image} shows infrared 2MASS images of the position of these X-ray sources. Based on the X-ray analysis we classify the sources as it is explained in the following:
\begin{table*}
\centering
\caption{List of the X-ray sources, which are members of 47~Tuc  \label{47tuxmember}}
\footnotesize
\centering
\addtolength{\tabcolsep}{-0.1cm}   
\begin{tabular}{ccccccccccl}
\hline\hline
NO   &RA              &DEC              &r1$\sigma$                       & \multicolumn{5}{c}{count-rate\,(0.2--5.~keV)}  &Var&Note$^{\dagger}$\\
                          &  (J2000)       &  (J2000)        & ($\arcsec$)                     & \multicolumn{5}{c}{(cts\,s$^{-1}$)}&&\\
                           &                &                 &                                 & OBS1&          OBS2&          OBS3&  OBS4&          OBS5&&\\
\hline
    263&           00 21 37.25  & -71 38 59.8     &         2.52&               --&               0.007$\pm$               0.001&               --&               --&               --&                  --&  CV/\,active binary \\
     267& 00 21 40.94  & -72 15 38.4   &0.98  & --&0.007$\pm$   0.001&0.008 $\pm$     0.002&--&0.006 $\pm$    0.001&  1.33$\pm$                0.29&CV \\
    320&            00 22 26.07  & -72 06 49.2     &              1.15&                --&               0.003$\pm$               0.001&               0.004$\pm$               0.289&               0.003$\pm$               0.001&               0.004$\pm$               0.001&                   1.30$\pm$               68.06&  CV \\
    340&          00 22 45.28  & -71 59 08.3&        1.20&               0.066$\pm$               0.008&               0.076$\pm$               0.002&               0.074$\pm$               0.003&               0.074$\pm$               0.002&               0.067$\pm$               0.002&                   1.15$\pm$                0.15&  Quiescent LMXB \\
    341&                 00 22 46.04  & -72 10 22.8   &   0.70&             0.011$\pm$               0.299&               0.008$\pm$               0.001&               0.009$\pm$               0.001&               0.007$\pm$               0.001&               0.008$\pm$               0.001&                   1.67$\pm$               26.23& CV \\
    376&           00 23 16.89  & -72 02 24.3  &              2.21&               --&               --&               --&               --&               0.002$\pm$               0.001&                  --& CV/\,active binary  \\
    378&          00 23 19.20  & -71 48 26.9 &          2.05&               --&               0.006$\pm$               0.001&               --&               0.003$\pm$               0.001&               --&                   1.76$\pm$                0.41&  CV/\,active binary   \\
    395&           00 23 31.36  & -72 11 00.4 &         2.48&               --&               --&               0.003$\pm$               0.001&               0.004$\pm$               0.001&               --&                   1.70$\pm$                0.51&CV/\,active binary  \\
    438&             00 24 03.66  & -71 55 48.0 &              1.60&            --&               0.002$\pm$               0.001&               --&               0.002$\pm$               0.000&               --&                   1.63$\pm$                0.49& Symbiotic star  \\
    480&    	00 24 43.35&  -72 18 25.50&       1.30&           0.022$\pm$               0.004&               0.032$\pm$               0.003&               0.022$\pm$               0.001&               0.028$\pm$               0.003&               0.021$\pm$               0.001&                   1.57$\pm$                0.15&  Unclassified  \\
    481&          00 24 44.21  & -72 08 19.0     &          1.16&               --&               0.006$\pm$               0.001&               0.007$\pm$               0.001&               0.007$\pm$               0.001&               0.008$\pm$               0.001&                   1.19$\pm$                0.26& Quiescent LMXB\\
    486&            00 24 45.96  & -72 08 59.0     &        2.95&               --&               0.004$\pm$               0.001&               --&               --&               -&                   --&  CV or active binary\\
    501&             00 24 56.97  & -72 06 51.8     &              1.5&              0.047$\pm$               0.005&               0.042$\pm$               0.002&               0.044$\pm$               0.002&               0.041$\pm$               0.002&               0.045$\pm$               0.002&                   1.14$\pm$                0.15&  CV \\
    522&              00 25 18.95  & -71 58 00.9     &             2.40&              --&               --&               --&               0.002$\pm$               0.000&               --&                  --& CV or active binary \\
    556&           00 25 43.48  & -72 18 52.1     &    1.52&               --&               --&               0.005$\pm$               0.001&               --&               0.006$\pm$               0.001&                   1.33$\pm$                0.29&  RS\,CVn\\
\hline
\hline
      \multicolumn{11}{l}{}\\
\end{tabular}
 \end{table*}
 
\subsubsection{Sources with an RGB counterpart}
Following X-ray sources have an RGB counterpart \citep{2014ApJ...780...94C, 2013A&A...550A..34C}, which is classified as a member of 47~Tuc.  They are candidates for different types of symbiotic stars or RS CVn active binaries in 47~Tuc:

\textbf{Src-No.\,438:} The brightest RGB counterpart belongs to this source (see Fig.~\ref{2mass-plot} and Fig.~\ref{opt-counterpart}).  The counterpart is a known asymptotic giant branch Star in 47 Tuc \citep{2014ApJ...780...94C} and located in 2$\sigma$ error circle of the X-ray source. The X-ray luminosity of the source is $L_{x}\gtrapprox$\,$10^{31}$\,erg\,s $^{-1}$ and the only significant hardness ratio is $HR2=0.19\pm0.21$, which means that the source is mainly detected in 0.6-2.3~keV.
 The infrared WISE and 2MASS counterparts suggest colours of $J-H=0.90\pm0.06$ and $K-W3=0.11\pm0.03$ for the source, which agrees well with the infrared colours of the symbiotic stars (see Sect.~\ref{infra-sec}). Therefore, the source is a candidate for a symbiotic star in 47~Tuc.

\textbf{Src-No.556:}  The sources with an X-ray luminosity of  $L_{x}\,\sim$4.$\times10^{31}$\,erg\,s$^{-1}$ can be a candidate for either a symbiotic star or an active binary (RS CVn type). According to the  hardness ratios the main part of the X-ray emission is <2.4 keV.  The source has a \chandra\, counterpart. According to \chandra\, position the red giant counterpart is located within 2$\sigma$ positional error and the \erosita\, position show the the red giant is overlapped with 2$\sigma$ X-ray positional error (see Fig.~\ref{infrared-image}).  The red~giant counterpart has a colour of $J-H<0.78$, which clarifies that the source has a low chance of being a symbiotic star (see Sect.~\ref{infra-sec}). It is more likely a candidate for RS~CVn contact binary.

\subsubsection{Sources with a counterpart in the main sequence}
Sources that have an optical or infrared counterpart on the main-sequence are marked with green circles in the colour magnitude diagrams  (see Fig.~\ref{opt-counterpart} and Fig.~\ref{2mass-plot}). These diagrams show that the sources are located on the main sequence of  47~Tuc. In general, with a main sequence counterpart, low luminosity X-ray sources are candidates for quiescent LMXBs, MSPs, CVs, or active binary systems.

\textbf{Src-No.\,263, Src-No.\,376, Src-No.\,395, src-No.\,486, Src-No.\,522:} They are  only detected in one or two observations. The HRs of these sources were not significant. The infrared and optical counterpart of these sources confirmed as a member of 47~Tuc within 1-2$\sigma$ X-ray positional error \citep{2011ApJS..193...23M, 2011A&A...530A..31L, 2017MNRAS.471.1446N}. According to their X-ray luminosities ($\sim10^{31}$\,erg\,s$^{-1}$) they can be candidates for either active binaries in a flare state or faint, variable, and non-magnetic CVs.  

\textbf{Src-No.\,267:}  The optical/infrared counterpart of the source, which is located which in ~1$\sigma$ X-ray positional error is already classified as a peculiar star \citep[Cl* NGC 104 LEE 2531;][]{2000A&AS..143....9W} and a member of 47~Tuc \citep{2011ApJS..193...23M}. The optical and infrared colour magnitude diagrams show that the counterpart located in horizontal branch of 47~Tuc (see Fig.~\ref{opt-counterpart} and Fig.~\ref{inf-count}). The HR diagrams shows that the source it is hard source and its luminosity ($L_{x}\sim$\,$10^{32}$\,erg\,s $^{-1}$) suggest that it can be candidate for either a non-magnetic CV or QLMXB. Considering the significant emission above >\,2.0~keV it can not be a candidate for an active binary. 

\textbf{Src-No.\,320:}  Source has no infrared counterpart. Its optical counterpart, which is located 2.4$\arcsec$ away from the X-ray position is considered as a post-main sequence star in 47~Tuc \citep{2011ApJS..193...23M} and the Gaia colour magnitude diagram shows it  on the main sequence (see Fig.~\ref{opt-counterpart}). The spectrum of the source is highly absorbed in soft X-rays and has the main emission >1.~keV (Fig.~\ref{spec.fig} and Table~\ref{spectral-Table}). The source shows no X-ray variability and has an absorbed $L_{x}\sim$\,$1.8\times10^{31}$\,erg\,s $^{-1}$.  The spectrum is too hard to make it an active binary candidate. We tried to fit the spectrum with an absorbed black-body model and also an absorbed nsa model (see Table~\ref{spectral-Table}). An absorbed black-body model suggest a radius of <\,0.1~km, and an absorbed nsa model an effective radius of <\,0.8~km, which are both typical for MSPs using these two models (see Sect.~\ref{member-sec}), while the temperatures of these two models are much higher than those of MSPs \citep[][]{2006ApJ...646.1104B}. Therefore the possibility that the compact object is a neutron star is very low. The source is most probably a CV candidate.

\textbf{Src-No.\,340:} { The optical/infrared counterpart of Src-No.\,340 is a main sequence star considering 2$\sigma$ \chandra\, and \erosita\, positional errors of the source (see Fig.~\ref{infrared-image}) and classified as a member of 47~Tuc \citep{2011ApJS..193...23M}.  We tried different models for the spectrum of the source. The best fit is with an absorbed black-body plus a power-law tail (Table~\ref{spectral-Table} and Fig.~\ref{spec.fig}).  The black-body model and the nsa model suggest a radius of <\,1.54~km and 13.5$^{+13.2}_{-5.6}$~km, respectively, which are both reasonable enough to assume the source as candidate for QLMXBs. Considering the counterpart, luminosity, and the X-ray energy range source is a candidate for QLMXBs.}

\textbf{Src-No.\,341:} The main emission of the source is detected between 0.5--5.0~keV with an X-ray luminosity of  $\sim$4.$\times10^{31}$\,erg\,s$^{-1}$.  There is no infrared counterpart for the source. There is a Gaia counterparts within the 1$\sigma$ positional error of the X-ray source source, which is located on the main sequence of 47~Tuc. This counterpart is also classified as a member of 47~Tuc \citep{2017MNRAS.471.1446N}.  We tried to fit the spectrum with a absorbed black-body plus a power-law model (for the hard tail component), which was failed. It was also unsuccessful while trying nsa model. Therefore, the possibility that Src-No.\,341 is either a MSPs with a hard power-law tail or a QLMXB is very low (see the discussion in Sect.~\ref{member-sec}). Moreover,  we could not fit the spectrum with two ionised plasma model components as  observed in the spectrum of bright active binaries. The hard {tail of the} spectrum (>\,2.0~keV) , which is best fit with a power-law model with a photon index <2.0 suggests that it is more likely a CV candidate rather than an active binary.

\textbf{Src-No.\,378:} The counterpart of the source, which is located within 1$\sigma$ positional error is classified as a member of 47~Tuc but not as an RGB star in available catalogues (see Sect.~\ref{47-Tuc-member}). However, The position of the source in infrared colour-magnitude diagram suggests that the counterpart is on RGB (see Fig.~\ref{2mass-plot}). Also, the position of the Gaia counterpart in optical colour magnitude diagrams shows that it is  about to leave the main sequence (see Fig.~\ref{opt-counterpart}).  The study of infrared colours of the source shows that $J-H<0.78$ and thus, the possibility of being a symbiotic star is very low (see Sect.~\ref{infra-sec}) The source was bright enough for neither spectral analysis nor HR study. X-ray luminosity of  $L_{x}\sim$\,$4\times10^{31}$\,erg\,s $^{-1}$ suggests that it can be  a CV or an active binary.

\textbf{Src-No.\,480:}  \erosita\, 2$\sigma$  positional error of  Src-No.480  shows no counterpart, while the \chandra\,  2$\sigma$  positional error shows a main sequence counterpart for the source classified as a member of 47~Tuc \citep{2011ApJS..193...23M}. There is also a red giant belong to 47~Tuc \citep{2011ApJS..193...23M}, which has an overlap with  2$\sigma$ positional error of both \erosita\, and \chandra\, (see Sect.~\ref{infrared-image}). It can not be clear if the emission is correlated with either the main sequence star or the red giant counterpart. We tried to fit the spectrum with different models. The best fit is obtained assuming two component \texttt{apec} model, which is similar to the spectra of the symbiotic stars. As Table~\ref{spectral-Table} shows the black-body plus power-law model does not fit well to the spectrum. Ineligible result was the same when we applied the nsa model. These facts make the nature of the source unclear if it is a symbiotic star, QLMXBs, or a CV. Therefore, we kept this source unclassified.

\textbf{Src-No.\,481:} The spectrum of the source is soft\,(see Fig.~\ref{spec.fig} and Table~\ref{spectral-Table}). The \erosita\, position shows only a \chandra\, counterpart, which also is considered to search for optical/infrared counterparts for the source. There was an optical counterpart within the 2$\sigma$ positional error\,(see Fig.\ref{infrared-image}). The counterpart is a star reported as a 47~Tuc members \citep{2015AJ....150..176C}.  We tried both black-body and nsa models for the spectrum (see Table~\ref{spectral-Table}). The radius of emitting surface using the absorbed black-body model is 3.6$^{+2.8}_{-1.3}$~km, and using the nsa model is 23$^{-11.}_{+13}$, which makes this source a candidate for a QLMXB \citep[e.g,][]{2003ApJ...598..501H}.

\textbf{Src-No.\,501:}  The source has a \chandra\, counterpart. The position and positional error of \chandra\, has been considered to search for the counterpart as well. Infrared/optical counterparts are located within 2$\sigma$ of the X-ray positional error of \erosita\, and  3$\sigma$ of \chandra\, X-ray source. The spectra of the source (Fig.~\ref{spec.fig} and Table~\ref{spectral-Table}) is fitted well with  an absorbed two component thermal model \citep[ collisionally ionized diffuse gas model, \texttt{apec},][]{2001ApJ...556L..91S}. According to the counterpart\,(, which is about to leave the main sequence) and luminosity\,($>10^{32}$\,erg\,s $^{-1}$), it is either a  CV or QLMXB candidate. The spectrum of the sources is fitted well with two apec models (see Fig.~\ref{spec.fig} and Table~\ref{spectral-Table}), while the attempt to fit it with an absorbed black-body and power-law model was problematic. We fitted an absorbed neutron star atmosphere model plus a power-law component to the spectrum. However, the temperature of neutron star atmosphere model was too high to yield a reasonable neutron star radius (see Sect.~\ref{member-sec}). Therefore, the source is rather a CV candidate.  

\subsection{other sources:}
\textbf{Src-No.\,552:} The source has a counterpart in optical and infrared located within 2$\sigma$ positional error of \erosita\, and 3$\sigma$ positional error of \chandra\, (see Fig.~\ref{infrared-image}).  The counterpart is already classified as the Mira star in the catalogue of long period variable stars of Small Magellanic Cloud\,(SMC) \citep{2011AcA....61..217S} and as optical/infrared colour magnitude diagrams show (see Fig.\ref{2mass-plot} and Fig\ref{opt-counterpart}), it is not in the population of 47~Tuc. The HRs show that the source should be observable in hard X-ray (>2.0keV, see Fig.\ref{hrs-plot}) and the luminosity of the source assuming a distance of 62.4~kpc for SMC is   $\sim10^{34}$\,erg\,s$^{-1}$, which makes a candidate for an X-ray binary in SMC.
\subsection{X-ray luminosity function (XLF)}
\label{xlf-sec}
\begin{figure}

\includegraphics[clip, trim={0.0cm  0.0cm  1.0cm  0.cm},width=0.47\textwidth]{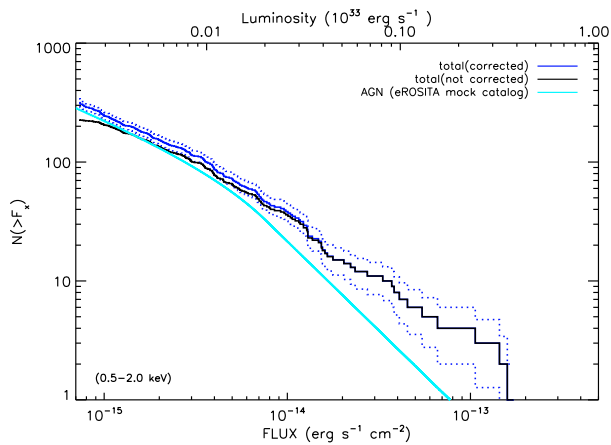}

\caption{ X-ray luminosity function in the field of 47~Tuc observed by \erosita. 
\label{xlf-image}}
\end{figure}
\chandra\, observation of 47~Tuc was limited to the central region with a radius of 2$\arcmin$.7 (See Sect.~\ref{intro}), while  \erosita\, observation gives an opportunity to study a large area surrounding this GC.
In order to provide a more comprehensive view of the population of X-ray binaries in 47~Tuc, we calculated the XLF of the central annular area with inner and outer radii of 1$\arcmin$.7 and 18$\arcmin$.8, respectively (see Fig.~\ref{rgb-image}). The inner region is excluded since there is only an unresolved emission from the X-ray sources located within  1$\arcmin$.7 central region of 47~Tuc in \erosita\, observations. The outer radius includes the area, which is covered by all observations and is therefore expected to have an almost uniform exposure time. 
In the first step, we ran source detection for the merged event files of all observations in the energy band of 0.5--2.0~keV. The sources, which have been classified as foreground stars and diffuse sources were removed from the list, so 226 sources remained within 18$\arcmin$.8.  In the second step, we corrected the XLF for incompleteness. It is expected that the sensitivity of the detectors is not uniform over the analysed region. To estimate the incompleteness for a flux range, we need to know the fraction of the area, in which the detectors were sensitive enough to detect a faint source and then correct the number of sources for this incompleteness. For this purpose, we created a sensitivity map of the combined event file of all observations in the energy range of 0.5--2.0 keV using the \texttt{eSASS} task, \texttt{apetool}. The sensitivity map gives the detection upper-limits for each pixel of the event file. Using the sensitivity map the cumulative XLF is corrected for  incompleteness by the following formula: 
\begin{equation}
N(>F_\mathrm{x})=A_\mathrm{tot} \sum_\mathrm{i=1}^{N_\mathrm{s}} \frac{1}{\omega(F_\mathrm{i})}\,,
\end{equation}
where $N(>F_\mathrm{x})$ is the number of sources with a flux higher than $F_\mathrm{x}$. For each source with a flux $F_\mathrm{i}$, the number is weighted by the normalised effective area $A_\mathrm{tot}/\omega(F_\mathrm{i})$, where $A_\mathrm{tot}$ is the total area, which have been used to calculate the XLF (i.e, annulus area = 1101.28 arcmin$^{2}$). $\omega(F_\mathrm{i})$ is the area of pixels, which are sensitive enough to detect sources with a flux $\geq F_\mathrm{x}$. $N_\mathrm{s}$ is the number of detected sources with a flux above $\geq F_\mathrm{x}$. In this way the detection of every source is weighted and the XLF gets corrected for the incompleteness. 

The XLF includes the members of 47~Tuc and background sources, mainly AGNs. We base the estimation of the number of background AGNs, which should be subtracted from the observed XLF, on the study of  \citet{2019MNRAS.487.2005C}. As  \citet{2019MNRAS.487.2005C} (their Figure~10) show, the simulation of AGN logN–logS distribution of \erosita\, in the flux range of $10^{-16}$ to $10^{-13}$\,erg\,s$^{-1}$\,cm$^{-2}$ (0.5--2.0~keV) is very well consistent with the results of \citet{2008MNRAS.388.1205G}. Therefore, to obtain the AGN logN–logS distribution for \erosita, we used the broken power-law model suggested by \citet{2008MNRAS.388.1205G} in the energy range of 0.2--5.0~keV:
\begin{equation}\label{cappellutieq}
\frac{dN}{df_{\chi}}=\begin{cases}
K\left(\frac{f_{\chi}}{f_{ref}}\right)^{\beta_{1}}\qquad  f_{\chi} < f_{b}\\
~\\
K^{\prime}\left(\frac{f_{\chi}}{f_{ref}}\right)^{\beta_{2}}\qquad  f_{\chi} \geq f_{b}\,,
\end{cases}
\end{equation}
where, $K^{\prime}=({f_{b}}/{f_{ref}})^{\beta_{1}-\beta_{2}}$, the break is $f_{b}$=$10^{-14}$\,erg\,s$^{-1}$\,cm$^{-2}$, $K$ is  $1.51\times10^{16}$deg$^{-2}$/erg\,s$^{-1}$\,cm$^{-2}$, and $\beta_{1}$ and $\beta_{2}$ are the power-law indexes for the fluxes lower and higher than the break, --1.58 and --2.50, respectively. The cyan line in the plot (Fig.~\ref{xlf-image}) shows the XLF of AGNs modified for the area of studied region (0.306 deg$^{2}$) taking into account the Galactic absorption in the direction of 47~Tuc (i.e, 5.5$\times10^{20}$\,cm$^{-2}$).The flux of AGNs in the study of \citet{2008MNRAS.388.1205G} is estimated assuming a power-law model with a photon index of 1.4. To calculate the flux of the sources in the energy range of 0.5--2.0~keV, we assumed the same model.  In Fig.~\ref{xlf-image}, the black line shows the observed source distribution, the dark blue line shows the distribution corrected for incompleteness, and the light blue line presents the AGN logN–logS distribution. 
One can see that there is very small fraction of brighter sources in excess over the background distribution that might  belong to 47 Tuc. For low luminosities, no significant excess is observed.

\section{Summary}
In this work we presented the results of analysis of five \erosita\, observations with the aim of the classification of X-ray sources in the field of this globular cluster. source detection has been separately performed for five observations of \erosita\, and 888 sources has been detected in the energy range of 0.2--5.0~keV. Using different methods of X-ray analysis consist of spectral and timing analyses, together with the multi-wavelength studies of the counterparts of X-ray sources in optical, near infrared and infrared surveys, a comprehensive study has been perform for the X-ray sources in the field of 47~Tuc, which resulted to the accurate classification of  15 X-ray sources as members of 47~Tuc. We identified  1 symbiotic stars,   2 quiescent low mass X-ray binaries, and   3 cataclysmic variables. There are  5 sources, which are candidates for either cataclysmic variables or contact binaries and  1 sources, which are candidates for RS~CVn contact binaries. Moreover,   126 AGNs and background galaxies and  25 Galactic foreground stars are classified in the field of 47~Tuc.   We identified 18 sources, which have soft X-ray emission <\,0.2~keV and 85 sources, which had hard X-ray emission >\,2.0~keV.  We could specifically classify one of the foreground stars as an flaring M~dwarf based on X-ray variability, spectral analysis, and its infrared/optical counterpart.  The XLF of 47~Tuc has been calculated. The result shows that there is no significant sign of population of X-ray sources that belongs to the globular cluster at low luminosities.
\begin{acknowledgements}
This  research  was  funded  by  the DLR research grant BWWI/DLR~500R1907. This study is based on observations obtained with \erosita, primary  instrument  aboard  SRG,  a  joint Russian-German  science  mission  supported  by  the  Russian  Space  Agency(Roskosmos), in the interests of the Russian Academy of Sciences represented by its Space Research Institute (IKI), and the Deutsches Zentrum für Luft- und Raumfahrt  (DLR).  The  SRG  spacecraft  was  built  by  Lavochkin  Association (NPOL)  and  its  subcontractors,  and  is  operated  by  NPOL  with  support  from IKI and the Max Planck Institute for Extraterrestrial Physics (MPE). The development and construction of the eROSITA X-ray instrument was led by MPE,with  contributions  from  the  Dr.  Karl  Remeis  Observatory  Bamberg \& ECAP (FAU Erlangen-Nürnberg), the University of Hamburg Observatory, the Leibniz  Institute  for  Astrophysics  Potsdam  (AIP),  and  the  Institute  for  Astronomy and Astrophysics of the University of Tübingen, with the support of DLR and the Max Planck Society. The eROSITA data shown here were processed using the eSASS software system developed by the German eROSITA consortium. This  research  has  made  use  of  the  SIMBAD  and  VIZIER  database,  operated at  CDS,  Strasbourg,  France,  and  of  the  NASA/IPAC  Extra-galactic  Database\,(NED),  which  is  operated  by  the  Jet  Propulsion  Laboratory,  California  Institute  of  Technology,  under  contract  with  the  National  Aeronautics  and  Space Administration. This publication makes use of data products from the Wide field Infrared Survey Explorer, which is a joint project of the University of California, Los Angeles, and the Jet Propulsion Laboratory/California Institute of Technology, funded by the National Aeronautics and Space Administration. This publication has made use of data products from the Two Micron All Sky Survey, which is a joint project of the University of Massachusetts and the Infrared Processing  and  Analysis  Center,  funded  by  the  National  Aeronautics  and  Space Administration  and  the  National  Science  Foundation.  Funding  for  SDSS  and SDSS-III has been provided by the Alfred P. Sloan Foundation, the Participating  Institutions,  the  National  Science  Foundation,  and  the  US  Department  of Energy Office of Science. The SDSS-III web site ishttp://www.sdss3.org/.SDSS-III  is  managed  by  the  Astrophysical  Research  Consortium  for  the  Participating  Institutions  of  the  SDSS-III  Collaboration  including  the  University of  Arizona,  the  Brazilian  Participation  Group,  Brookhaven  National  Laboratory, University of Cambridge, University of Florida, the French Participation Group, the German Participation Group, the Instituto de Astrofisica de Canarias, the Michigan State/Notre Dame/JINA Participation Group, Johns Hopkins University,  Lawrence  Berkeley  National  Laboratory,  Max  Planck  Institute  for Astrophysics, New Mexico State University, New York University, Ohio State University, Pennsylvania State University, University of Portsmouth, Princeton University, the Spanish Participation Group, University of Tokyo, University of Utah, Vanderbilt University, University of Virginia, University of Washington, and Yale University. This research has made use of SAO Image DS9, developed by Smithsonian Astrophysical Observatory.
\end{acknowledgements}

\bibliographystyle{aa} 
\bibliography{bibtex.bib}

\begin{thebibliography}{98}
\expandafter\ifx\csname natexlab\endcsname\relax\def\natexlab#1{#1}\fi

\bibitem[{{Akras} {et~al.}(2019){Akras}, {Leal-Ferreira}, {Guzman-Ramirez}, \&
  {Ramos-Larios}}]{2019MNRAS.483.5077A}
{Akras}, S., {Leal-Ferreira}, M.~L., {Guzman-Ramirez}, L., \& {Ramos-Larios},
  G. 2019, \mnras, 483, 5077

\bibitem[{{Altamirano} {et~al.}(2008){Altamirano}, {Casella}, {Patruno},
  {Wijnands}, \& {van der Klis}}]{2008ApJ...674L..45A}
{Altamirano}, D., {Casella}, P., {Patruno}, A., {Wijnands}, R., \& {van der
  Klis}, M. 2008, \apjl, 674, L45

\bibitem[{{Auriere} {et~al.}(1989){Auriere}, {Koch-Miramond}, \&
  {Ortolani}}]{1989A&A...214..113A}
{Auriere}, M., {Koch-Miramond}, L., \& {Ortolani}, S. 1989, \aap, 214, 113

\bibitem[{{Bacher} {et~al.}(2005){Bacher}, {Kimeswenger}, \&
  {Teutsch}}]{2005MNRAS.362..542B}
{Bacher}, A., {Kimeswenger}, S., \& {Teutsch}, P. 2005, \mnras, 362, 542

\bibitem[{{Bahramian} {et~al.}(2014){Bahramian}, {Heinke}, {Sivakoff},
  {Altamirano}, {Wijnands}, {Homan}, {Linares}, {Pooley}, {Degenaar}, \&
  {Gladstone}}]{2014ApJ...780..127B}
{Bahramian}, A., {Heinke}, C.~O., {Sivakoff}, G.~R., {et~al.} 2014, \apj, 780,
  127

\bibitem[{{Bailer-Jones} {et~al.}(2019){Bailer-Jones}, {Fouesneau}, \&
  {Andrae}}]{2019MNRAS.490.5615B}
{Bailer-Jones}, C. A.~L., {Fouesneau}, M., \& {Andrae}, R. 2019, \mnras, 490,
  5615

\bibitem[{{Bailer-Jones} {et~al.}(2021){Bailer-Jones}, {Rybizki}, {Fouesneau},
  {Demleitner}, \& {Andrae}}]{2021yCat.1352....0B}
{Bailer-Jones}, C.~A.~L., {Rybizki}, J., {Fouesneau}, M., {Demleitner}, M., \&
  {Andrae}, R. 2021, VizieR Online Data Catalog, I/352

\bibitem[{{Bailer-Jones} {et~al.}(2018){Bailer-Jones}, {Rybizki}, {Fouesneau},
  {Mantelet}, \& {Andrae}}]{2018AJ....156...58B}
{Bailer-Jones}, C.~A.~L., {Rybizki}, J., {Fouesneau}, M., {Mantelet}, G., \&
  {Andrae}, R. 2018, \aj, 156, 58

\bibitem[{{Balman}(2012)}]{2012MmSAI..83..585B}
{Balman}, S. 2012, \memsai, 83, 585

\bibitem[{{Bassa} {et~al.}(2004){Bassa}, {Pooley}, {Homer}, {Verbunt},
  {Gaensler}, {Lewin}, {Anderson}, {Margon}, {Kaspi}, \& {van der
  Klis}}]{2004ApJ...609..755B}
{Bassa}, C., {Pooley}, D., {Homer}, L., {et~al.} 2004, \apj, 609, 755

\bibitem[{{Bassa} {et~al.}(2008){Bassa}, {Pooley}, {Verbunt}, {Homer},
  {Anderson}, \& {Lewin}}]{2008A&A...488..921B}
{Bassa}, C.~G., {Pooley}, D., {Verbunt}, F., {et~al.} 2008, \aap, 488, 921

\bibitem[{{Becker} \& {Tr{\"u}mper}(1999)}]{1999A&A...341..803B}
{Becker}, W. \& {Tr{\"u}mper}, J. 1999, \aap, 341, 803

\bibitem[{{Belloni} {et~al.}(2019){Belloni}, {Giersz}, {Rivera Sandoval},
  {Askar}, \& {Cieciel{\r{a}}g}}]{2019MNRAS.483..315B}
{Belloni}, D., {Giersz}, M., {Rivera Sandoval}, L.~E., {Askar}, A., \&
  {Cieciel{\r{a}}g}, P. 2019, \mnras, 483, 315

\bibitem[{{Bernardini} \& {Cackett}(2014)}]{2014MNRAS.439.2771B}
{Bernardini}, F. \& {Cackett}, E.~M. 2014, \mnras, 439, 2771

\bibitem[{{Bhattacharya} {et~al.}(2017){Bhattacharya}, {Heinke}, {Chugunov},
  {Freire}, {Ridolfi}, \& {Bogdanov}}]{2017MNRAS.472.3706B}
{Bhattacharya}, S., {Heinke}, C.~O., {Chugunov}, A.~I., {et~al.} 2017, \mnras,
  472, 3706

\bibitem[{{Bogdanov} {et~al.}(2006){Bogdanov}, {Grindlay}, {Heinke}, {Camilo},
  {Freire}, \& {Becker}}]{2006ApJ...646.1104B}
{Bogdanov}, S., {Grindlay}, J.~E., {Heinke}, C.~O., {et~al.} 2006, \apj, 646,
  1104

\bibitem[{{Brown} {et~al.}(1998){Brown}, {Bildsten}, \&
  {Rutledge}}]{1998ApJ...504L..95B}
{Brown}, E.~F., {Bildsten}, L., \& {Rutledge}, R.~E. 1998, \apjl, 504, L95

\bibitem[{{Brunner} {et~al.}(2021){Brunner}, {Liu}, {Lamer}, {Georgakakis},
  {Merloni}, {Brusa}, {Bulbul}, {Dennerl}, {Friedrich}, {Liu}, {Maitra},
  {Nandra}, {Ramos-Ceja}, {Sanders}, {Stewart}, {Boller}, {Buchner}, {Clerc},
  {Comparat}, {Dwelly}, {Eckert}, {Finoguenov}, {Freyberg}, {Ghirardini},
  {Gueguen}, {Haberl}, {Kreykenbohm}, {Krumpe}, {Osterhage}, {Pacaud},
  {Predehl}, {Reiprich}, {Robrade}, {Salvato}, {Santangelo}, {Schrabback},
  {Schwope}, \& {Wilms}}]{2021arXiv210614517B}
{Brunner}, H., {Liu}, T., {Lamer}, G., {et~al.} 2021, arXiv e-prints,
  arXiv:2106.14517

\bibitem[{{Buccheri} {et~al.}(1983){Buccheri}, {Bennett}, {Bignami}, {Bloemen},
  {Boriakoff}, {Caraveo}, {Hermsen}, {Kanbach}, {Manchester}, {Masnou},
  {Mayer-Hasselwander}, {{\"O}zel}, {Paul}, {Sacco}, {Scarsi}, \&
  {Strong}}]{1983A&A...128..245B}
{Buccheri}, R., {Bennett}, K., {Bignami}, G.~F., {et~al.} 1983, \aap, 128, 245

\bibitem[{{Buccheri} {et~al.}(1988){Buccheri}, {di Gesu}, {Maccarone}, \&
  {Sacco}}]{1988A&A...201..194B}
{Buccheri}, R., {di Gesu}, V., {Maccarone}, M.~C., \& {Sacco}, B. 1988, \aap,
  201, 194

\bibitem[{{Campana} {et~al.}(1998){Campana}, {Colpi}, {Mereghetti}, {Stella},
  \& {Tavani}}]{1998A&ARv...8..279C}
{Campana}, S., {Colpi}, M., {Mereghetti}, S., {Stella}, L., \& {Tavani}, M.
  1998, \aapr, 8, 279

\bibitem[{{Carretta} {et~al.}(2013){Carretta}, {Gratton}, {Bragaglia},
  {D'Orazi}, \& {Lucatello}}]{2013A&A...550A..34C}
{Carretta}, E., {Gratton}, R.~G., {Bragaglia}, A., {D'Orazi}, V., \&
  {Lucatello}, S. 2013, \aap, 550, A34

\bibitem[{{Chen} {et~al.}(2018){Chen}, {Richer}, {Caiazzo}, \&
  {Heyl}}]{2018ApJ...867..132C}
{Chen}, S., {Richer}, H., {Caiazzo}, I., \& {Heyl}, J. 2018, \apj, 867, 132

\bibitem[{{Cheng} {et~al.}(2019){Cheng}, {Li}, {Li}, {Xu}, \&
  {Fang}}]{2019ApJ...876...59C}
{Cheng}, Z., {Li}, Z., {Li}, X., {Xu}, X., \& {Fang}, T. 2019, \apj, 876, 59

\bibitem[{{Cheng} {et~al.}(2018){Cheng}, {Li}, {Xu}, {Li}, {Zhu}, \&
  {Fang}}]{2018ApJ...869...52C}
{Cheng}, Z., {Li}, Z., {Xu}, X., {et~al.} 2018, \apj, 869, 52

\bibitem[{{Clark}(1975)}]{1975ApJ...199L.143C}
{Clark}, G.~W. 1975, \apjl, 199, L143

\bibitem[{{Cohen} {et~al.}(2015){Cohen}, {Hempel}, {Mauro}, {Geisler},
  {Alonso-Garcia}, \& {Kinemuchi}}]{2015AJ....150..176C}
{Cohen}, R.~E., {Hempel}, M., {Mauro}, F., {et~al.} 2015, \aj, 150, 176

\bibitem[{{Cominsky} {et~al.}(1977){Cominsky}, {Forman}, {Jones}, \&
  {Tananbaum}}]{1977ApJ...211L...9C}
{Cominsky}, L., {Forman}, W., {Jones}, C., \& {Tananbaum}, H. 1977, \apjl, 211,
  L9

\bibitem[{{Comparat} {et~al.}(2019){Comparat}, {Merloni}, {Salvato}, {Nandra},
  {Boller}, {Georgakakis}, {Finoguenov}, {Dwelly}, {Buchner}, {Del Moro},
  {Clerc}, {Wang}, {Zhao}, {Prada}, {Yepes}, {Brusa}, {Krumpe}, \&
  {Liu}}]{2019MNRAS.487.2005C}
{Comparat}, J., {Merloni}, A., {Salvato}, M., {et~al.} 2019, \mnras, 487, 2005

\bibitem[{{Cordero} {et~al.}(2014){Cordero}, {Pilachowski}, {Johnson},
  {McDonald}, {Zijlstra}, \& {Simmerer}}]{2014ApJ...780...94C}
{Cordero}, M.~J., {Pilachowski}, C.~A., {Johnson}, C.~I., {et~al.} 2014, \apj,
  780, 94

\bibitem[{{Cutri} \& {et al.}(2014)}]{2014yCat.2328....0C}
{Cutri}, R.~M. \& {et al.} 2014, VizieR Online Data Catalog, II/328

\bibitem[{{Cutri} {et~al.}(2003){Cutri}, {Skrutskie}, {van Dyk}, {Beichman},
  {Carpenter}, {Chester}, {Cambresy}, {Evans}, {Fowler}, {Gizis}, {Howard},
  {Huchra}, {Jarrett}, {Kopan}, {Kirkpatrick}, {Light}, {Marsh}, {McCallon},
  {Schneider}, {Stiening}, {Sykes}, {Weinberg}, {Wheaton}, {Wheelock}, \&
  {Zacarias}}]{2003yCat.2246....0C}
{Cutri}, R.~M., {Skrutskie}, M.~F., {van Dyk}, S., {et~al.} 2003, VizieR Online
  Data Catalog, II/246

\bibitem[{{Dai} {et~al.}(2015){Dai}, {Griffin}, {Kochanek}, {Nugent}, \&
  {Bregman}}]{2015ApJS..218....8D}
{Dai}, X., {Griffin}, R.~D., {Kochanek}, C.~S., {Nugent}, J.~M., \& {Bregman},
  J.~N. 2015, \apjs, 218, 8

\bibitem[{{Dempsey} {et~al.}(1993{\natexlab{a}}){Dempsey}, {Linsky}, {Fleming},
  \& {Schmitt}}]{1993ApJS...86..599D}
{Dempsey}, R.~C., {Linsky}, J.~L., {Fleming}, T.~A., \& {Schmitt}, J.~H.~M.~M.
  1993{\natexlab{a}}, \apjs, 86, 599

\bibitem[{{Dempsey} {et~al.}(1997){Dempsey}, {Linsky}, {Fleming}, \&
  {Schmitt}}]{1997ApJ...478..358D}
{Dempsey}, R.~C., {Linsky}, J.~L., {Fleming}, T.~A., \& {Schmitt}, J.~H.~M.~M.
  1997, \apj, 478, 358

\bibitem[{{Dempsey} {et~al.}(1993{\natexlab{b}}){Dempsey}, {Linsky}, {Schmitt},
  \& {Fleming}}]{1993ApJ...413..333D}
{Dempsey}, R.~C., {Linsky}, J.~L., {Schmitt}, J.~H.~M.~M., \& {Fleming}, T.~A.
  1993{\natexlab{b}}, \apj, 413, 333

\bibitem[{{Di Salvo} \& {Sanna}(2020)}]{2020arXiv201009005D}
{Di Salvo}, T. \& {Sanna}, A. 2020, arXiv e-prints, arXiv:2010.09005

\bibitem[{{Dotter} {et~al.}(2008){Dotter}, {Chaboyer}, {Jevremovi{\'c}},
  {Kostov}, {Baron}, \& {Ferguson}}]{2008ApJS..178...89D}
{Dotter}, A., {Chaboyer}, B., {Jevremovi{\'c}}, D., {et~al.} 2008, {The
  Dartmouth Stellar Evolution Database}

\bibitem[{{Edelson} \& {Malkan}(2012)}]{2012ApJ...751...52E}
{Edelson}, R. \& {Malkan}, M. 2012, \apj, 751, 52

\bibitem[{{Edmonds} {et~al.}(2003){Edmonds}, {Gilliland}, {Heinke}, \&
  {Grindlay}}]{2003ApJ...596.1177E}
{Edmonds}, P.~D., {Gilliland}, R.~L., {Heinke}, C.~O., \& {Grindlay}, J.~E.
  2003, \apj, 596, 1177

\bibitem[{{Flesch}(2019)}]{2019arXiv191205614F}
{Flesch}, E.~W. 2019, arXiv e-prints, arXiv:1912.05614

\bibitem[{{Gaia Collaboration}(2020)}]{2020yCat.1350....0G}
{Gaia Collaboration}. 2020, VizieR Online Data Catalog, I/350

\bibitem[{{Gaia Collaboration} {et~al.}(2018){Gaia Collaboration}, {Brown},
  {Vallenari}, {Prusti}, {de Bruijne}, {Babusiaux}, {Bailer-Jones}, {Biermann},
  {Evans}, {Eyer}, {Jansen}, {Jordi}, {Klioner}, {Lammers}, {Lindegren},
  {Luri}, {Mignard}, {Panem}, {Pourbaix}, {Randich}, {Sartoretti}, {Siddiqui},
  {Soubiran}, {van Leeuwen}, {Walton}, {Arenou}, {Bastian}, {Cropper},
  {Drimmel}, {Katz}, {Lattanzi}, {Bakker}, {Cacciari}, {Casta{\~n}eda},
  {Chaoul}, {Cheek}, {De Angeli}, {Fabricius}, {Guerra}, {Holl}, {Masana},
  {Messineo}, {Mowlavi}, {Nienartowicz}, {Panuzzo}, {Portell}, {Riello},
  {Seabroke}, {Tanga}, {Th{\'e}venin}, {Gracia-Abril}, {Comoretto},
  {Garcia-Reinaldos}, {Teyssier}, {Altmann}, {Andrae}, {Audard},
  {Bellas-Velidis}, {Benson}, {Berthier}, {Blomme}, {Burgess}, {Busso},
  {Carry}, {Cellino}, {Clementini}, {Clotet}, {Creevey}, {Davidson}, {De
  Ridder}, {Delchambre}, {Dell'Oro}, {Ducourant},
  {Fern{\'a}ndez-Hern{\'a}ndez}, {Fouesneau}, {Fr{\'e}mat}, {Galluccio},
  {Garc{\'\i}a-Torres}, {Gonz{\'a}lez-N{\'u}{\~n}ez}, {Gonz{\'a}lez-Vidal},
  {Gosset}, {Guy}, {Halbwachs}, {Hambly}, {Harrison}, {Hern{\'a}ndez},
  {Hestroffer}, {Hodgkin}, {Hutton}, {Jasniewicz}, {Jean-Antoine-Piccolo},
  {Jordan}, {Korn}, {Krone-Martins}, {Lanzafame}, {Lebzelter}, {L{\"o}ffler},
  {Manteiga}, {Marrese}, {Mart{\'\i}n-Fleitas}, {Moitinho}, {Mora}, {Muinonen},
  {Osinde}, {Pancino}, {Pauwels}, {Petit}, {Recio-Blanco}, {Richards},
  {Rimoldini}, {Robin}, {Sarro}, {Siopis}, {Smith}, {Sozzetti}, {S{\"u}veges},
  {Torra}, {van Reeven}, {Abbas}, {Abreu Aramburu}, {Accart}, {Aerts},
  {Altavilla}, {{\'A}lvarez}, {Alvarez}, {Alves}, {Anderson}, {Andrei},
  {Anglada Varela}, {Antiche}, {Antoja}, {Arcay}, {Astraatmadja}, {Bach},
  {Baker}, {Balaguer-N{\'u}{\~n}ez}, {Balm}, {Barache}, {Barata}, {Barbato},
  {Barblan}, {Barklem}, {Barrado}, {Barros}, {Barstow}, {Bartholom{\'e}
  Mu{\~n}oz}, {Bassilana}, {Becciani}, {Bellazzini}, {Berihuete}, {Bertone},
  {Bianchi}, {Bienaym{\'e}}, {Blanco-Cuaresma}, {Boch}, {Boeche}, {Bombrun},
  {Borrachero}, {Bossini}, {Bouquillon}, {Bourda}, {Bragaglia}, {Bramante},
  {Breddels}, {Bressan}, {Brouillet}, {Br{\"u}semeister}, {Brugaletta},
  {Bucciarelli}, {Burlacu}, {Busonero}, {Butkevich}, {Buzzi}, {Caffau},
  {Cancelliere}, {Cannizzaro}, {Cantat-Gaudin}, {Carballo}, {Carlucci},
  {Carrasco}, {Casamiquela}, {Castellani}, {Castro-Ginard}, {Charlot},
  {Chemin}, {Chiavassa}, {Cocozza}, {Costigan}, {Cowell}, {Crifo}, {Crosta},
  {Crowley}, {Cuypers}, {Dafonte}, {Damerdji}, {Dapergolas}, {David}, {David},
  {de Laverny}, {De Luise}, {De March}, {de Martino}, {de Souza}, {de Torres},
  {Debosscher}, {del Pozo}, {Delbo}, {Delgado}, {Delgado}, {Di Matteo},
  {Diakite}, {Diener}, {Distefano}, {Dolding}, {Drazinos}, {Dur{\'a}n},
  {Edvardsson}, {Enke}, {Eriksson}, {Esquej}, {Eynard Bontemps}, {Fabre},
  {Fabrizio}, {Faigler}, {Falc{\~a}o}, {Farr{\`a}s Casas}, {Federici},
  {Fedorets}, {Fernique}, {Figueras}, {Filippi}, {Findeisen}, {Fonti},
  {Fraile}, {Fraser}, {Fr{\'e}zouls}, {Gai}, {Galleti}, {Garabato},
  {Garc{\'\i}a-Sedano}, {Garofalo}, {Garralda}, {Gavel}, {Gavras}, {Gerssen},
  {Geyer}, {Giacobbe}, {Gilmore}, {Girona}, {Giuffrida}, {Glass}, {Gomes},
  {Granvik}, {Gueguen}, {Guerrier}, {Guiraud}, {Guti{\'e}rrez-S{\'a}nchez},
  {Haigron}, {Hatzidimitriou}, {Hauser}, {Haywood}, {Heiter}, {Helmi}, {Heu},
  {Hilger}, {Hobbs}, {Hofmann}, {Holland}, {Huckle}, {Hypki}, {Icardi},
  {Jan{\ss}en}, {Jevardat de Fombelle}, {Jonker}, {Juh{\'a}sz}, {Julbe},
  {Karampelas}, {Kewley}, {Klar}, {Kochoska}, {Kohley}, {Kolenberg},
  {Kontizas}, {Kontizas}, {Koposov}, {Kordopatis}, {Kostrzewa-Rutkowska},
  {Koubsky}, {Lambert}, {Lanza}, {Lasne}, {Lavigne}, {Le Fustec}, {Le
  Poncin-Lafitte}, {Lebreton}, {Leccia}, {Leclerc}, {Lecoeur-Taibi},
  {Lenhardt}, {Leroux}, {Liao}, {Licata}, {Lindstr{\o}m}, {Lister}, {Livanou},
  {Lobel}, {L{\'o}pez}, {Managau}, {Mann}, {Mantelet}, {Marchal}, {Marchant},
  {Marconi}, {Marinoni}, {Marschalk{\'o}}, {Marshall}, {Martino}, {Marton},
  {Mary}, {Massari}, {Matijevi{\v{c}}}, {Mazeh}, {McMillan}, {Messina},
  {Michalik}, {Millar}, {Molina}, {Molinaro}, {Moln{\'a}r}, {Montegriffo},
  {Mor}, {Morbidelli}, {Morel}, {Morris}, {Mulone}, {Muraveva}, {Musella},
  {Nelemans}, {Nicastro}, {Noval}, {O'Mullane}, {Ord{\'e}novic},
  {Ord{\'o}{\~n}ez-Blanco}, {Osborne}, {Pagani}, {Pagano}, {Pailler},
  {Palacin}, {Palaversa}, {Panahi}, {Pawlak}, {Piersimoni}, {Pineau}, {Plachy},
  {Plum}, {Poggio}, {Poujoulet}, {Pr{\v{s}}a}, {Pulone}, {Racero}, {Ragaini},
  {Rambaux}, {Ramos-Lerate}, {Regibo}, {Reyl{\'e}}, {Riclet}, {Ripepi}, {Riva},
  {Rivard}, {Rixon}, {Roegiers}, {Roelens}, {Romero-G{\'o}mez}, {Rowell},
  {Royer}, {Ruiz-Dern}, {Sadowski}, {Sagrist{\`a} Sell{\'e}s}, {Sahlmann},
  {Salgado}, {Salguero}, {Sanna}, {Santana-Ros}, {Sarasso}, {Savietto},
  {Schultheis}, {Sciacca}, {Segol}, {Segovia}, {S{\'e}gransan}, {Shih},
  {Siltala}, {Silva}, {Smart}, {Smith}, {Solano}, {Solitro}, {Sordo}, {Soria
  Nieto}, {Souchay}, {Spagna}, {Spoto}, {Stampa}, {Steele},
  {Steidelm{\"u}ller}, {Stephenson}, {Stoev}, {Suess}, {Surdej}, {Szabados},
  {Szegedi-Elek}, {Tapiador}, {Taris}, {Tauran}, {Taylor}, {Teixeira},
  {Terrett}, {Teyssandier}, {Thuillot}, {Titarenko}, {Torra Clotet}, {Turon},
  {Ulla}, {Utrilla}, {Uzzi}, {Vaillant}, {Valentini}, {Valette}, {van Elteren},
  {Van Hemelryck}, {van Leeuwen}, {Vaschetto}, {Vecchiato}, {Veljanoski},
  {Viala}, {Vicente}, {Vogt}, {von Essen}, {Voss}, {Votruba}, {Voutsinas},
  {Walmsley}, {Weiler}, {Wertz}, {Wevers}, {Wyrzykowski}, {Yoldas},
  {{\v{Z}}erjal}, {Ziaeepour}, {Zorec}, {Zschocke}, {Zucker}, {Zurbach}, \&
  {Zwitter}}]{2018A&A...616A...1G}
{Gaia Collaboration}, {Brown}, A.~G.~A., {Vallenari}, A., {et~al.} 2018, \aap,
  616, A1

\bibitem[{{Gendre} {et~al.}(2003){Gendre}, {Barret}, \&
  {Webb}}]{2003A&A...403L..11G}
{Gendre}, B., {Barret}, D., \& {Webb}, N. 2003, \aap, 403, L11

\bibitem[{{Georgakakis} {et~al.}(2008){Georgakakis}, {Nandra}, {Laird}, {Aird},
  \& {Trichas}}]{2008MNRAS.388.1205G}
{Georgakakis}, A., {Nandra}, K., {Laird}, E.~S., {Aird}, J., \& {Trichas}, M.
  2008, \mnras, 388, 1205

\bibitem[{{Gratton} {et~al.}(2019){Gratton}, {Bragaglia}, {Carretta},
  {D'Orazi}, {Lucatello}, \& {Sollima}}]{2019A&ARv..27....8G}
{Gratton}, R., {Bragaglia}, A., {Carretta}, E., {et~al.} 2019, \aapr, 27, 8

\bibitem[{{Gratton} {et~al.}(2013){Gratton}, {Lucatello}, {Sollima},
  {Carretta}, {Bragaglia}, {Momany}, {D'Orazi}, {Cassisi}, {Pietrinferni}, \&
  {Salaris}}]{2013A&A...549A..41G}
{Gratton}, R.~G., {Lucatello}, S., {Sollima}, A., {et~al.} 2013, \aap, 549, A41

\bibitem[{{Grindlay} {et~al.}(2001){Grindlay}, {Heinke}, {Edmonds}, \&
  {Murray}}]{2001Sci...292.2290G}
{Grindlay}, J.~E., {Heinke}, C., {Edmonds}, P.~D., \& {Murray}, S.~S. 2001,
  Science, 292, 2290

\bibitem[{{Haberl} {et~al.}(2002){Haberl}, {Motch}, \&
  {Zickgraf}}]{2002A&A...387..201H}
{Haberl}, F., {Motch}, C., \& {Zickgraf}, F.~J. 2002, \aap, 387, 201

\bibitem[{{Hansen} {et~al.}(2013){Hansen}, {Kalirai}, {Anderson}, {Dotter},
  {Richer}, {Rich}, {Shara}, {Fahlman}, {Hurley}, {King}, {Reitzel}, \&
  {Stetson}}]{2013Natur.500...51H}
{Hansen}, B.~M.~S., {Kalirai}, J.~S., {Anderson}, J., {et~al.} 2013, \nat, 500,
  51

\bibitem[{{Hasinger} {et~al.}(1994){Hasinger}, {Johnston}, \&
  {Verbunt}}]{1994A&A...288..466H}
{Hasinger}, G., {Johnston}, H.~M., \& {Verbunt}, F. 1994, \aap, 288, 466

\bibitem[{{Heinke} {et~al.}(2010){Heinke}, {Altamirano}, {Cohn}, {Lugger},
  {Budac}, {Servillat}, {Linares}, {Strohmayer}, {Markwardt}, {Wijnands},
  {Swank}, {Knigge}, {Bailyn}, \& {Grindlay}}]{2010ApJ...714..894H}
{Heinke}, C.~O., {Altamirano}, D., {Cohn}, H.~N., {et~al.} 2010, \apj, 714, 894

\bibitem[{{Heinke} {et~al.}(2005){Heinke}, {Grindlay}, {Edmonds}, {Cohn},
  {Lugger}, {Camilo}, {Bogdanov}, \& {Freire}}]{2005ApJ...625..796H}
{Heinke}, C.~O., {Grindlay}, J.~E., {Edmonds}, P.~D., {et~al.} 2005, \apj, 625,
  796

\bibitem[{{Heinke} {et~al.}(2003){Heinke}, {Grindlay}, {Lugger}, {Cohn},
  {Edmonds}, {Lloyd}, \& {Cool}}]{2003ApJ...598..501H}
{Heinke}, C.~O., {Grindlay}, J.~E., {Lugger}, P.~M., {et~al.} 2003, \apj, 598,
  501

\bibitem[{{Heinke} {et~al.}(2020){Heinke}, {Ivanov}, {Koch}, {Andrews},
  {Chomiuk}, {Cohn}, {Crothers}, {de Boer}, {Ivanova}, {Kong}, {Leigh},
  {Lugger}, {Nelson}, {Parr}, {Rosolowsky}, {Ruiter}, {Sarazin}, {Shaw},
  {Sivakoff}, \& {van den Berg}}]{2020MNRAS.492.5684H}
{Heinke}, C.~O., {Ivanov}, M.~G., {Koch}, E.~W., {et~al.} 2020, \mnras, 492,
  5684

\bibitem[{{Hertz} \& {Grindlay}(1983)}]{1983ApJ...267L..83H}
{Hertz}, P. \& {Grindlay}, J.~E. 1983, \apjl, 267, L83

\bibitem[{{HI4PI Collaboration} {et~al.}(2016){HI4PI Collaboration}, {Ben
  Bekhti}, {Fl{\"o}er}, {Keller}, {Kerp}, {Lenz}, {Winkel}, {Bailin},
  {Calabretta}, {Dedes}, {Ford}, {Gibson}, {Haud}, {Janowiecki}, {Kalberla},
  {Lockman}, {McClure-Griffiths}, {Murphy}, {Nakanishi}, {Pisano}, \&
  {Staveley-Smith}}]{2016A&A...594A.116H}
{HI4PI Collaboration}, {Ben Bekhti}, N., {Fl{\"o}er}, L., {et~al.} 2016, \aap,
  594, A116

\bibitem[{{Homan} {et~al.}(2018){Homan}, {van den Berg}, {Heinke}, {Pooley},
  {Degenaar}, {van den Eijnden}, {Bahramian}, {Gendreau}, \&
  {Arzoumanian}}]{2018ATel11598....1H}
{Homan}, J., {van den Berg}, M., {Heinke}, C., {et~al.} 2018, The Astronomer's
  Telegram, 11598, 1

\bibitem[{{Jonker} {et~al.}(2007){Jonker}, {Steeghs}, {Chakrabarty}, \&
  {Juett}}]{2007ApJ...665L.147J}
{Jonker}, P.~G., {Steeghs}, D., {Chakrabarty}, D., \& {Juett}, A.~M. 2007,
  \apjl, 665, L147

\bibitem[{{Kuulkers} {et~al.}(2006){Kuulkers}, {Norton}, {Schwope}, \&
  {Warner}}]{2006csxs.book..421K}
{Kuulkers}, E., {Norton}, A., {Schwope}, A., \& {Warner}, B. 2006, {X-rays from
  cataclysmic variables}, Vol.~39, 421--460

\bibitem[{{Lane} {et~al.}(2011){Lane}, {Kiss}, {Lewis}, {Ibata}, {Siebert},
  {Bedding}, {Sz{\'e}kely}, \& {Szab{\'o}}}]{2011A&A...530A..31L}
{Lane}, R.~R., {Kiss}, L.~L., {Lewis}, G.~F., {et~al.} 2011, \aap, 530, A31

\bibitem[{{Luna} {et~al.}(2013){Luna}, {Sokoloski}, {Mukai}, \&
  {Nelson}}]{2013A&A...559A...6L}
{Luna}, G.~J.~M., {Sokoloski}, J.~L., {Mukai}, K., \& {Nelson}, T. 2013, \aap,
  559, A6

\bibitem[{{Maccacaro} {et~al.}(1988){Maccacaro}, {Gioia}, {Wolter}, {Zamorani},
  \& {Stocke}}]{1988ApJ...326..680M}
{Maccacaro}, T., {Gioia}, I.~M., {Wolter}, A., {Zamorani}, G., \& {Stocke},
  J.~T. 1988, \apj, 326, 680

\bibitem[{{Marks} \& {Kroupa}(2010)}]{2010MNRAS.406.2000M}
{Marks}, M. \& {Kroupa}, P. 2010, \mnras, 406, 2000

\bibitem[{{Marocco} {et~al.}(2021){Marocco}, {Eisenhardt}, {Fowler},
  {Kirkpatrick}, {Meisner}, {Schlafly}, {Stanford}, {Garcia}, {Caselden},
  {Cushing}, {Cutri}, {Faherty}, {Gelino}, {Gonzalez}, {Jarrett}, {Koontz},
  {Mainzer}, {Marchese}, {Mobasher}, {Schlegel}, {Stern}, {Teplitz}, \&
  {Wright}}]{2021ApJS..253....8M}
{Marocco}, F., {Eisenhardt}, P. R.~M., {Fowler}, J.~W., {et~al.} 2021, \apjs,
  253, 8

\bibitem[{{McDonald} {et~al.}(2011){McDonald}, {Boyer}, {van Loon}, {Zijlstra},
  {Hora}, {Babler}, {Block}, {Gordon}, {Meade}, {Meixner}, {Misselt},
  {Robitaille}, {Sewi{\l}o}, {Shiao}, \& {Whitney}}]{2011ApJS..193...23M}
{McDonald}, I., {Boyer}, M.~L., {van Loon}, J.~T., {et~al.} 2011, \apjs, 193,
  23

\bibitem[{{Mereghetti} {et~al.}(2016){Mereghetti}, {Kuiper}, {Tiengo},
  {Hessels}, {Hermsen}, {Stovall}, {Possenti}, {Rankin}, {Esposito}, {Turolla},
  {Mitra}, {Wright}, {Stappers}, {Horneffer}, {Oslowski}, {Serylak}, \&
  {Grie{\ss}meier}}]{2016ApJ...831...21M}
{Mereghetti}, S., {Kuiper}, L., {Tiengo}, A., {et~al.} 2016, \apj, 831, 21

\bibitem[{{Merloni} {et~al.}(2012){Merloni}, {Predehl}, {Becker},
  {B{\"o}hringer}, {Boller}, {Brunner}, {Brusa}, {Dennerl}, {Freyberg},
  {Friedrich}, {Georgakakis}, {Haberl}, {Hasinger}, {Meidinger}, {Mohr},
  {Nandra}, {Rau}, {Reiprich}, {Robrade}, {Salvato}, {Santangelo}, {Sasaki},
  {Schwope}, {Wilms}, \& {German eROSITA Consortium}}]{2012arXiv1209.3114M}
{Merloni}, A., {Predehl}, P., {Becker}, W., {et~al.} 2012, arXiv e-prints,
  arXiv:1209.3114

\bibitem[{{Monroe} {et~al.}(2016){Monroe}, {Prochaska}, {Tejos}, {Worseck},
  {Hennawi}, {Schmidt}, {Tumlinson}, \& {Shen}}]{2016AJ....152...25M}
{Monroe}, T.~R., {Prochaska}, J.~X., {Tejos}, N., {et~al.} 2016, \aj, 152, 25

\bibitem[{{Mukai}(2017)}]{2017PASP..129f2001M}
{Mukai}, K. 2017, \pasp, 129, 062001

\bibitem[{{Narloch} {et~al.}(2017){Narloch}, {Kaluzny}, {Poleski}, {Rozyczka},
  {Pych}, \& {Thompson}}]{2017MNRAS.471.1446N}
{Narloch}, W., {Kaluzny}, J., {Poleski}, R., {et~al.} 2017, \mnras, 471, 1446

\bibitem[{{Paresce} {et~al.}(1992){Paresce}, {de Marchi}, \&
  {Ferraro}}]{1992Natur.360...46P}
{Paresce}, F., {de Marchi}, G., \& {Ferraro}, F.~R. 1992, \nat, 360, 46

\bibitem[{{Pooley} {et~al.}(2003){Pooley}, {Lewin}, {Anderson}, {Baumgardt},
  {Filippenko}, {Gaensler}, {Homer}, {Hut}, {Kaspi}, {Makino}, {Margon},
  {McMillan}, {Portegies Zwart}, {van der Klis}, \&
  {Verbunt}}]{2003ApJ...591L.131P}
{Pooley}, D., {Lewin}, W. H.~G., {Anderson}, S.~F., {et~al.} 2003, \apjl, 591,
  L131

\bibitem[{{Predehl} {et~al.}(2021){Predehl}, {Andritschke}, {Arefiev},
  {Babyshkin}, {Batanov}, {Becker}, {B{\"o}hringer}, {Bogomolov}, {Boller},
  {Borm}, {Bornemann}, {Br{\"a}uninger}, {Br{\"u}ggen}, {Brunner}, {Brusa},
  {Bulbul}, {Buntov}, {Burwitz}, {Burkert}, {Clerc}, {Churazov}, {Coutinho},
  {Dauser}, {Dennerl}, {Doroshenko}, {Eder}, {Emberger}, {Eraerds},
  {Finoguenov}, {Freyberg}, {Friedrich}, {Friedrich}, {F{\"u}rmetz},
  {Georgakakis}, {Gilfanov}, {Granato}, {Grossberger}, {Gueguen}, {Gureev},
  {Haberl}, {H{\"a}lker}, {Hartner}, {Hasinger}, {Huber}, {Ji}, {Kienlin},
  {Kink}, {Korotkov}, {Kreykenbohm}, {Lamer}, {Lomakin}, {Lapshov}, {Liu},
  {Maitra}, {Meidinger}, {Menz}, {Merloni}, {Mernik}, {Mican}, {Mohr},
  {M{\"u}ller}, {Nandra}, {Nazarov}, {Pacaud}, {Pavlinsky}, {Perinati},
  {Pfeffermann}, {Pietschner}, {Ramos-Ceja}, {Rau}, {Reiffers}, {Reiprich},
  {Robrade}, {Salvato}, {Sanders}, {Santangelo}, {Sasaki}, {Scheuerle},
  {Schmid}, {Schmitt}, {Schwope}, {Shirshakov}, {Steinmetz}, {Stewart},
  {Str{\"u}der}, {Sunyaev}, {Tenzer}, {Tiedemann}, {Tr{\"u}mper}, {Voron},
  {Weber}, {Wilms}, \& {Yaroshenko}}]{2021A&A...647A...1P}
{Predehl}, P., {Andritschke}, R., {Arefiev}, V., {et~al.} 2021, \aap, 647, A1

\bibitem[{{Primini} {et~al.}(1993){Primini}, {Forman}, \&
  {Jones}}]{1993ApJ...410..615P}
{Primini}, F.~A., {Forman}, W., \& {Jones}, C. 1993, \apj, 410, 615

\bibitem[{{Ramsay} {et~al.}(2004){Ramsay}, {Cropper}, {Wu}, {Mason},
  {C{\'o}rdova}, \& {Priedhorsky}}]{2004MNRAS.350.1373R}
{Ramsay}, G., {Cropper}, M., {Wu}, K., {et~al.} 2004, \mnras, 350, 1373

\bibitem[{{Rengarajan} \& {Verma}(1983)}]{1983MNRAS.205..447R}
{Rengarajan}, T.~N. \& {Verma}, R.~P. 1983, \mnras, 205, 447

\bibitem[{{Ridolfi} {et~al.}(2016){Ridolfi}, {Freire}, {Torne}, {Heinke}, {van
  den Berg}, {Jordan}, {Kramer}, {Bassa}, {Sarkissian}, {D'Amico}, {Lorimer},
  {Camilo}, {Manchester}, \& {Lyne}}]{2016MNRAS.462.2918R}
{Ridolfi}, A., {Freire}, P.~C.~C., {Torne}, P., {et~al.} 2016, \mnras, 462,
  2918

\bibitem[{{Rutledge} {et~al.}(2002){Rutledge}, {Bildsten}, {Brown}, {Pavlov},
  \& {Zavlin}}]{2002ApJ...578..405R}
{Rutledge}, R.~E., {Bildsten}, L., {Brown}, E.~F., {Pavlov}, G.~G., \&
  {Zavlin}, V.~E. 2002, \apj, 578, 405

\bibitem[{{Salvato} {et~al.}(2018){Salvato}, {Buchner}, {Budav{\'a}ri},
  {Dwelly}, {Merloni}, {Brusa}, {Rau}, {Fotopoulou}, \&
  {Nandra}}]{2018MNRAS.473.4937S}
{Salvato}, M., {Buchner}, J., {Budav{\'a}ri}, T., {et~al.} 2018, \mnras, 473,
  4937

\bibitem[{{Sanna} {et~al.}(2018){Sanna}, {Bahramian}, {Bozzo}, {Heinke},
  {Altamirano}, {Wijnands}, {Degenaar}, {Maccarone}, {Riggio}, {Di Salvo},
  {Iaria}, {Burgay}, {Possenti}, {Ferrigno}, {Papitto}, {Sivakoff}, {D'Amico},
  \& {Burderi}}]{2018A&A...610L...2S}
{Sanna}, A., {Bahramian}, A., {Bozzo}, E., {et~al.} 2018, \aap, 610, L2

\bibitem[{{Sanna} {et~al.}(2017){Sanna}, {Papitto}, {Burderi}, {Bozzo},
  {Riggio}, {Di Salvo}, {Ferrigno}, {Rea}, \& {Iaria}}]{2017A&A...598A..34S}
{Sanna}, A., {Papitto}, A., {Burderi}, L., {et~al.} 2017, \aap, 598, A34

\bibitem[{{Scargle}(1982)}]{1982ApJ...263..835S}
{Scargle}, J.~D. 1982, \apj, 263, 835

\bibitem[{{Schlafly} \& {Finkbeiner}(2011)}]{2011ApJ...737..103S}
{Schlafly}, E.~F. \& {Finkbeiner}, D.~P. 2011, \apj, 737, 103

\bibitem[{{Secrest} {et~al.}(2015){Secrest}, {Dudik}, {Dorland}, {Zacharias},
  {Makarov}, {Fey}, {Frouard}, \& {Finch}}]{2015ApJS..221...12S}
{Secrest}, N.~J., {Dudik}, R.~P., {Dorland}, B.~N., {et~al.} 2015, \apjs, 221,
  12

\bibitem[{{Smith} {et~al.}(2001){Smith}, {Brickhouse}, {Liedahl}, \&
  {Raymond}}]{2001ApJ...556L..91S}
{Smith}, R.~K., {Brickhouse}, N.~S., {Liedahl}, D.~A., \& {Raymond}, J.~C.
  2001, \apjl, 556, L91

\bibitem[{{Soszy{\'n}ski} {et~al.}(2011){Soszy{\'n}ski}, {Udalski},
  {Szyma{\'n}ski}, {Kubiak}, {Pietrzy{\'n}ski}, {Wyrzykowski}, {Ulaczyk},
  {Poleski}, {Koz{\l}owski}, \& {Pietrukowicz}}]{2011AcA....61..217S}
{Soszy{\'n}ski}, I., {Udalski}, A., {Szyma{\'n}ski}, M.~K., {et~al.} 2011,
  \actaa, 61, 217

\bibitem[{{Suleimanov} {et~al.}(2019){Suleimanov}, {Doroshenko}, \&
  {Werner}}]{2019MNRAS.482.3622S}
{Suleimanov}, V.~F., {Doroshenko}, V., \& {Werner}, K. 2019, \mnras, 482, 3622

\bibitem[{{van Teeseling} {et~al.}(1996){van Teeseling}, {Beuermann}, \&
  {Verbunt}}]{1996A&A...315..467V}
{van Teeseling}, A., {Beuermann}, K., \& {Verbunt}, F. 1996, \aap, 315, 467

\bibitem[{{Verbunt} \& {Hasinger}(1998)}]{1998A&A...336..895V}
{Verbunt}, F. \& {Hasinger}, G. 1998, \aap, 336, 895

\bibitem[{{Wenger} {et~al.}(2000){Wenger}, {Ochsenbein}, {Egret}, {Dubois},
  {Bonnarel}, {Borde}, {Genova}, {Jasniewicz}, {Lalo{\"e}}, {Lesteven}, \&
  {Monier}}]{2000A&AS..143....9W}
{Wenger}, M., {Ochsenbein}, F., {Egret}, D., {et~al.} 2000, \aaps, 143, 9

\bibitem[{{West} {et~al.}(2011){West}, {Morgan}, {Bochanski}, {Andersen},
  {Bell}, {Kowalski}, {Davenport}, {Hawley}, {Schmidt}, {Bernat}, {Hilton},
  {Muirhead}, {Covey}, {Rojas-Ayala}, {Schlawin}, {Gooding}, {Schluns},
  {Dhital}, {Pineda}, \& {Jones}}]{2011AJ....141...97W}
{West}, A.~A., {Morgan}, D.~P., {Bochanski}, J.~J., {et~al.} 2011, \aj, 141, 97

\bibitem[{{Wijnands} {et~al.}(2005){Wijnands}, {Heinke}, {Pooley}, {Edmonds},
  {Lewin}, {Grindlay}, {Jonker}, \& {Miller}}]{2005ApJ...618..883W}
{Wijnands}, R., {Heinke}, C.~O., {Pooley}, D., {et~al.} 2005, \apj, 618, 883

\bibitem[{{Wolf} {et~al.}(2018){Wolf}, {Onken}, {Luvaul}, {Schmidt}, {Bessell},
  {Chang}, {Da Costa}, {Mackey}, {Martin-Jones}, {Murphy}, {Preston}, {Scalzo},
  {Shao}, {Smillie}, {Tisserand}, {White}, \& {Yuan}}]{2018PASA...35...10W}
{Wolf}, C., {Onken}, C.~A., {Luvaul}, L.~C., {et~al.} 2018, \pasa, 35, e010

\bibitem[{{Worpel} {et~al.}(2016){Worpel}, {Schwope}, {Granzer}, {Reinsch},
  {Schwarz}, \& {Traulsen}}]{2016A&A...592A.114W}
{Worpel}, H., {Schwope}, A.~D., {Granzer}, T., {et~al.} 2016, \aap, 592, A114

\bibitem[{{Wright} {et~al.}(2010){Wright}, {Eisenhardt}, {Mainzer}, {Ressler},
  {Cutri}, {Jarrett}, {Kirkpatrick}, {Padgett}, {McMillan}, {Skrutskie},
  {Stanford}, {Cohen}, {Walker}, {Mather}, {Leisawitz}, {Gautier}, {McLean},
  {Benford}, {Lonsdale}, {Blain}, {Mendez}, {Irace}, {Duval}, {Liu}, {Royer},
  {Heinrichsen}, {Howard}, {Shannon}, {Kendall}, {Walsh}, {Larsen}, {Cardon},
  {Schick}, {Schwalm}, {Abid}, {Fabinsky}, {Naes}, \&
  {Tsai}}]{2010AJ....140.1868W}
{Wright}, E.~L., {Eisenhardt}, P. R.~M., {Mainzer}, A.~K., {et~al.} 2010, \aj,
  140, 1868

\bibitem[{{Yokogawa} {et~al.}(2000){Yokogawa}, {Paul}, {Ozaki}, {Nagase},
  {Chakrabarty}, \& {Takeshima}}]{2000ApJ...539..191Y}
{Yokogawa}, J., {Paul}, B., {Ozaki}, M., {et~al.} 2000, \apj, 539, 191

\bibitem[{{Zavlin} {et~al.}(1996){Zavlin}, {Pavlov}, \&
  {Shibanov}}]{1996A&A...315..141Z}
{Zavlin}, V.~E., {Pavlov}, G.~G., \& {Shibanov}, Y.~A. 1996, \aap, 315, 141

\end{thebibliography}
\begin{appendices}
\appendix
\onecolumn
\section{Image of infrared 2MASS counterparts of 47~Tuc members}
\label{infrared-image}
The Infrared 2MASS\,($k_{s}$ band) images of the X-ray sources, which are classified as members of 47~Tuc (see Sect.\ref{diss-sec}). Images show  2$\sigma$ positional error of \erosita\, X-ray sources (black). If a source has a \chandra\, counterpart, 2$\sigma$ \chandra\, positional error is shown with blue circle.  Infrared 2MASS counterpart is in red circles and blue crosses show the position of optical Gaia counterpart. Since the typical $3\sigma$ positional error of Gaia counterparts\,($\sim$0.03$\arcsec$) is negligible in comparison to the X-ray and infrared positional errors\,($\sim$0.3$\arcsec$) therefore, they are shown by crosses. The scale of 5.0$\arcsec$ is shown for all images.
\begin{figure}[hbt!]

\centering 
  \subfloat[Src-No.\,263]{\includegraphics[clip, trim={0.0cm 2.cm 0.cm 0.0cm},width=0.26\textwidth]{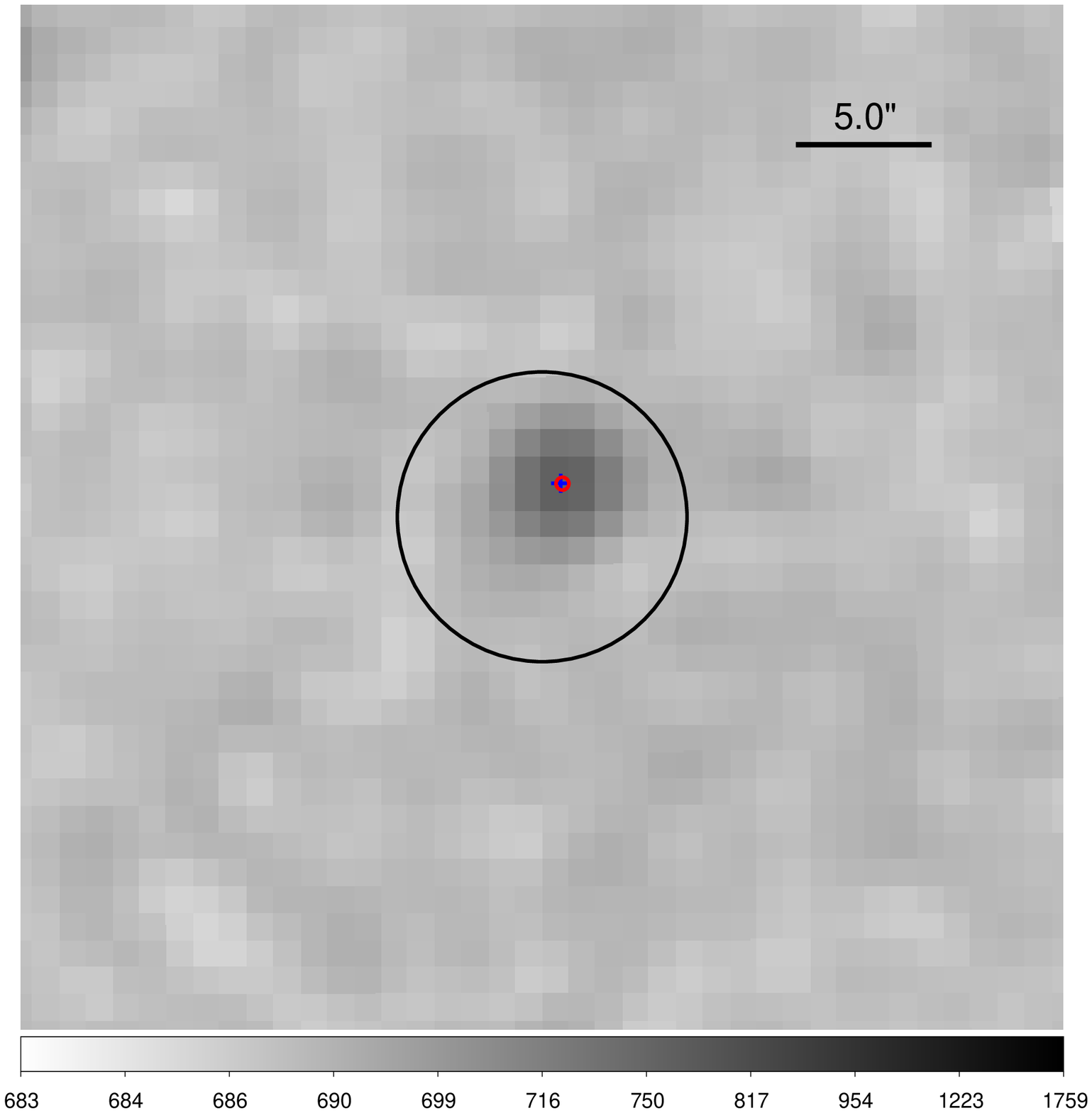}}
   \subfloat[Src-No.\,267]{\includegraphics[clip, trim={0.0cm 2.cm 0.cm 0.0cm},width=0.26\textwidth]{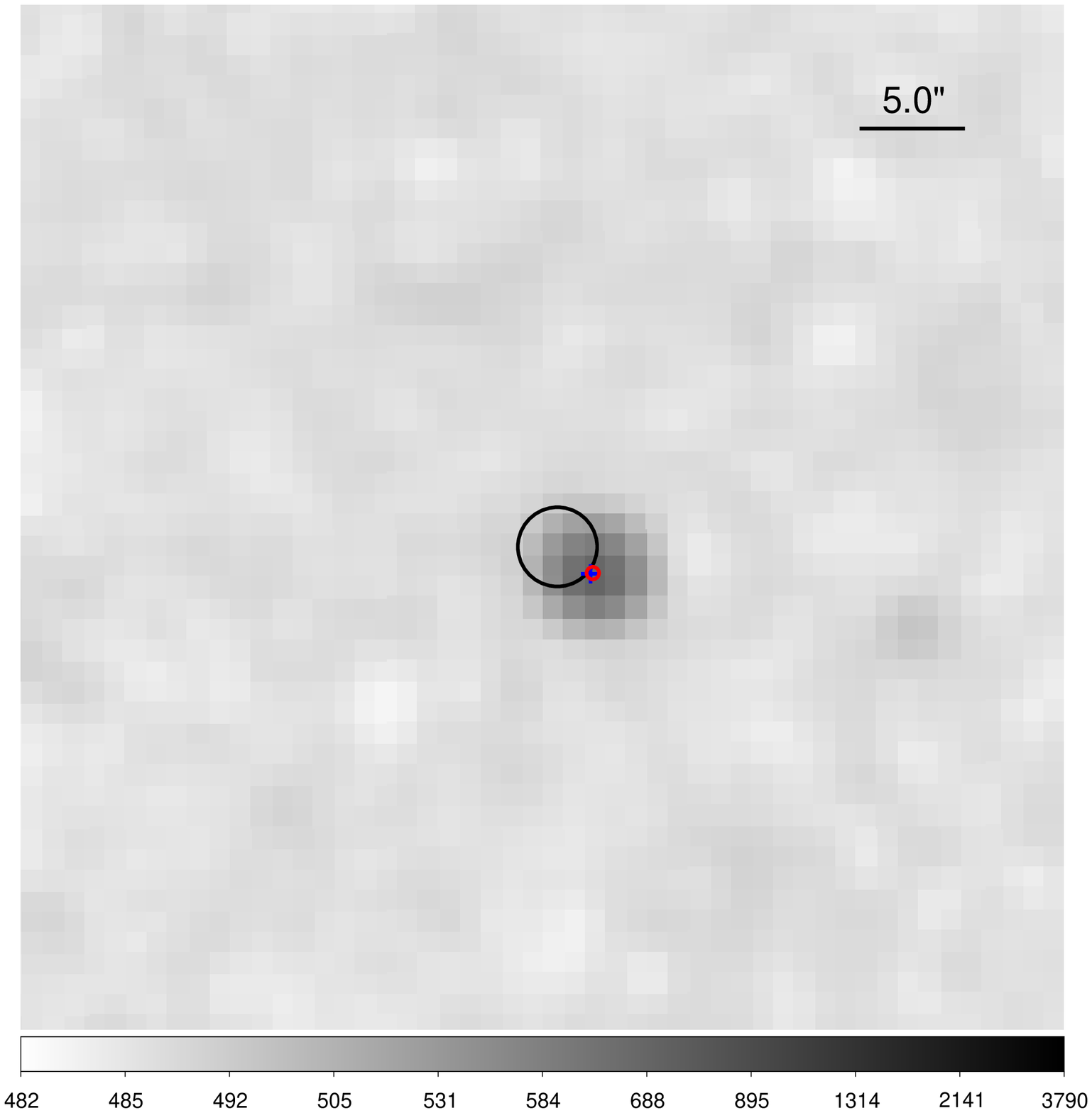}}
  \subfloat[Src-No.\,320]{\includegraphics[clip, trim={0.0cm 2.cm 0.cm 0.0cm},width=0.26\textwidth]{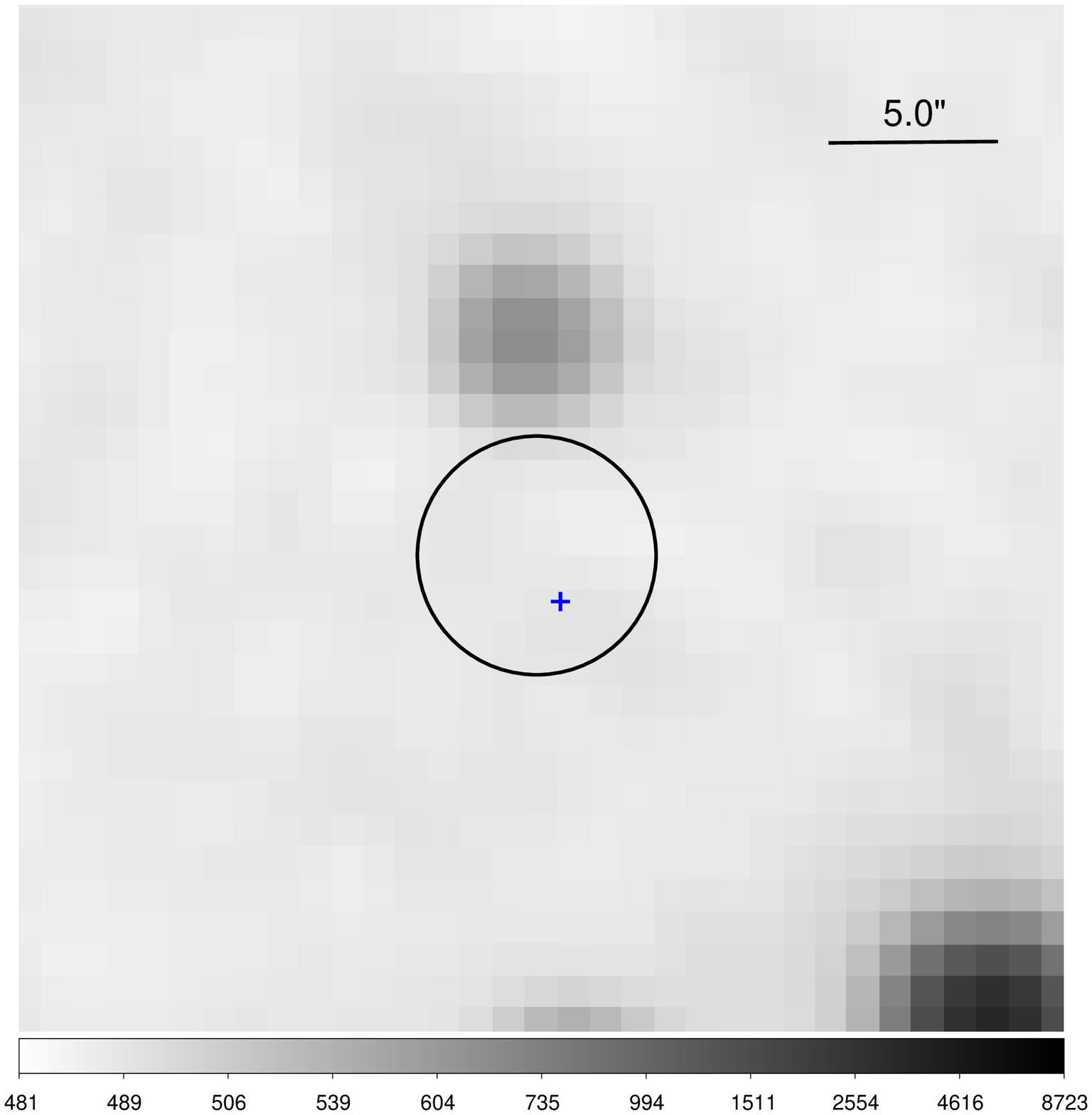}}\\
  \subfloat[Src-No.\,340]{\includegraphics[clip, trim={0.0cm 2.cm 0.cm 0.0cm},width=0.26\textwidth]{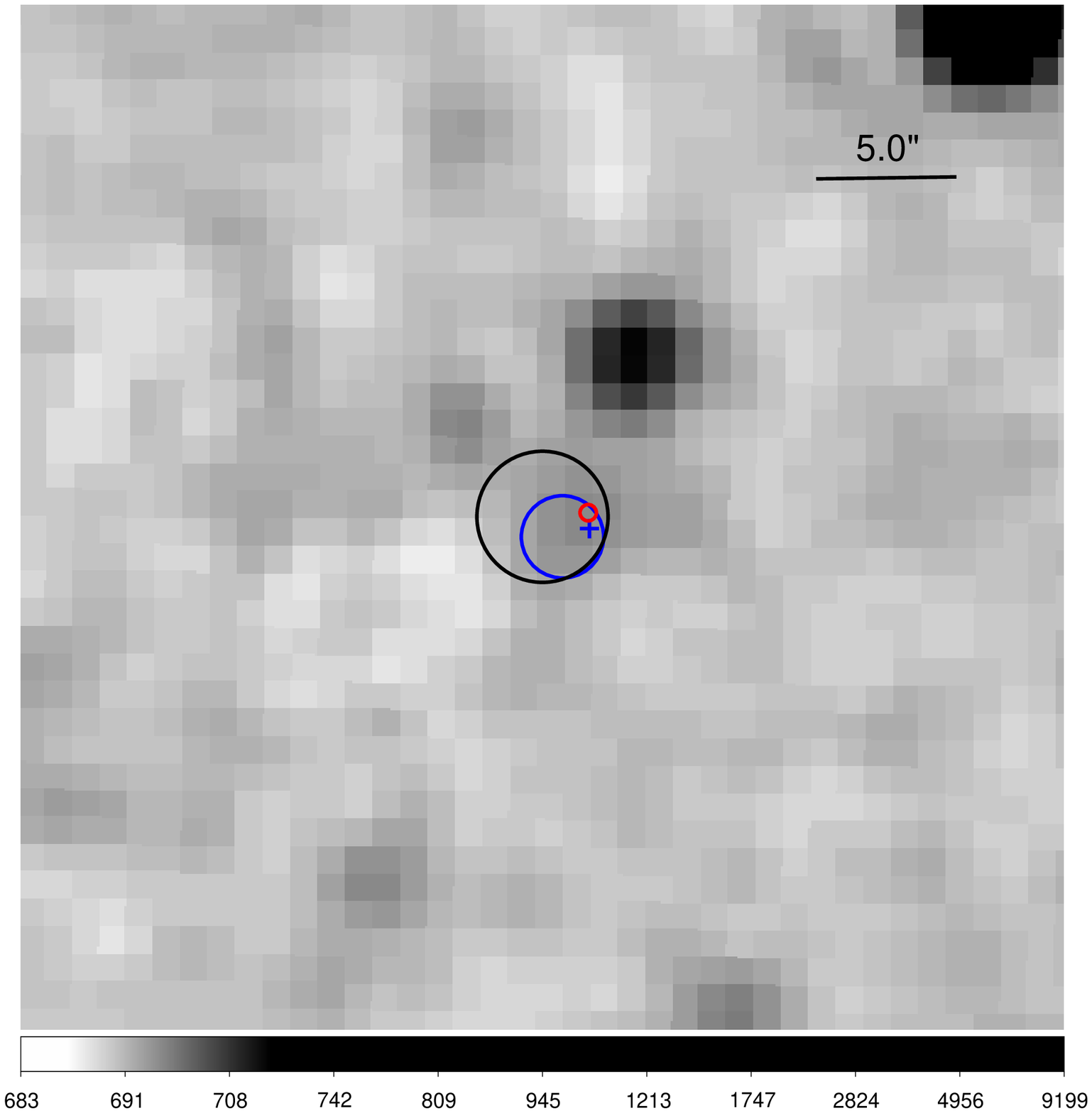}}
  \subfloat[Src-No.\,341]{\includegraphics[clip, trim={0.0cm 2.cm 0.cm 0.0cm},width=0.26\textwidth]{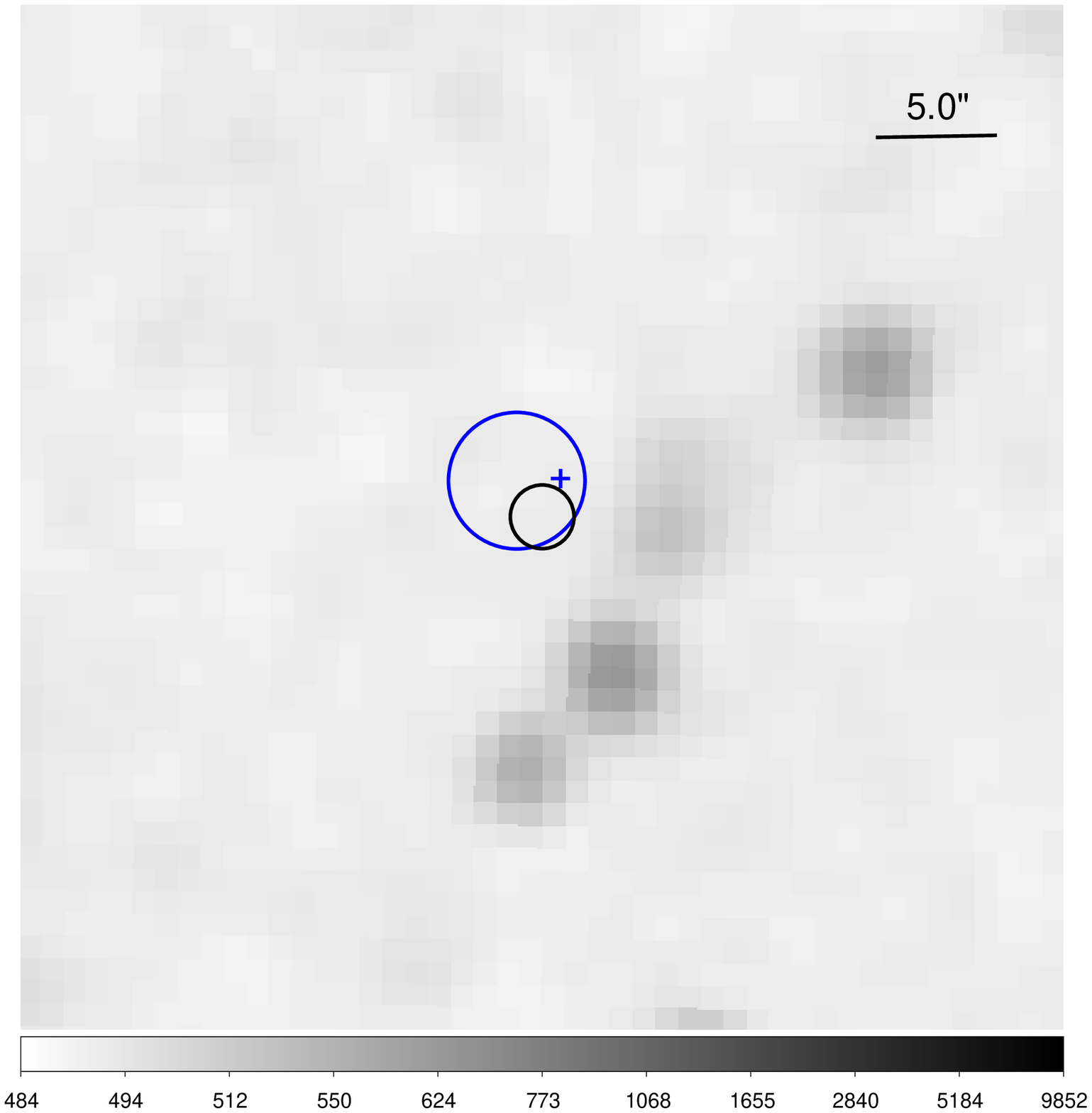}}
  \subfloat[Src-No.\,376]{\includegraphics[clip, trim={0.0cm 2.cm 0.cm 0.0cm},width=0.26\textwidth]{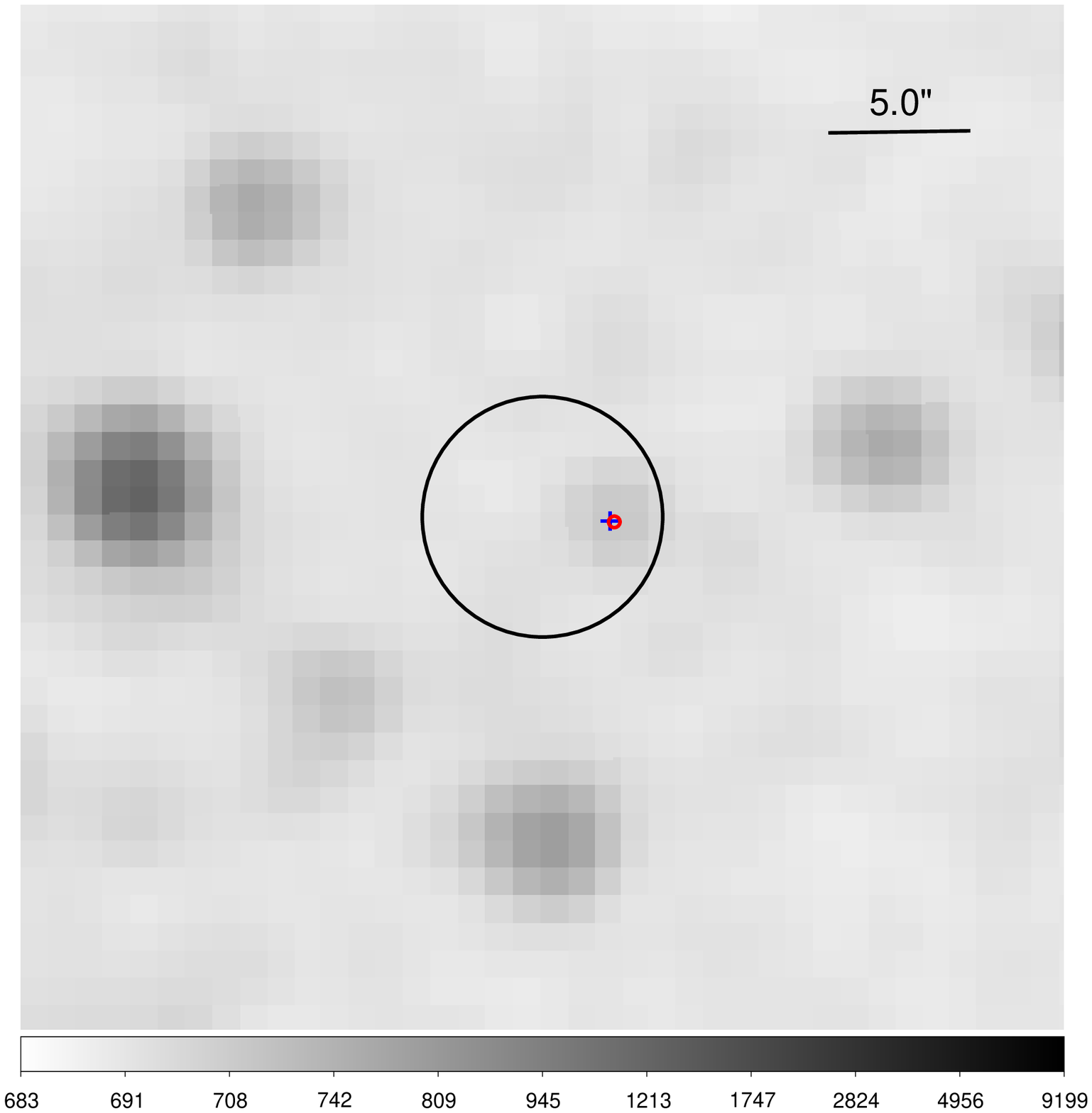}}\\
  \subfloat[Src-No.\,378]{\includegraphics[clip, trim={0.0cm 2.cm 0.cm 0.0cm},width=0.26\textwidth]{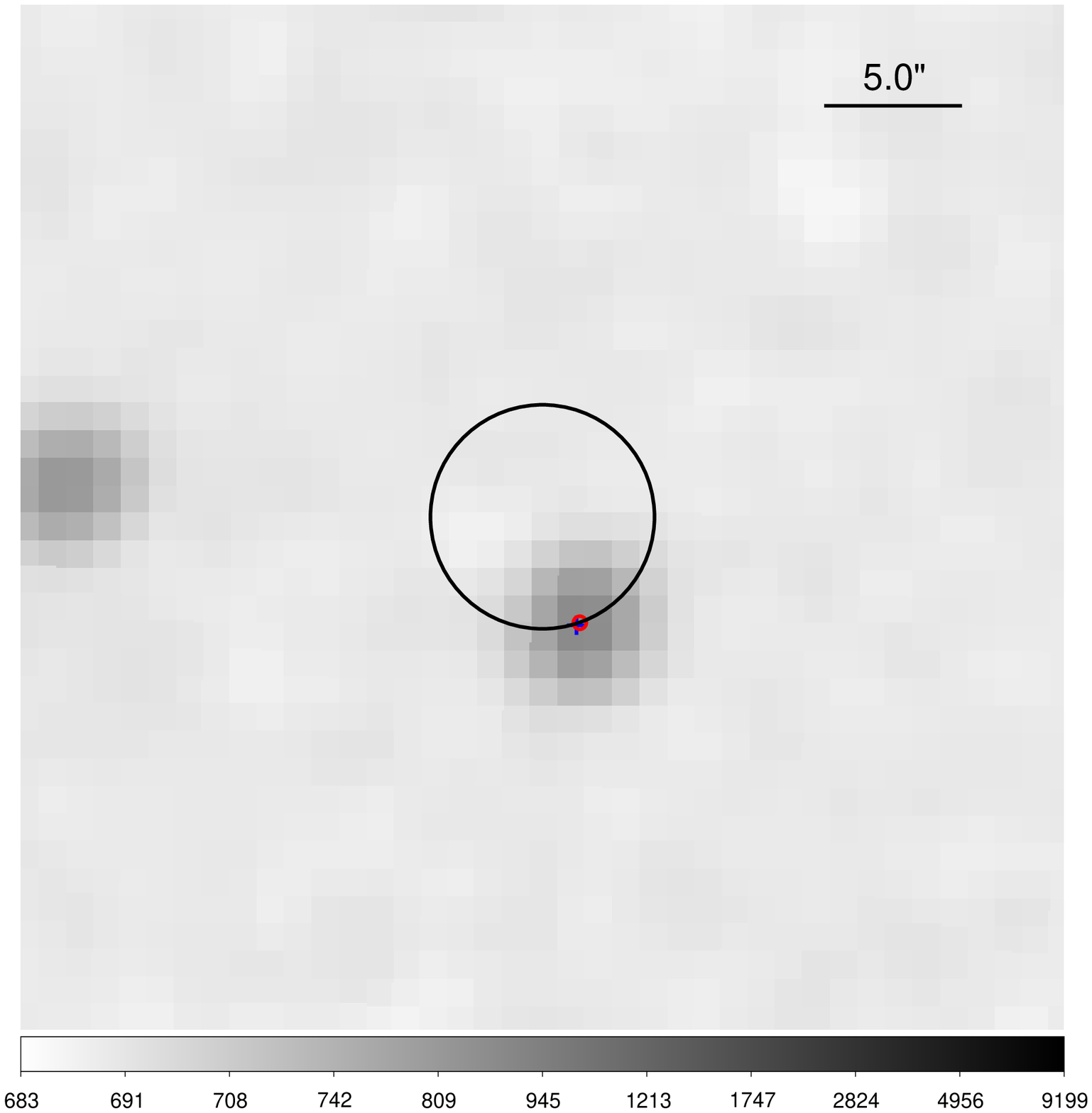}}
  \subfloat[Src-No.\,395]{\includegraphics[clip, trim={0.0cm 2.cm 0.cm 0.0cm},width=0.26\textwidth]{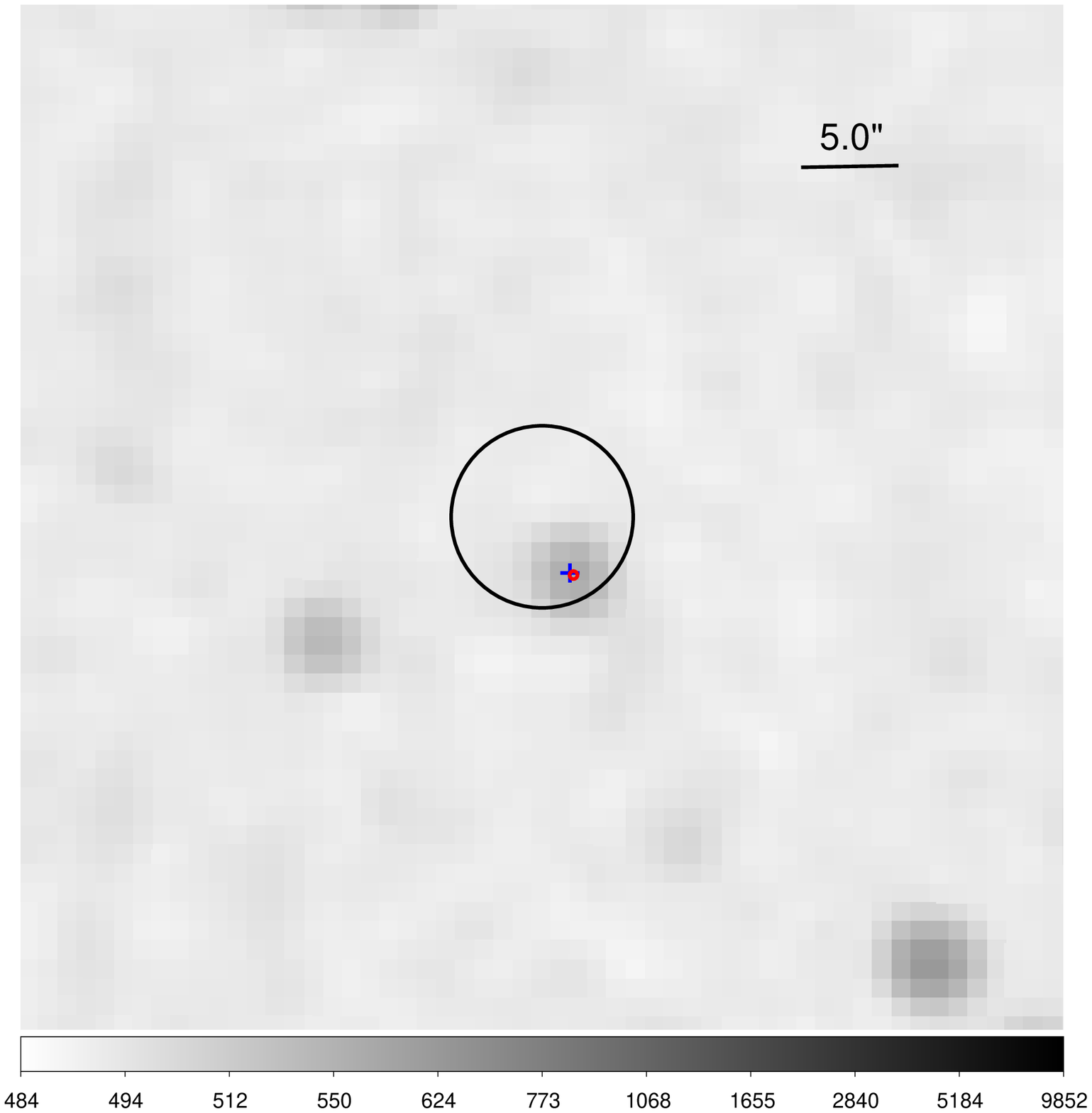}}
 \subfloat[Src-No.\,438]{\includegraphics[clip, trim={0.0cm 2.cm 0.cm 0.0cm},width=0.26\textwidth]{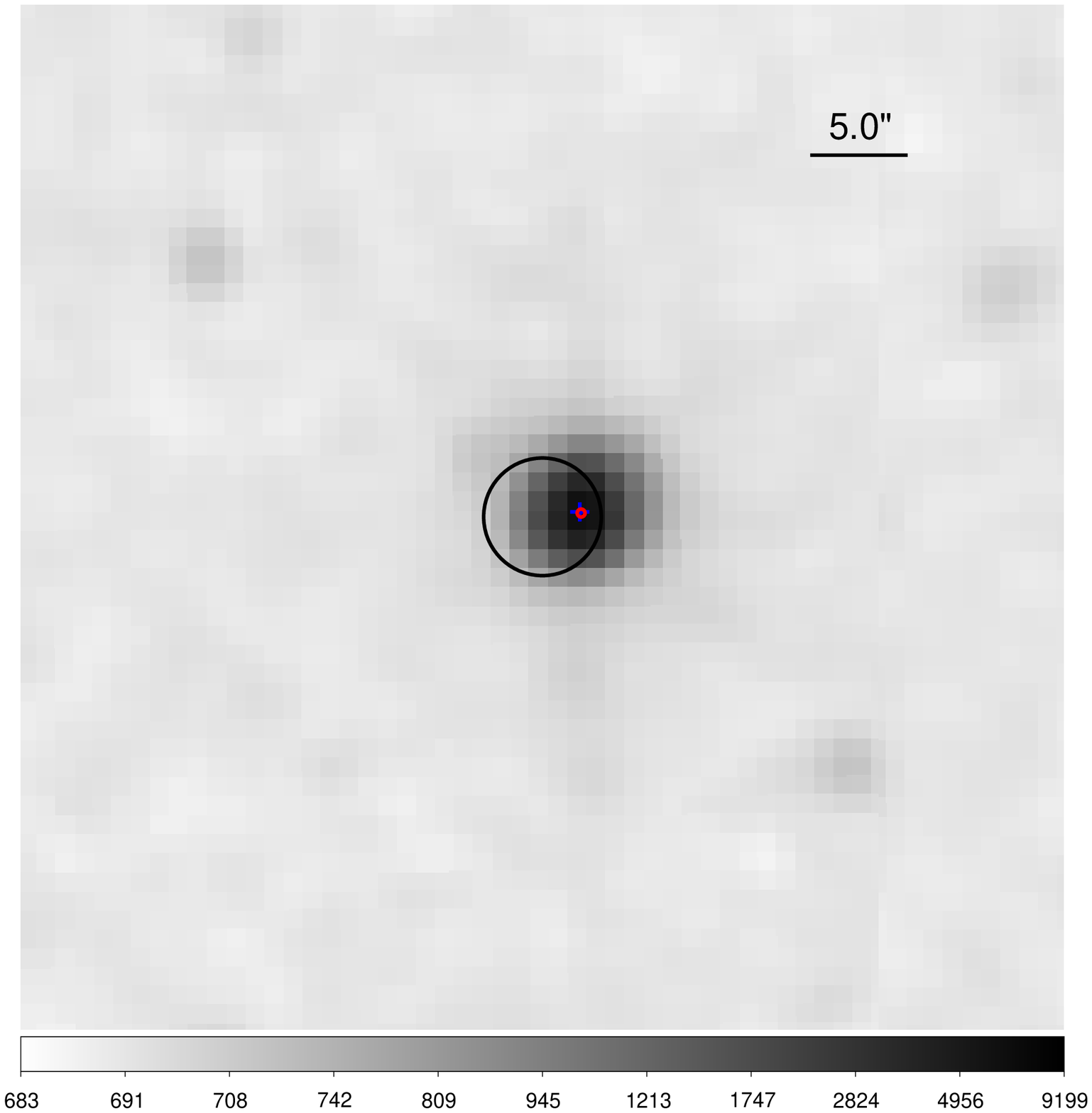}}\\

  \end{figure}
\pagebreak
\onecolumn
\clearpage
\begin{figure}
\centering
\vspace{-0.5cm}

 \subfloat[Src-No.\,480]{\includegraphics[clip, trim={0.0cm 2.cm 0.cm 0.0cm},width=0.26\textwidth]{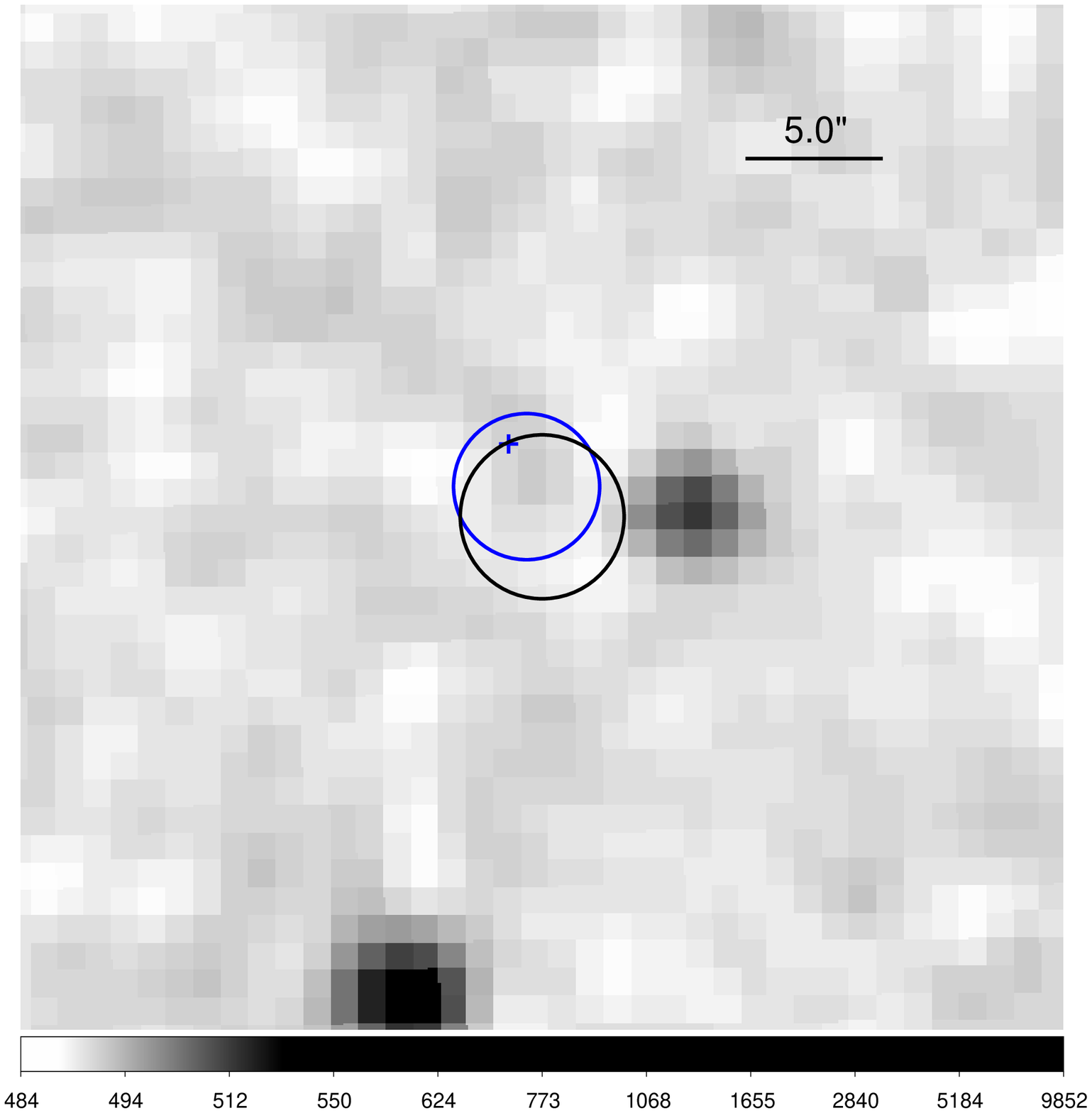}}
\subfloat[Src-No.\,481]{\includegraphics[clip, trim={0.0cm 2.cm 0.cm 0.0cm},width=0.26\textwidth]{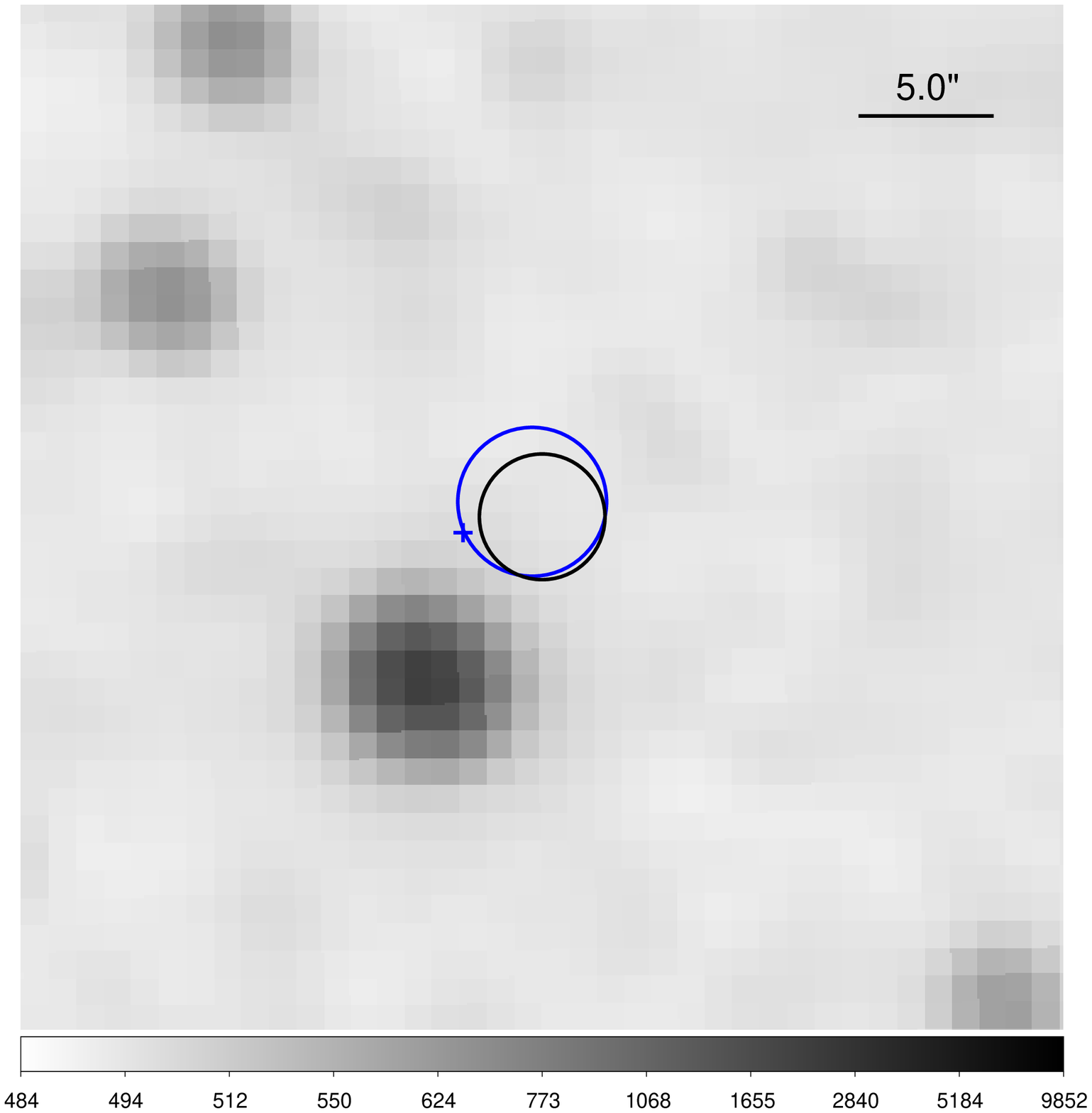}}
 \subfloat[Src-No.\,486]{\includegraphics[clip, trim={0.0cm 2.cm 0.cm 0.0cm},width=0.26\textwidth]{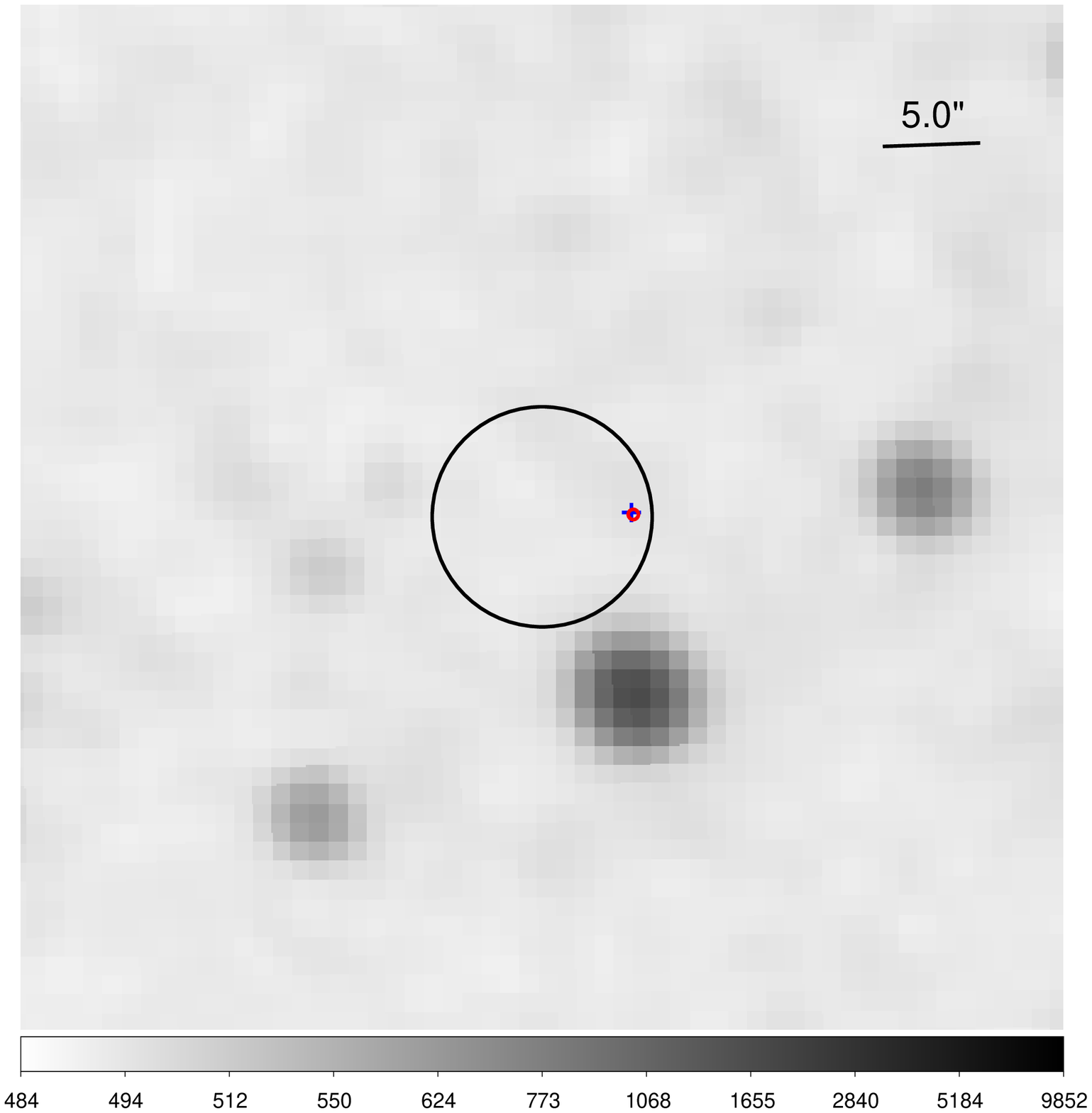}}\\
 \subfloat[Src-No.\,501]{\includegraphics[clip, trim={0.0cm 2.cm 0.cm 0.0cm},width=0.26\textwidth]{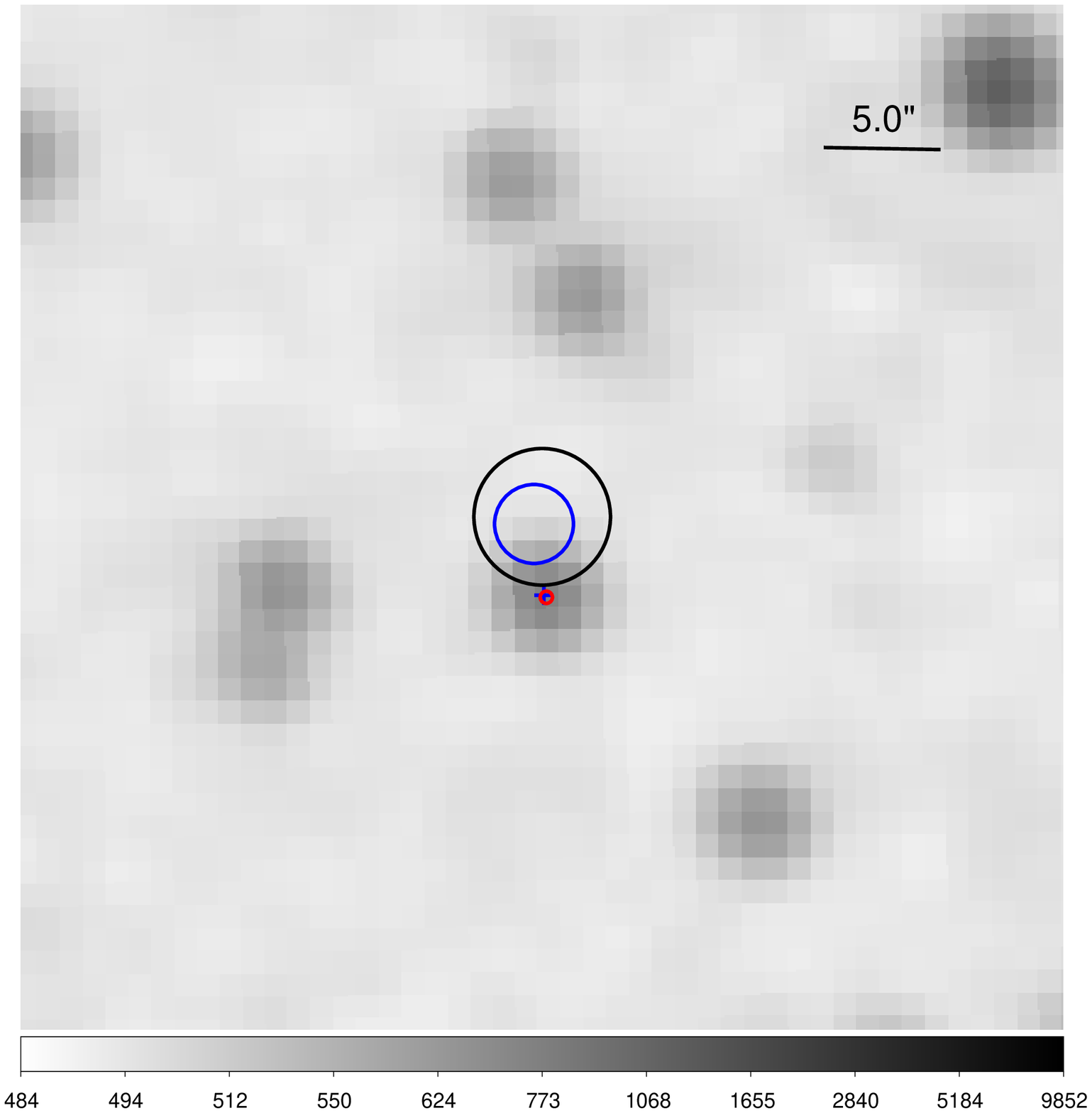}}
   \subfloat[Src-No.\,522]{\includegraphics[clip, trim={0.0cm 2.cm 0.cm 0.0cm},width=0.26\textwidth]{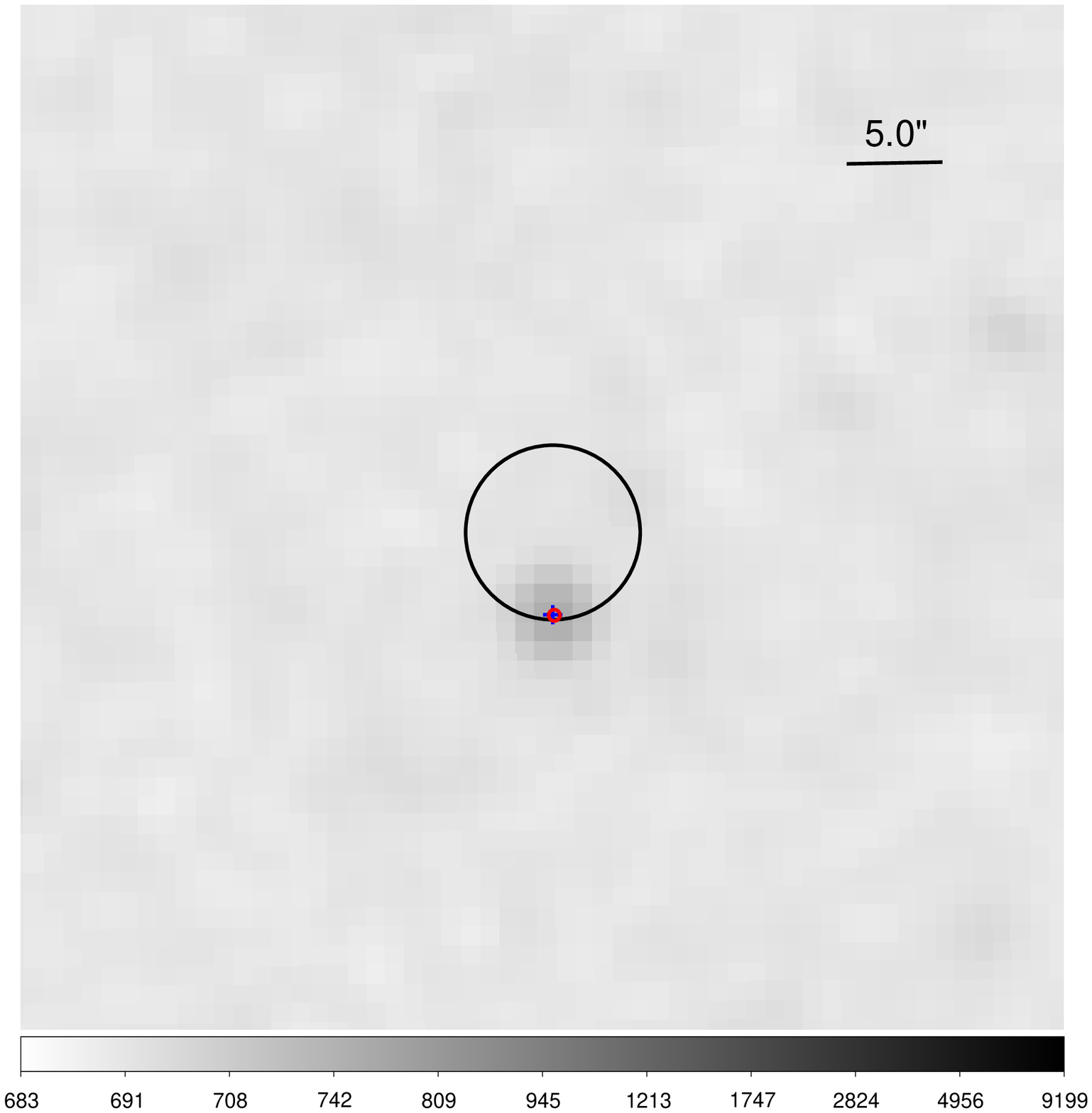}}
   \subfloat[Src-No.\,552]{\includegraphics[clip, trim={0.0cm 2.cm 0.cm 0.0cm},width=0.26\textwidth]{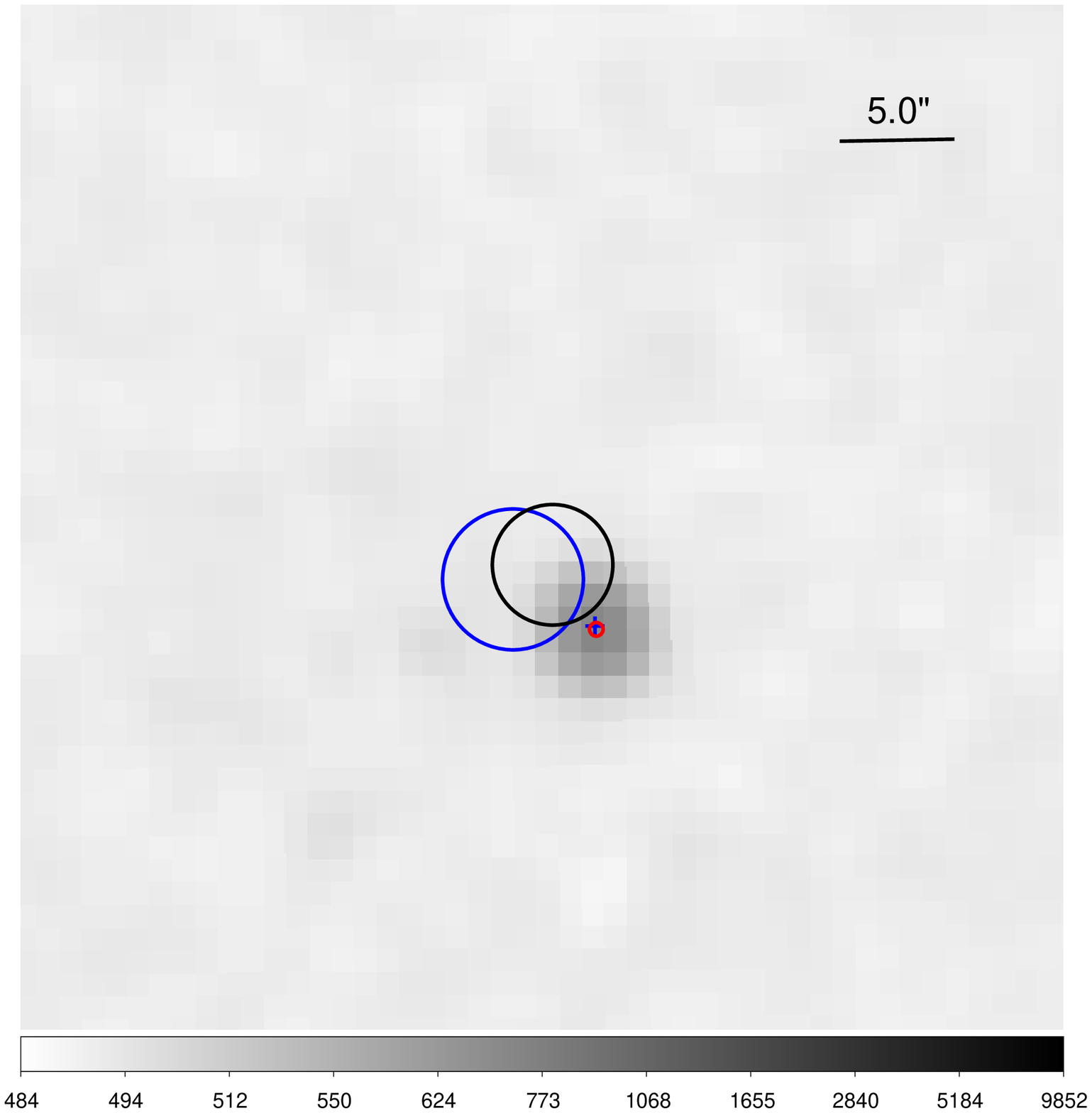}}\\

      \subfloat[Src-No.\,556]{\includegraphics[clip, trim={0.0cm 2.cm 0.cm 0.0cm},width=0.26\textwidth]{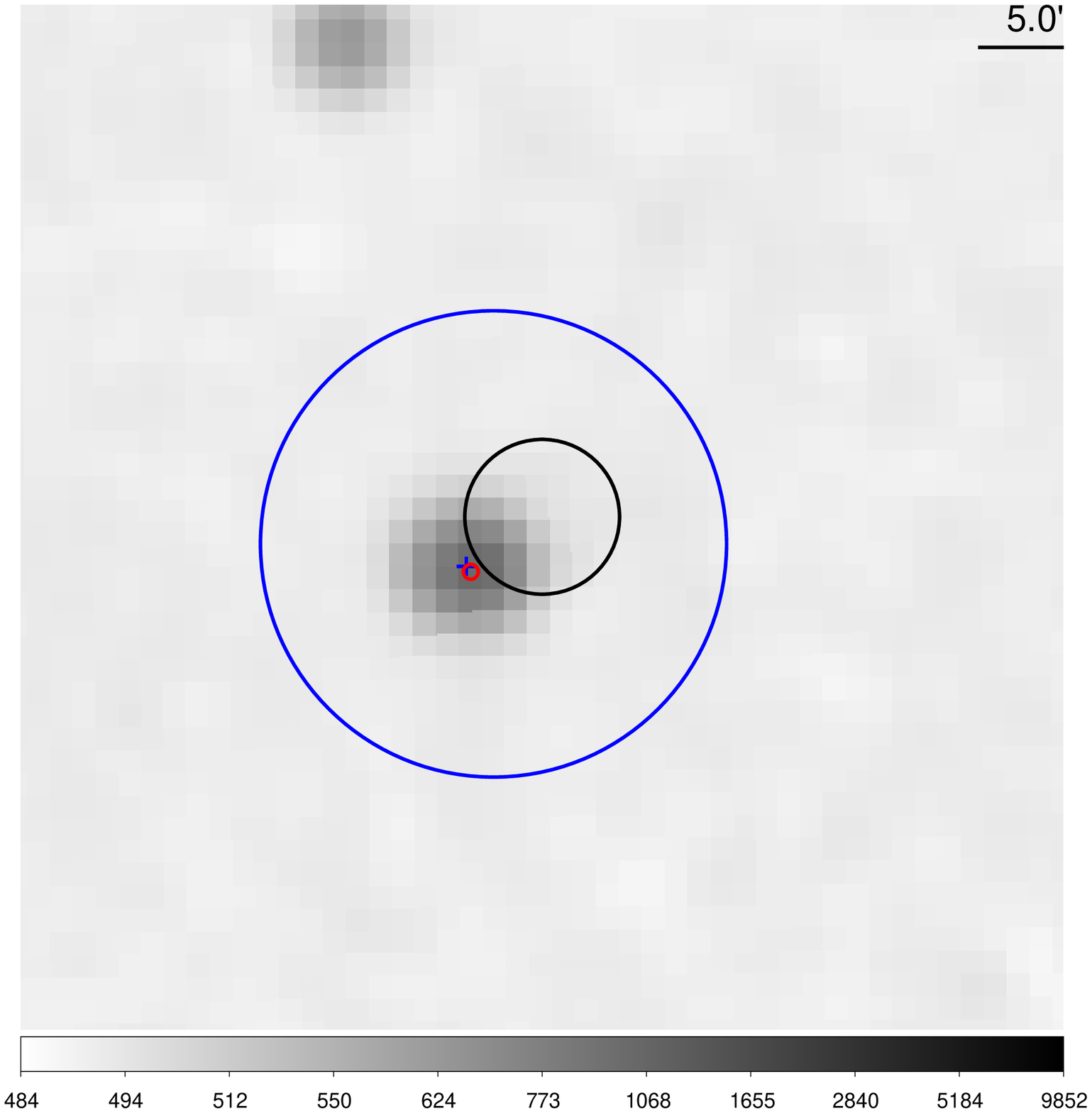}}
  
  \end{figure}
\pagebreak
\clearpage

\onecolumn
\begin{landscape}
\raggedright
\section{Source catalogue}
\setlength{\tabcolsep}{1.1mm}
\scriptsize{
\begin{longtable}{cccccccccclllcll}
    \caption{X-ray sources in the FOV of 47 Tuc.$^{\dagger}$:  \label{catalogue-x-ray}}\tabularnewline
\hline
\scriptsize
NO & eROSITA SRC-Name &RA              &DEC              &r1$\sigma$                       & \multicolumn{5}{c}{Flux (0.2--5.~keV)}  &\multicolumn{3}{l}{Hardness ratio} &Var&Classification$^{*}$&CATA$^{**}$\\
                       &   &  (J2000)       &  (J2000)        & ($\arcsec$)                     & \multicolumn{5}{c}{($10^{-14}$ erg\,s$^{-1}$\,cm$^{-2}$)}&&&&&&\\
               &            &                &                 &                                 & OBS1&          OBS2&          OBS3&  OBS4&          OBS5& HR1&HR2&HR3&&&\\
\hline
\endfirsthead
\multicolumn {16}{r}{Continued from previous page}\\
\hline 
 NO & eROSITA SRC-Name &RA              &DEC              &r1$\sigma$                       & \multicolumn{5}{c}{Flux (0.2--5.~keV)} &\multicolumn{3}{l}{Hardness ratio} &Var& Classification$^{*}$&CATA$^{**}$\\
                &     &  (J2000)       &  (J2000)        & ($\arcsec$)                     & \multicolumn{5}{c}{($10^{-14}$ erg\,s$^{-1}$\,cm$^{-2}$)}&&&&&&\\
            &         &                &                 &                                 & OBS1&          OBS2&          OBS3&  OBS4&          OBS5&HR1&HR2&HR3&&&\\
\hline
\endhead
\hline
\multicolumn {16}{r}{}\\

\endfoot

\hline
\endlastfoot
        1&       SRGJ001544.20-720140.6     &       00 15 44.20  & -72 01 40.6     &       4.76&                --&                3.55$\pm$                0.93&                --&                --&                --&                0.26$\pm$                0.46&                0.26$\pm$                0.27&               -0.11$\pm$                0.31  &                --&          &           			   \\
      2&       SRGJ001548.65-720510.1     &       00 15 48.65  & -72 05 10.1     &       2.53&                --&                5.13$\pm$                0.90&                --&                --&                --&                0.57$\pm$                0.25&                0.22$\pm$                0.14&               -0.73$\pm$                0.24  &                --&          &           CalPV	   	\\
      3&       SRGJ001551.31-715137.4     &       00 15 51.31  & -71 51 37.4     &       5.55&                --&                3.11$\pm$                0.87&                --&                --&                --&               -0.17$\pm$                0.24&               -0.24$\pm$                0.32&               -0.19$\pm$                0.63  &                --&          &                              	   \\
      4&       SRGJ001557.06-714549.5     &       00 15 57.06  & -71 45 49.5     &       3.35&                --&                2.47$\pm$                0.83&                --&                --&                --&                1.00$\pm$                5.69&               -1.00$\pm$               12.52&                1.00$\pm$                0.23  &                --&          &                              	   \\
\end{longtable}}
\end{landscape}

\onecolumn

\pagebreak

\footnotesize{
\begin{longtable}{cccccccc}
    \caption{Infrared magnitudes of counterparts of X-ray sources of 47 tuc in different energy filters 2MASS and WISE surveys.$\dagger$
\label{inf-count} }\tabularnewline
\hline\hline\
No&$W1$ mag&$W2$ mag&$W3$ mag&$W4$ mag&$J$ mag&$H$ mag&$K_{s}$ mag\\

\hline
\endfirsthead
\multicolumn {8}{c}{}\\
\hline\hline
No&$W1$ mag&$W2$ mag&$W3$ mag&$W4$ mag&$J$ mag&$H$ mag&$K_{s}$ mag\\
\hline
\endhead
\hline
\multicolumn {8}{c}{}\\

\endfoot

\hline
\endlastfoot
        3&               18.02$\pm$                0.19&               17.10$\pm$                0.31&<               12.94&<                9.10&                --&                --&                -- \\
      4&               14.57$\pm$                0.03&               14.62$\pm$                0.05&<               12.72&<                9.31&               15.15$\pm$                0.05&               14.72$\pm$                0.08&               14.71$\pm$                0.08 \\
      5&               17.20$\pm$                0.10&               16.86$\pm$                0.24&<               12.20&<                8.58&                --&                --&                -- \\
      6&               17.32$\pm$                0.10&               17.04$\pm$                0.29&<               13.08&<                8.85&                --&                --&                -- \\
      9&               16.29$\pm$                0.05&               15.98$\pm$                0.12&<               12.21&<                9.02&                --&                --&                -- \\
     10&               16.63$\pm$                0.07&               16.52$\pm$                0.19&<               12.71&<                9.39&                --&                --&                -- \\
     11&               16.68$\pm$                0.06&               16.12$\pm$                0.12&<               12.55&<                9.05&                --&                --&                -- \\
     12&               14.02$\pm$                0.03&               13.75$\pm$                0.03&<               12.17&<                9.48&               16.00$\pm$                0.10&               15.43$\pm$                0.13&               14.60$\pm$                0.13 \\
\end{longtable}}

\onecolumn

\pagebreak

\onecolumn

\footnotesize{

\begin{longtable}{ccccccc}
    \caption{Optical magnitudes of counterparts of X-ray sources of 47 Tuc in Gaia and SkyMapper Southern Sky Survey.  \label{opt-count-table}}\tabularnewline
\hline
NO                  & \multicolumn{3}{c}{Gaia magnitudes\citep{2020yCat.1350....0G}}  &\multicolumn{2}{l}{SkyMapper magnitudes\citep{2018PASA...35...10W}} &Gaia distance \citep{2018AJ....156...58B}\\
& $G$ mag  & $G_{BP}$ mag & $G_{RP}$ mag & $g$ mag  &  $r$ mag& \\                        
& 300~nm & 400-500 nm & 600-750 nm & 467 nm & 616 nm & pc \\               
\hline
\endfirsthead
\multicolumn {7}{c}{Continued from previous page}\\
\hline 
 NO                  & \multicolumn{3}{c}{Gaia magnitudes\citep{2020yCat.1350....0G}}  &\multicolumn{2}{l}{SkyMapper magnitudes \citep{2018PASA...35...10W}} &Gaia distance \citep{2018AJ....156...58B}\\
& $G$ mag  & $G_{BP}$ mag & $G_{RP}$ mag & $g$ mag  &  $r$ mag& \\                        
& 300 nm  & 400-500 nm & 600-750 nm & 353-358 nm & 379-384 nm & pc \\   
\hline
\endhead
\hline
\multicolumn {7}{c}{Continued on the next page}\\

\endfoot

\hline
        4&              16.331$\pm$               0.001&              16.698$\pm$               0.004&              15.767$\pm$               0.003&               --&               --&   \\
           9&              17.392$\pm$               0.001&              17.728$\pm$               0.008&              16.882$\pm$               0.009&               --&               --&   \\
          30&              13.912$\pm$               0.000&              14.366$\pm$               0.001&              13.296$\pm$               0.001&               --&               --&   \\
          33&              17.319$\pm$               0.001&              17.870$\pm$               0.010&              16.621$\pm$               0.006&               --&               --&   \\
          41&              10.024$\pm$               0.010&              10.252$\pm$               0.030&               9.714$\pm$               0.030&              10.382$\pm$               0.003&              10.274$\pm$               0.003&   361.985$\pm$               2.893   \\
          45&              17.207$\pm$               0.006&              16.986$\pm$               0.017&              15.938$\pm$               0.010&               --&               --&   \\
          50&              17.067$\pm$               0.001&              17.392$\pm$               0.007&              16.558$\pm$               0.006&              17.640$\pm$               0.086&              17.061$\pm$               0.056&   \\
          71&               9.792$\pm$               0.001&              10.081$\pm$               0.002&               9.373$\pm$               0.001&               --&               --&  160.834$\pm$               2.976 \\
\hline
 
\end{longtable}
} 

\end{appendices}


\end{document}